\documentclass[a4paper,11pt]{article}

\usepackage{jheppub}
\usepackage[T1]{fontenc}
\usepackage[utf8]{inputenc}
\usepackage{lineno}
\usepackage{booktabs}
\usepackage{tabularx}
\usepackage{tabu}
\usepackage{tabularray}
\usepackage{multirow}
\usepackage{multicol}
\usepackage{longtable}
\usepackage{makecell}
\usepackage{xspace}
\usepackage{wrapfig}
\usepackage[new]{old-arrows}

\newcolumntype{D}{>{\centering \hsize=.2\hsize \arraybackslash}X}
\newcolumntype{C}{>{\centering \hsize=.4\hsize \arraybackslash}X}

\newcolumntype{?}{!{\vrule width 1.2pt}}

\usepackage{amsmath}
\usepackage{amsthm}
\usepackage{amsfonts}
\usepackage{slashed}
\usepackage{mathtools}

\usepackage{graphicx}
\usepackage{caption}
\usepackage{subcaption}
\usepackage{tikz}
\usepackage[compat=1.1.0]{tikz-feynman}
\usepackage{tikz-3dplot}
\usepackage{axodraw2}
\usepackage{verbatim}
\usepackage{slashed}
\usepackage{array}
\usepackage{enumitem}
\usepackage{algorithm2e}

\SetKwComment{Comment}{/* }{ */}
\RestyleAlgo{ruled}

\usetikzlibrary{calc}
\usetikzlibrary{arrows.meta}
\usetikzlibrary{decorations.markings}
\usetikzlibrary{decorations.pathmorphing}
\usetikzlibrary{shapes.geometric}
\pgfdeclarelayer{nodelayer}
\pgfdeclarelayer{edgelayer}
\pgfsetlayers{main,edgelayer,nodelayer}
\tikzstyle{none}=[]

\tikzstyle{blank}=[fill=white, shape=circle, draw=white, inner sep=0.8pt]
\tikzstyle{dot}=[fill=black, shape=circle, draw=black, inner sep=0.8pt]
\tikzstyle{fat}=[fill=blue, shape=circle, draw, line width=1pt, inner sep=0.8pt, minimum size=4mm]
\tikzstyle{blob}=[fill=gray, shape=circle, draw=black, line width=1pt, inner sep=0.8pt, minimum size=1cm]
\tikzstyle{star}=[fill=black, shape=star, inner sep=0.8pt, minimum size=5mm]
\tikzstyle{arrow}=[{-{Classical TikZ Rightarrow[length=2mm,width=1.5mm]}}, draw={rgb,255: red,61; green,171; blue,83}, line width=1pt, preaction={{draw=white,line width=2pt}}, line cap=rect]

\tikzstyle{gluoncoil}=[-, decorate, decoration={coil,aspect=1.4,segment length=2.5mm}]
\tikzstyle{fermion}=[-, draw={rgb,255: red,176; green,36; blue,39}, line width=1pt, preaction={{draw=white,line width=2pt}}, line cap=rect, postaction=decorate, decoration={markings,mark=at position .60 with {\arrow{Stealth[round,width=5pt]}}}]
\tikzstyle{scalar}=[-, line width=1pt, dashed, draw]
\tikzstyle{fermionarrow}=[-, postaction=decorate, decoration={markings,mark=at position .60 with {\arrow{Stealth[round,width=5pt]}}}]
\tikzstyle{ct scalar}=[-, line width=1pt, dashed, draw={rgb,255: red,102; green,102; blue,102}, postaction=decorate, decoration={markings,mark=at position .50 with {\node[style=star,minimum size=5mm]{};}}]
\tikzstyle{ct fermion}=[-, line width=1pt, preaction={{draw=white,line width=2pt}}, line cap=rect, postaction=decorate, decoration={markings,mark=at position 0.25 with {\arrow{Stealth[round,width=5pt]}}, mark=at position 0.75 with {\arrow{Stealth[round,width=5pt]}}, mark=at position .50 with {\node[style=star,minimum size=5mm]{};}}]

\newcommand{\pysecdec}{py{\textsc{SecDec}}\xspace}
\newcommand{\cuda}{{\textsc{Cuda}}\xspace}

\title{\boldmath Positive Integrands from Feynman Integrals in the Minkowski Regime}

\author[a]{S. P. Jones,}
\author[b]{A. Olsson,}
\author[a]{T. W. Stone}
\affiliation[a]{Institute for Particle Physics Phenomenology, Durham University, South Road, Durham DH1 3LE, U.K.}
\affiliation[b]{Institute for Theoretical Physics, Karlsruhe Institute of Technology (KIT), Wolfgang-Gaede-Str. 1, 76131 Karlsruhe, Germany}

\emailAdd{stephen.jones@durham.ac.uk}
\emailAdd{anton.olsson@kit.edu}
\emailAdd{thomas.w.stone@durham.ac.uk}

\preprint{{\small  KA-TP-20-2025, IPPP/25/41, P3H-25-046}}

\abstract{
We present a method for rewriting dimensionally regulated Feynman parameter integrals in the Minkowski regime as a sum of real, positive integrands multiplied by complex prefactors. 
This representation eliminates the need for contour deformation, allowing for direct numerical or analytic evaluation of the integrals. 
We develop an algorithm to construct such representations for a broad class of integrals and demonstrate its generalisation through selected examples. 
Our approach is applied to integrals up to three loops, including cases with internal masses and off-shell external legs. 
The resulting expressions are suitable for evaluation using existing techniques, such as sector decomposition, where we observe performance gains of up to four orders of magnitude in certain cases.
}

\begin{document}
\maketitle
\flushbottom
\allowdisplaybreaks

\section{Introduction}
\label{sec:introduction}
Knowledge of scattering amplitudes is critically important for producing precise theoretical predictions for observables at current high-energy particle colliders (such as the LHC, HL-LHC~\cite{Huss:2025nlt}) and future colliders including FCC-ee, FCC-hh, CLIC, CEPC and ILC.
Often, a challenging step in the computation of multi-loop amplitudes is the evaluation of Feynman integrals.

Interest in the study of Feynman integrals in parameter space has undergone somewhat of a revival in recent years, leading to new results and research directions including integration-by-parts reduction in parameter space (parametric annihilators)~\cite{Bitoun:2017nre,Chen:2019mqc,Chen:2019fzm,Chen:2020wsh,Artico:2023jrc}, counting of master integrals (Euler characteristics)~\cite{Smirnov:2010hn,Bitoun:2018afx}, syzygy relations in parameter space~\cite{Agarwal:2020dye}, Landau discriminants~\cite{Mizera:2021icv,Fevola:2023kaw,Fevola:2023fzn}, all order studies of UV/IR divergences~\cite{Arkani-Hamed:2022cqe} and finite integrals~\cite{vonManteuffel:2014qoa,Gambuti:2023eqh,delaCruz:2024xsm}, singularities of integrals~\cite{Helmer:2025ljj,Arkani-Hamed:2022cqe}, generalised hypergeometric systems~\cite{Chestnov:2022alh}, the analytic continuation/threshold structure of the S-matrix~\cite{Mizera:2021fap, Hannesdottir:2022bmo, Mizera:2023tfe}, graphical methods for the expansion of Feynman integrals~\cite{Gardi:2022khw,Gardi:2024axt}, and efficient methods for numerically computing integrals in parameter space using tropical geometry (FeynTrop)~\cite{Borinsky:2020rqs,Borinsky:2023jdv}.

In parameter space, one salient aspect of Feynman integrals is the UV and IR properties originating from behaviour near to boundaries of integration.
These properties can be captured using tools from algebraic/tropical geometry, for example Newton polytopes~\cite{Bogner:2007cr,Smirnov:2008aw,Kaneko:2009qx,Kaneko:2010kj,Schlenk:2016epj,Heinrich:2021dbf}.
However, the behaviour and analytic properties of Feynman integrals near to physical thresholds is also of great interest.
At and above physical thresholds, poles are present on the positive real axis of the Feynman parameters.
In the context of the direct evaluation of Feynman integrals in parameter space, a ubiquitous procedure for handling these poles is that of contour deformation~\cite{Soper:1998ye,Soper:1999xk,Binoth:2005ff,Nagy:2006xy,Anastasiou:2006hc,Anastasiou:2007qb,Lazopoulos:2007ix,Lazopoulos:2007bv,Anastasiou:2008rm,Gong:2008ww,Becker:2010ng,Mizera:2021icv,Hannesdottir:2022bmo,Borinsky:2023jdv}.
Various explicit implementations of contour deformation are utilised in numerical packages for the evaluation of Feynman integrals including \pysecdec~\cite{Borowka:2014aaa,Borowka:2015mxa,Borowka:2017idc,Borowka:2018goh,Jahn:2020tpj,Heinrich:2021dbf,Heinrich:2023til} and FIESTA~\cite{Smirnov:2013eza,Smirnov:2015mct,Smirnov:2021rhf}, which rely on sector decomposition~\cite{Binoth:2000ps} for the resolution of singularities near the integration boundary, and FeynTrop~\cite{Borinsky:2023jdv}, which uses tropical geometry.
Alternatively, the entire contour can be shifted away from the real axis by a small amount, and an extrapolation back to the physical value can be performed~\cite{deDoncker:2004bf,Yuasa:2011ff,deDoncker:2017gnb,Baglio:2020ini,deDoncker:2024lmk}.
One downside of the first approach is that it can drastically increase the complexity of the resulting integrands by deforming them into the complex plane, introducing an associated Jacobian determinant.
Furthermore, both approaches are arbitrary, depending either on the exact contour chosen and/or the values of the small deformation parameter.
These approaches are also very sensitive to end-point and/or pinch singularities, which trap the contour against the integration boundaries or between poles within the integration domain, leading to large numerical fluctuations — a situation often encountered when the numerical values of kinematic invariants or masses span several orders of magnitude.
Finally, methods depending on contour deformation can fail for integrals with a leading Landau singularity within the domain of integration~\cite{Gardi:2024axt}.

Methods have been explored to remove the need for contour deformation for momentum-space integrals, for example, in the context of loop-tree duality \cite{Catani:2008xa,Capatti:2019edf,Kermanschah:2021wbk,Kermanschah:2024utt}, or to minimise its impact~\cite{Pittau:2021jbs,Pittau:2024ffn}, it is a natural question to ask whether it is possible to do the same in Feynman parameter space (see Ref.~\cite{Binoth:2002xh} for an early approach).
In this work, we focus on the vanishing of the second Symanzik $\mathcal{F}$-polynomial within the domain of integration of the Feynman parameters~\cite{Jones:2024gmw}.
We view this not as isolated poles along the real axes but instead as a codimension-1 hypersurface in the space of the Feynman parameters.
We then consider the set of integrals that span the positive orthant obtained by taking the co-ordinate axes and the $\mathcal{F}=0$ hypersurface as a boundary of integration\footnote{A subset of the resulting integrals correspond to generalised cuts~\cite{Britto:2023rig}.}.
We show that, for a wide variety of integrals, it is possible to straightforwardly obtain the parameter space integrands associated to these resolved integrals and provide an algorithm for one class of integrals.
We demonstrate that the resulting integrals are purely real and can be evaluated entirely without contour deformation.
Selected massless and massive Feynman integrals up to 3-loops are considered, focusing on examples with on-shell and off-shell external legs, massless and massive internal lines, non-planar integrals and elliptic integrals.
Specific considerations relevant to the resolution of each integral are discussed.
For many such integrals, the resolved representation is dramatically simpler to integrate, resulting in an acceleration of between 1 and 4 orders of magnitude.
Although we present an example of the numerical performance benefits, which are significant, the goal of this paper is not to find the most efficient possible resolution, nor do we explore optimisations of our numerical implementation. 
We anticipate that further significant performance optimisations are possible, relying on the specific properties of the resolved integrands.

In Section~\ref{sec:preliminaries}, we introduce our notation for parameter space integrals and provide a brief review of existing approaches involving contour deformation.
In Section~\ref{sec:method}, we describe the general resolution procedure and provide an algorithm applicable to a class of Feynman integrals.
In Sections~\ref{sec:examples} and ~\ref{sec:performance}, we demonstrate the application of our procedure to selected examples and numerically profile the result. In Section~\ref{sec:eval}, we discuss potential implications for the evaluation of parametric integrals in the Minkowski regime using existing tools.

\section{Preliminaries}
\label{sec:preliminaries}
In this section, we will briefly introduce our notation for Feynman integrals in Feynman parameter space.
We will review the singularities present in dimensionally regulated parameter integrals and summarise the method of contour deformation, which is typically used to evaluate parameter integrals outside the Euclidean regime. 

\subsection{Integrands in Parameter Space}
\label{ssec:parameter}
An $L$-loop dimensionally regularised Feynman integral with $D=4-2\epsilon$ can be written in terms of Feynman parameters as,
\begin{equation}
J(\mathbf{s})=\frac{\left(-1\right)^{\nu}\Gamma\left(\nu-L D/2\right)}{\prod_{i=1}^{N}\Gamma\left(\nu_{i}\right)}\lim_{\delta\to0^+}\!\int_{\mathbb{R}_{\geq0}^{N}}\prod\limits_{i=1}^{N}\mathrm{d}x_{i} x_{i}^{\nu_{i}-1}\frac{\mathcal{U}\!\left(\mathbf{x}\right)^{\nu-(L+1)D/2}}{\left(\mathcal{F}\!\left(\mathbf{x};\mathbf{s}\right)-i\delta\right)^{\nu-LD/2}}\delta(1-\alpha(\mathbf{x})),
    \label{eq:fp2}
\end{equation}
where $\mathcal{U}(\mathbf{x})$ and $\mathcal{F}(\mathbf{x};\mathbf{s})$ are homogeneous polynomials in the Feynman parameters, $\mathbf{x}=(x_1,\ldots,x_N)$, with degree $L$ and $L+1$, respectively.
We denote propagator powers by $\nu_i$ and their sum by $\nu = \sum_{i=1}^N \nu_i$.
The integral depends on external kinematic invariants, $s_j$, and internal masses, $m_i$, via $\mathbf{s}=(s_1,\ldots,s_M,m^2_1,...,m^2_N)$.
The argument of the Dirac delta function defines a hyperplane that bounds the integral in the positive domain for at least one $x_i \geq 0$, see Refs.~\cite{Cheng:1987ga,Panzer:2015ida,Weinzierl:2022eaz}.
Common choices of the function $\alpha(\mathbf{x})$ include the $N$-dimensional simplex $\alpha(\mathbf{x}) = \sum_{i=1}^N x_i$, or lower dimensional simplices $\alpha(\mathbf{x}) = \sum_S x_i$, with $S$ a non-empty subset of the Feynman parameters.
The parameter $\delta>0$ appearing in the denominator of Eq.~\eqref{eq:fp2} imposes the causal Feynman propagator prescription.

The functions $\mathcal{U}(\mathbf{x})$ and $\mathcal{F}(\mathbf{x};\mathbf{s})$ are the first and second Symanzik polynomials, respectively.
They can be constructed from the spanning trees of a Feynman graph as,
\begin{align}
\mathcal{U}(\mathbf{x}) &=\sum_{T^1}^{}\prod_{e\notin T^1}^{} x_e, \qquad \mathcal{F}_0(\mathbf{x};\mathbf{s}) = \sum_{T^2}^{} (-s_{T^2}^{}) \prod_{e\notin T^2}^{} x_e, \label{eq:uf} \\
\mathcal{F}(\mathbf{x};\mathbf{s}) & = \mathcal{F}_0(\mathbf{x};\mathbf{s}) +\mathcal{U}(\mathbf{x})\sum_{e}^{}m_e^2 x_e,& & & 
\end{align}
where the notation $T^1$ and $T^2$ denotes a spanning tree and a spanning 2-tree of the graph $G$, respectively. The symbol $s_{T^2}^{}$ represents the square of the total momentum flowing between the components of the spanning 2-tree $T^2$.

From Eq.~\eqref{eq:uf}, we observe that $\mathcal{U}(\mathbf{x})$ and $\mathcal{F}_0(\mathbf{x};\mathbf{s})$ are at most linear in any single Feynman parameter, while $\mathcal{F}(\mathbf{x};\mathbf{s})$ can be linear or quadratic
depending on whether the associated propagator is massless or massive.
Furthermore, we note that $\mathcal{U}(\mathbf{x})$ consists purely of positive monomials of Feynman parameters with positive coefficients, it is therefore manifestly positive in $\mathbb{R}_{>0}^{N}$.
The $\mathcal{F}(\mathbf{x};\mathbf{s})$ polynomial is built from monomials with both positive and negative sign depending on the value of the invariants $\mathbf{s}=\{s_{T^2}, m_e^2\}$.

Both $\mathcal{U}(\mathbf{x})$ and $\mathcal{F}(\mathbf{x};\mathbf{s})$ may vanish on the intersection of the hyperplane defined by the Dirac-delta function with the coordinate hyperplanes, or, equivalently, for subsets of $\mathbf{x}$ vanishing.
Depending on the propagator powers, $\nu$, and the space-time dimension, $D$, this can give rise to UV and IR divergences.
When the invariants $\mathbf{s}$ are chosen such that $\mathcal{F}(\mathbf{x};\mathbf{s})$ contains monomials with different signs, the  $\mathcal{F}(\mathbf{x};\mathbf{s})$ polynomial can additionally vanish and/or change sign on hypersurfaces within the integration domain, $\mathbb{R}_{>0}^{N}$.
The vanishing of $\mathcal{F}(\mathbf{x};\mathbf{s})$ within the integration domain is usually 
associated with kinematic (pseudo-)thresholds.

We distinguish between a \textit{manifestly same-sign kinematic regime}, where all monomials in $\mathcal{F}(\mathbf{x};\mathbf{s})$ share the same sign and a more general \textit{same-sign regime}\footnote{The terms \textit{Euclidean} and \textit{pseudo-Euclidean} are sometimes used in the literature to refer to this regime.}, where $\mathcal{F}(\mathbf{x};\mathbf{s})$
retains a definite sign throughout the integration domain, $\mathbb{R}_{>0}^{N}$, despite individual
monomials having mixed signs.
In the most general situation $\mathcal{F}(\mathbf{x};\mathbf{s})$ vanishes on hypersurfaces in $\mathbb{R}_{>0}^{N}$ and has no fixed definite sign in the domain of integration, we will refer to this as a \textit{mixed-sign (or Minkowski) regime}.

\subsection{Landau Equations}
\label{ssec:landau}

The necessary, but not sufficient, conditions for a dimensionally regulated parameter integral to have a singularity are described by the Landau equations~\cite{Landau:1959fi,Bjorken:1959fd,10.1143/PTP.22.128,Cutkosky:1960sp,Coleman:1965xm}. In parameter space they can be written as~\cite{Eden:1966dnq},
\begin{equation}
    \mathcal{F}(\mathbf{x};\mathbf{s}) = 0, \qquad x_k \frac{\partial \mathcal{F}(\mathbf{x};\mathbf{s})}{\partial x_k} = 0, \quad \text{for all } k \in \{1,\dots,N\}.
    \label{eq:landau}
\end{equation}

As discussed above, the $\mathcal{F}$-polynomial can vanish and potentially give rise to singularities when some subset of Feynman parameters vanish, $x_i \rightarrow 0$, i.e. at the boundary of the integration domain, causing each individual monomial in $\mathcal{F}$ to vanish.
Such singularities occur independently of the sign of the individual monomials in $\mathcal{F}$ and can be identified algorithmically, for example, by using the geometric approach to sector decomposition~\cite{Bogner:2007cr,Smirnov:2008aw,Kaneko:2009qx,Kaneko:2010kj,Schlenk:2016epj,Heinrich:2021dbf} in which they appear as facets of a Newton polytope.

The cases where both $\mathcal{F}(\mathbf{x};\mathbf{s})$ and all $\partial \mathcal{F}(\mathbf{x};\mathbf{s})/\partial x_k$ vanish away from the boundary of the integration domain can occur when monomials of different sign are present in $\mathcal{F}(\mathbf{x};\mathbf{s})$ as well as all partial derivatives w.r.t. the Feynman parameters.
In these cases the monomials can cancel against each other within the domain of integration and lead to solutions of the Landau equations.

Significant progress in analysing and solving the Landau equations has been made recently, see for example Refs.~\cite{Brown:2009ta,Panzer:2014caa,Fevola:2023kaw,Fevola:2023fzn,Dlapa:2023cvx,Jiang:2024eaj,Helmer:2024wax,Caron-Huot:2024brh,Hannesdottir:2024cnn,Hannesdottir:2024hke,He:2024fij,Correia:2025yao}.
In Ref.~\cite{Gardi:2024axt}, the analysis of a set of solutions of the Landau equations in parameter space, which are present for generic kinematics, led to the idea that integrals can be split or \textit{dissected} on hypersurfaces which solve the Landau equations within the integration domain.
The Landau solutions are mapped to the boundaries of the new integration
domains, making all singularities detectable using geometric decomposition.
In this work, we will explore this idea in a slightly different context, instead focusing on the hypersurface defined by $\mathcal{F}(\mathbf{x};\mathbf{s})=0$ (the variety of $\mathcal{F}$), but not requiring that all partial derivatives vanish.
Generally, such hypersurfaces are not solutions of the Landau equations and therefore do not give rise to (non-spurious) dimensionally regulated singularities of the Feynman integral. However, as described in Section~\ref{ssec:contourdef}, these varieties do introduce a substantial (and computationally expensive) obstacle.

\subsection{Contour Deformation}
\label{ssec:contourdef}

When evaluating integrals in the mixed-sign (Minkowski) regime, the $\mathcal{F}(\mathbf{x};\mathbf{s})$ polynomial can vanish within the domain of integration due to cancellation between monomials.
When this occurs, the $i\delta$ appearing in Eq.~\eqref{eq:fp2} acts as a deformation of the integration contour into the complex-plane and provides a prescription for evaluating the integral in a theory with causal propagators.
In the mixed-sign regime, the integral can become complex valued. The values of the kinematic invariants, $\mathbf{s}$, at which the integral transitions from a same-sign to a mixed-sign regime is sometimes called a \textit{threshold} or \textit{pseudo-threshold}. Such points can correspond, for example, to internal propagators becoming on-shell.

In parameter space, for values of the kinematics below all physical thresholds, the variety of $\mathcal{F}$ does not enter the positive orthant $\mathbb{R}_{>0}^N$ but may touch the boundary of integration for some $x_i \rightarrow 0$.
At a physical threshold, the variety of $\mathcal{F}$ enters the positive orthant; often, on the surface defined by the argument of the Dirac delta function, it does this either as an endpoint singularity (moving into the domain close to a boundary where some $x_i \rightarrow 0$) or as a pinch singularity (appearing inside the domain not necessarily close to a boundary).
At a threshold, the Landau equations of Eq.~\eqref{eq:landau} are satisfied.
For values of the kinematic invariants above this threshold, the typical situation is that the contour is no longer pinched or trapped against a boundary of integration, but solutions of $\mathcal{F}(\mathbf{x};\mathbf{s})=0$ are still present in the positive orthant, this is the Minkowski regime.
The $i\delta$ prescription defines an analytic continuation of the integral and specifies the branch on which multivalued functions, such as square-roots and logarithms which result from the integration over the Feynman parameters, should be evaluated.

The location (and shape) of the variety of $\mathcal{F}(\mathbf{x};\mathbf{s})$ (the hypersurface defined by $\mathcal{F}(\mathbf{x};\mathbf{s})=0$) depends on the signs and magnitudes of the kinematic invariants, $\mathbf{s}$.
In the same-sign regime, the variety of $\mathcal{F}(\mathbf{x};\mathbf{s})$ is located in the non-positive domain of $\mathbf{x}$ and only touches the integration domain when some $x_i$ vanish.
In the mixed-sign regime, the variety enters the domain of integration also away from the integration boundary.
To correctly define the causal integral, the integration contour can be deformed into the complex-plane by shifting each Feynman parameter by a vanishingly small imaginary part $i\tau_k$:
\begin{equation}
    x_k \to z_k = x_k - i\tau_k.
\label{eq:deformation-shift}
\end{equation}
This shift results in the $\mathcal{F}$-polynomial transforming as,
\begin{equation}
    \mathcal{F}(\mathbf{x};\mathbf{s}) \to \mathcal{F}(\mathbf{z};\mathbf{s}) = \mathcal{F}(\mathbf{x};\mathbf{s}) - i\sum_k \tau_k \frac{\partial \mathcal{F}(\mathbf{x};\mathbf{s})}{\partial x_k} + \mathcal{O}(\tau^2).
\label{eq:contour-deformation}
\end{equation}
The individual $\tau_k$ can then be chosen in accordance with the causal $i\delta$ prescription, i.e. to ensure that the deformed $\mathcal{F}(\mathbf{z};\mathbf{s})$ develops a negative imaginary part where $\mathcal{F}(\mathbf{x};\mathbf{s})$ vanishes.
If the $\tau_k$ are sufficiently small, the original and deformed contours can be
connected without enclosing any poles, ensuring by Cauchy’s theorem that the
deformation leaves the integral invariant.
An example of a choice of shift parameters that achieves all of the above (in the integration domain $x_i \in [0,1]$) is given by~\cite{Soper:1998yp,Soper:1999xk,Binoth:2002xh,Binoth:2005ff,Nagy:2006xy,Anastasiou:2006hc,Anastasiou:2007qb,Lazopoulos:2007ix,Lazopoulos:2007bv,Anastasiou:2008rm,Gong:2008ww,Becker:2010ng,Borowka:2014aaa},
\begin{equation}
    \tau_k = \lambda_k \, x_k (1-x_k) \frac{\partial \mathcal{F}(\mathbf{x})}{\partial x_k},
\label{eq:shift-example}
\end{equation}
where $\lambda_k$ are arbitrary parameters chosen small enough that the loop does not enclose any poles.
Inserting the deformation into Eq.~\eqref{eq:contour-deformation} gives,
\begin{equation}
    \mathcal{F}(\mathbf{z};\mathbf{s}) = \mathcal{F}(\mathbf{x};\mathbf{s}) - i\sum_k \lambda_k \, x_k (1-x_k) \left( \frac{\partial \mathcal{F}(\mathbf{x};\mathbf{s})}{\partial x_k} \right)^2 + \mathcal{O}(\tau^2).
\end{equation}
For sufficiently small $\lambda_k$, such that we can neglect terms of $\mathcal{O}(\tau^3)$ which contribute to the imaginary part with a positive sign, this choice gives a negative imaginary part to $\mathcal{F}(\mathbf{z};\mathbf{s})$ except at $x_k = 0$ and $x_k=1$ (chosen as the boundary of integration) and where all $\partial\mathcal{F}(\mathbf{x};\mathbf{s})/\partial x_k$ vanish.
Note that Eq.~\eqref{eq:contour-deformation} suggests an interpretation of the Landau equations in the situation where $\mathcal{F}(\mathbf{x};\mathbf{s})$ and $\partial\mathcal{F}(\mathbf{x};\mathbf{s})/\partial x_k \ \forall\  k$ simultaneously vanish; in this case, the infinitesimal deformation of the integration contour into the complex-plane according to Eq.~\eqref{eq:deformation-shift} does not avoid the variety of $\mathcal{F}(\mathbf{x};\mathbf{s})$.

The advantage of the deformation procedure given in Eq.~\eqref{eq:deformation-shift} and Eq.~\eqref{eq:shift-example} is that the deformed integral is exactly equal to the original Feynman integral.
However, a significant drawback of this procedure is that for certain kinematic configurations the integrand can become highly oscillatory with a large positive contribution from one part of the contour cancelled by a large negative contribution from elsewhere along the contour.
Furthermore, the change of variables from $\mathbf{x} \rightarrow \mathbf{z}$ introduces an $(N-1) \times (N-1)$ Jacobian determinant (after integrating out the projective Dirac $\delta$-functional) depending in a non-trivial way on the integration variables and kinematics.
In high-dimensional cases, this Jacobian can be significantly more complicated than the original integrand.
Furthermore, the contour depends on the arbitrary parameters $\lambda_k$ and it is not always trivial to pick valid values for which the variance of the integrand, which depends on the choice of $\lambda_k$, remains small, see e.g. Ref.~\cite{Winterhalder:2021ngy}.
An additional issue arises for certain integrals, such as the crown integral studied in Section~\ref{sec:crown}, which has a leading Landau singularity for generic values of the kinematics.
In this case, deforming the contour of integration fails to give a result entirely, as the deformation of the contour (which is proportional to $\partial\mathcal{F}(\mathbf{x};\mathbf{s})/\partial x_k$) vanishes for $\mathcal{F}(\mathbf{x};\mathbf{s})=0$, precisely where it is needed, leading to a pinched contour.

Alternatively, the Feynman integral can be directly evaluated with the $\delta$ of the $i\delta$ prescription set to several small positive (non-zero) numbers.
The correct result can then be obtained by performing an extrapolation of these evaluations to $\delta \rightarrow 0^+$, see Refs.~\cite{deDoncker:2004bf,Yuasa:2011ff,deDoncker:2017gnb,Baglio:2020ini,deDoncker:2024lmk}.
The drawback of this approach is that the value of the integral with non-zero $\delta$ only approximates the true result and any error due to the extrapolation to $\delta \rightarrow 0^+$ needs to be carefully assessed.
Furthermore, the value of $\delta$ must always be chosen small enough that no additional poles are crossed in the complex-plane, this can force the integration contour close to the singular surface making the integrand oscillatory. 

In the present manuscript, we describe an alternative strategy for evaluating Feynman integrals in parameter space in the mixed-sign regime entirely without using a contour deformation.

\section{Method}
\label{sec:method}

In this section, we propose a procedure for evaluating Feynman parametrised integrals in the mixed-sign (Minkowski) regime, i.e. where $\mathcal{F}(\mathbf{x};\mathbf{s})$ has no definite sign for $\mathbf{x} \in \mathbb{R}_{>0}^{N}$, without deforming the integration contour into the complex-plane.
After introducing the general concept in Section~\ref{ssec:method_overview}, we present a constructive algorithm valid for a specific class of Feynman integrals in Section~\ref{ssec:method_ub}.
We show that when this form can be achieved, the resulting integrands are strictly non-negative and the analytic continuation of the resulting integrals becomes trivial.
In Section~\ref{ssec:method_beyond_ub} we briefly discuss the application of the concept to general Feynman integrals.
Examples which can be solved with the method of Section~\ref{ssec:method_ub} and using generalisations of the algorithm are presented in Section~\ref{sec:examples}.

\subsection{Overview}
\label{ssec:method_overview}

The central idea of this work is to construct transformations of the Feynman parameters such that the variety of the $\mathcal{F}$-polynomial (the hypersurface defined by $\mathcal{F}(\mathbf{x};\mathbf{s})=0$) is mapped to the boundary of the integration domain\footnote{This idea is inspired by the techniques introduced in Ref.~\cite{Gardi:2024axt} to address leading Landau singularities. In that work, integrals were dissected in such a way that the solutions of the Landau equations, i.e. derivatives of $\mathcal{F}$, were mapped to integration boundaries.}.
After this, $\mathcal{F}$ only vanishes on the integration boundary. Any singularities resulting from this can be algorithmically dealt with using existing methods, such as sector decomposition.

The procedure involves splitting the integration domain into regions where $\mathcal{F}>0$ and $\mathcal{F}<0$ and integrating these regions separately. 
We may also further sub-divide the positive and negative regions for technical or computational ease (see the massive examples in Section~\ref{ssec:massive}). 
For the regions where $\mathcal{F}$ is negative, we factor out a minus sign from $\mathcal{F}$, ensuring the Feynman $i\delta$-prescription is respected; after this, we will have only non-negative integrands. 
The resulting general decomposition of the original Feynman integral, $J(\mathbf{s})$, is
\begin{equation}
    J(\mathbf{s})=\sum_{n_+=\ \!\!1}^{N_+}J^{+,n_+}(\mathbf{s})+\lim_{\delta\to0^+}\left(-1-i\delta\right)^{-\left(\nu-LD/2\right)}\sum_{n_-=\ \!\!1}^{N_-}J^{-,n_-}(\mathbf{s}),
    \label{eq:decomp}
\end{equation}
where we have allowed for both the positive and negative regions to be subdivided into $N_+$ and $N_-$ sub-regions respectively. 
After this decomposition, the imaginary part of the original integral $J(\mathbf{s})$ is fully determined by the contribution from the negative region(s) in Eq.~\eqref{eq:decomp} and, furthermore, is a result of the $\lim_{\delta\to0^+}\left(-1-i\delta\right)^{-\left(\nu-LD/2\right)} = \exp[i\pi (\nu-LD/2)]$ factor multiplying the purely real $J^-(\mathbf{s})$ contribution(s). 
We remark that, in many cases, the division can be chosen such that there is only a single positive and a single negative region (see the algorithm in Section~\ref{ssec:method_ub} and the massless examples in Section~\ref{ssec:massless}). 

It has long been known that for integrals in a kinematic regime where $\mathcal{F}$ is strictly non-positive in the integration domain, a minus sign can be factored out and the resulting non-negative (Euclidean) integrand can be evaluated. 
As a trivial example, consider the massless bubble in the physical $p^2=s>0$ regime,
\begin{align}
J_{\mathrm{bub},\ \!\!m=0}(\mathbf{s})=&\lim_{\delta\to0^+}\Gamma\left(\epsilon\right)\int_{\mathbb{R}^2_{\geq0}}\!\!\mathrm{d}x_1\mathrm{d}x_2\,\frac{\left(x_1+x_2\right)^{-2+2\epsilon}}{\left(-sx_1x_2-i\delta\right)^{\epsilon}}\,\delta(1-\alpha(\mathbf{x}))\nonumber\\
=&\lim_{\delta\to0^+}(-1-i\delta)^{-\epsilon} \,\Gamma\left(\epsilon\right)\int_{\mathbb{R}^2_{\geq0}}\!\!\mathrm{d}x_1\mathrm{d}x_2\,\frac{\left(x_1+x_2\right)^{-2+2\epsilon}}{\left(sx_1x_2\right)^{\epsilon}}\,\delta(1-\alpha(\mathbf{x}))\nonumber\\
=&\lim_{\delta\to0^+}(-1-i\delta)^{-\epsilon}J_{\mathrm{bub},\ \!\!m=0}^{-}(\mathbf{s}),
\label{negfactoring}
\end{align}
where, in the final line we have translated this simple manipulation into the language of our decomposition given in Eq.~\eqref{eq:decomp}. 
Our procedure essentially generalises this idea by mapping a generic integral into integrals which are already positive (where $\mathcal{F}>0$ originally) plus cases like this where $\mathcal{F}<0$ and we can factor out the prescription to generate positive integrands.

In order to construct the integrands appearing in Eq.~\eqref{eq:decomp}, we must ensure that our transformations do not spoil the non-negativity of $\mathcal{U}$ (nor introduce zeroes of $\mathcal{U}$ within the integration domain) as well as avoiding transformations with Jacobian determinants which break the positivity of the resulting integrand within (but not necessarily on the boundary of) the integration domain. Furthermore, we must check that the transformations applied do not miss any regions from the original integration domain as well as prohibiting transformations that map regions from outside the original integration domain into the new domain.

These demands may initially appear quite constraining and hence, one might assume that the transformations are potentially difficult to construct.
However, we will show that for a large number of massless integrals (a sample of which are detailed in Section~\ref{ssec:massless}), the procedure is algorithmic.
Additionally, geometric visualisations can prove extremely useful for building an understanding of how the transformation should appear, which aids its construction.
We adopt this approach on a case-by-case basis for the resolution of the massive integrals in Section~\ref{ssec:massive}.
In its current form, this approach suffers from the disadvantage that it is difficult to visualise beyond four propagators, although this does not present an in-principle obstacle to the concept.

The interplay between the positive and negative contributions in our dissection also allows us to obtain an understanding of the structure of the original integral from a new perspective.
For example, if we have a finite integral with a complex-valued leading order in the $\epsilon$ expansion, we must necessarily have a pole in $\epsilon$ in the total negative contribution which generates an imaginary $\epsilon^0$ term when multiplied with the expansion of $\left(-1-i\delta\right)^{-\left(\nu-LD/2\right)}$. 
In order for the full integral to be finite, the total positive contribution must have the exact same pole in $\epsilon$ such that the poles cancel in the full integral to leave a finite leading order.

\subsection{Algorithm for Univariate Bisectable Integrals}
\label{ssec:method_ub}

Here we describe an algorithmic procedure for resolving a class of Feynman integrals which we call \textit{univariate bisectable in} $\mathbf{s}_R$.
The algorithm will succeed in mapping an integral in a given mixed-sign (Minkowski) regime, $\mathbf{s}_R$, defined by a set of inequalities depending on external kinematics, to a single integral in which $\mathcal{F}(\mathbf{x};\mathbf{s})$ is non-negative and a single integral in which $\mathcal{F}(\mathbf{x};\mathbf{s})$ is non-positive if there is a single variable for which the $\mathcal{F}(\mathbf{x};\mathbf{s}) = 0$ hypersurface divides the integration domain in two. 
In Section~\ref{sec:examples} we discuss several non-trivial examples for which this algorithm succeeds and we also solve cases for which this algorithm is not sufficient.

We begin by considering a generic Feynman integral of the form Eq.~\eqref{eq:fp},
\begin{align}
 J(\mathbf{s}) &= \frac{\left(-1\right)^{\nu}\Gamma\left(\nu-L D/2\right)}{\prod_{i=1}^{N}\Gamma\left(\nu_{i}\right)} \lim_{\delta\to0^+} I(\mathbf{s};\delta), \label{eq:fp} \\
I(\mathbf{s};\delta) &= \int_{\mathbb{R}_{\geq0}^{N}} \mathrm{d}\mathbf{x} \  \mathcal{I}(\mathbf{x};\mathbf{s};\delta)\, \delta(1-\alpha(\mathbf{x}))= \int_{\mathbb{R}_{\geq0}^{N}}\prod\limits_{i=1}^{N} \frac{\mathrm{d}x_{i}}{x_i} x_{i}^{\nu_{i}}\frac{\mathcal{U}\!\left(\mathbf{x}\right)^{\nu-(L+1)D/2}}{\left(\mathcal{F}\!\left(\mathbf{x};\mathbf{s}\right)-i\delta\right)^{\nu-LD/2}} \delta(1-\alpha(\mathbf{x})). \label{feynparamint}
\end{align}
with $\mathrm{d} \mathbf{x} = \prod_{i=1}^N \mathrm{d} x_i$.
Our goal is to cast the integral into the form of Eq.~\eqref{eq:decomp} for a specific kinematic region.

We begin by defining some convenient notation, let $\mathbf{x} = \{x_1, \ldots, x_N\}$ be the complete set of Feynman parameters and $\mathbf{x}_{\neq i} = \mathbf{x}\setminus \{x_i\} = \{x_1, \ldots, x_{i-1},x_{i+1},\ldots x_N\}$ be the set excluding a single parameter $x_i$.
Let $\mathbf{s}_R=\{ \mathbf{s}_\mathrm{min} < \mathbf{s} < \mathbf{s}_\mathrm{max} \}$ be a kinematic region defined by a system of inequalities, this notation should be interpreted as placing minimum and maximum limits on each independent kinematic invariant/mass on which the integral depends (i.e. after applying momentum conservation to eliminate any dependent invariants).
The success of the algorithm depends both on the Feynman integral and on the choice of kinematic region, a conservative choice of input region would be from one (pseudo-)threshold to the next (pseudo-)threshold in each variable without crossing any intermediate thresholds.

\begin{algorithm}
\caption{Univariate Bisection (UB)}\label{alg:univariate}
\KwIn{$\mathcal{I}(\mathbf{x};\mathbf{s};\delta)$, $\mathbf{s}_R$}
\KwOut{$\mathcal{I}^{+}(\mathbf{x};\mathbf{s})$, $\mathcal{I}^{-}(\mathbf{x};\mathbf{s})$}
 \ForEach{$x_i \in \mathbf{x}$}{
  Let $r = \mathtt{Reduce}[ \{\mathcal{F}(\mathbf{x};\mathbf{s})<0\} \cup \{0<\mathbf{x}\} \cup \mathbf{s}_R, \,x_i]$\;
  \uIf{$r \sim$ \eqref{eq:ub_a}}{
  Let $\mathcal{I}^-(\mathbf{x};\mathbf{s}) = \mathcal{J}(\mathbf{x}_{\neq i},y_i)\  \mathcal{I}(\mathbf{x}_{\neq i},y_i;-\mathbf{s};0)$\\
  Let $\mathcal{I}^+(\mathbf{x};\mathbf{s}) = \mathcal{J}^\prime(\mathbf{x}_{\neq i},y_i^\prime)\  \mathcal{I}(\mathbf{x}_{\neq i},y_i^\prime;\mathbf{s};0)$ \\
  \Return{$\mathcal{I}^+(\mathbf{x};\mathbf{s})$, $\mathcal{I}^-(\mathbf{x};\mathbf{s})$}
   }
   \uElseIf{$r \sim$ \eqref{eq:ub_b}}{
  Let $\mathcal{I}^-(\mathbf{x};\mathbf{s}) = \mathcal{J}^\prime(\mathbf{x}_{\neq i},y_i^\prime)\  \mathcal{I}(\mathbf{x}_{\neq i},y_i^\prime;-\mathbf{s};0)$\\
  Let $\mathcal{I}^+(\mathbf{x};\mathbf{s}) = \mathcal{J}(\mathbf{x}_{\neq i},y_i)\  \mathcal{I}(\mathbf{x}_{\neq i},y_i;\mathbf{s};0)$ \\
  \Return{$\mathcal{I}^+(\mathbf{x};\mathbf{s})$, $\mathcal{I}^-(\mathbf{x};\mathbf{s})$}
   }
 }
\Return{$\neg \mathrm{UB}$ in $\mathbf{s}_R$}
\end{algorithm}

In Algorithm~\ref{alg:univariate} we state the univariate bisection procedure.
The input to the algorithm is the integrand of the Feynman integral to be resolved and the kinematic regime of interest.
The algorithm iterates over each Feynman parameter and attempts to find a valid bisection.
The $\mathtt{Reduce}$ procedure attempts to solve the system of inequalities in the variable $x_i$ and is implemented, for example, in the \texttt{Mathematica} computer algebra system.
The forms of the reduced system, $r$, for which a bisection is valid are given by either of 
\begin{align}
&\{0<x_i<f\left(\mathbf{x}_{\neq i}\right)\}& &\cup& &\{0< \mathbf{x}_{\neq i}\}& &\cup& &\mathbf{s}_R,& \label{eq:ub_a} \\
&\{f\left(\mathbf{x}_{\neq i}\right)<x_i\}& &\cup&   &\{0< \mathbf{x}_{\neq i}\}& &\cup& &\mathbf{s}_R,& \label{eq:ub_b}
\end{align}
where $f(\mathbf{x}_{\neq i})$ is a rational (if $\mathcal{F}(\mathbf{x};\mathbf{s})$ is linear in $x_i$) or, in general, an algebraic (if $\mathcal{F}(\mathbf{x};\mathbf{s})$ is quadratic in $x_i$) function with unit degree of homogeneity.
If a valid bisection can be found then we construct a transformation for the bisection parameter, $x_i$, that maps the variety to an integration boundary.
If the reduced system $r$ is of form Eq.~\eqref{eq:ub_a} then we map $\mathcal{F}=0$ to $x_i\rightarrow\infty$ while keeping the boundary at $x_i=0$ fixed.
If $r$ is of the form Eq.~\eqref{eq:ub_b} then we instead map the variety $\mathcal{F}=0$ to $x_i=0$ while keeping the boundary at $x_i\rightarrow\infty$ fixed. 
These mappings can be achieved by substituting $x_i$ with $y_i$ or $y_i^\prime$, given by
\begin{align}
y_i =&\, \frac{x_i}{x_i+  x_j}f\left(\mathbf{x}_{\neq i}\right), \label{eq:sub_a} \\
y_i^\prime =&\, x_i+f\left(\mathbf{x}_{\neq i}\right). \label{eq:sub_b}
\end{align}
These transformations are derived by solving $x_i'=\frac{x_ix_j}{f\left(\mathbf{x}_{\neq i}\right)-x_i}$ (for Eq.~\eqref{eq:sub_a}) and $x_i'=x_i-f\left(\mathbf{x}_{\neq i}\right)$ (for Eq.~\eqref{eq:sub_b}) for $x_i$ and then relabelling $x_i'$ back to $x_i$ post-transformation.
In the mapping of Eq.~\eqref{eq:sub_a} the variable $x_j \neq x_i$ appearing in the denominator is an arbitrary Feynman parameter.
The function $\mathcal{J}(\mathbf{x}_{\neq i},y_i)$ appearing in the algorithm is the Jacobian determinant resulting from the change of variables from $x_i$ to $y_i$ (similarly for $\mathcal{J}^\prime(\mathbf{x}_{\neq i},y_i^\prime)=1$).
When defining $\mathcal{I}^-(\mathbf{x};\mathbf{s})$ we factor a minus sign out of the $\mathcal{F}$-polynomial, we indicate this in our algorithm by calling $\mathcal{I}$ with argument $- \mathbf{s}$. This is valid since only $\mathcal{F}$ depends on the
kinematics and it is linear in the squared invariants.
If the algorithm succeeds then it will return the non-negative integrands $\mathcal{I}^+(\mathbf{x};\mathbf{s})$ and $\mathcal{I}^-(\mathbf{x};\mathbf{s})$, the result for the original integral is then given by,
\begin{align}
 J^\pm(\mathbf{s}) &= \frac{\left(-1\right)^{\nu}\Gamma\left(\nu-L D/2\right)}{\prod_{i=1}^{N}\Gamma\left(\nu_{i}\right)} \int_{\mathbb{R}_{\geq0}^{N}} \mathrm{d}\mathbf{x} \  \mathcal{I}^\pm(\mathbf{x};\mathbf{s}) \, \delta(1-\alpha^\pm(\mathbf{x})) \label{eq:pm_decomp}\\
    J(\mathbf{s}) &= J^{+}(\mathbf{s})+\lim_{\delta\to0^+}\left(-1-i\delta\right)^{-\left(\nu-LD/2\right)} J^{-}(\mathbf{s}).
    \label{eq:ub_decomp}
\end{align}
If the algorithm fails then the integral is not univariate bisectable in the kinematic region $\mathbf{s}_R$, a different (more restrictive) choice of region may be necessary or the structure of integral itself may prevent any univariate bisection from being obtained, we will comment on this case further in Section~\ref{ssec:method_beyond_ub}.

Several aspects of the above algorithm are arbitrary and the
outcome often depends on the exact ordering of Feynman parameters in $\mathbf{x}$.
Firstly, there may be multiple possible bisection parameters for a given Feynman integral; in order to obtain the simplest possible integrands, it may be beneficial to examine multiple resolutions with different bisection parameters.
It is sometimes even possible to select different bisection parameters for constructing $\mathcal{I}^-(\mathbf{x};\mathbf{s})$ and $\mathcal{I}^+(\mathbf{x};\mathbf{s})$.
Secondly, the $x_j$ appearing in Eq.~\eqref{eq:sub_a} can be replaced by a more general function, e.g. a constant, however, we prefer the transformation to be homogeneous such that the resulting integrand retains its homogeneity. 
Any valid linear function of the Feynman parameters would also achieve this.

The univariate bisection algorithm is agnostic to the choice of hypersurface $\alpha(\mathbf{x})$ in the $\delta$-function of Eq.~\eqref{feynparamint}, such that a choice of $\alpha^\pm(\mathbf{x})$ in Eq.~\eqref{eq:pm_decomp} can be made post-resolution. However, we must ensure transformations are constructed such that some initial $\delta$-functional that bounds the integral in the positive $\mathbf{x}$ domain can be chosen.
If the homogeneous transformation given in Eq.~\eqref{eq:sub_a} is used, the integral remains projective and the existence of a valid initial $\delta$-functional is automatically guaranteed (distinct choices of $\alpha^+(\mathbf{x})$ and $\alpha^-(\mathbf{x})$ may then be made, if so desired).
This follows by viewing the original integral as defined over $\mathbb{RP}^{N-1}_{\geq 0}$ with the regions induced by bisection interpreted as projective subsets of $\mathbb{RP}^{N-1}_{\geq 0}$. This structure is preserved because the regions are bounded by coordinate hyperplanes and the projective variety defined by the zero locus of the homogeneous $\mathcal{F}$-polynomial. After being mapped back to $\mathbb{RP}^{N-1}_{\geq 0}$ via homogeneous transformations, each resulting region supports a well-defined projective integral which can be expressed as an integral over $\mathbb{R}^{N}_{\geq 0} $ with any standard choice of $ \delta $-functional constraint, such as the $N $-dimensional simplex or other admissible choices.
In the numerical profiling of the examples of Section~\ref{ssec:massless}, we will choose $\delta(1-x_N)$, with $x_N$ the last Feynman parameter.
In contrast to this, for the integrals which can not be bisected in a single variable, presented in Section~\ref{ssec:massive}, we will often make a specific choice of the $\delta$-functional and integrate over one variable before considering the dissection of the integral.

\subsection{Beyond Univariate Bisectable Integrals}
\label{ssec:method_beyond_ub}

In Section~\ref{ssec:method_ub}, we have presented a simple algorithm for resolving integrals using a bisection in only a single variable.
As we will show in Section~\ref{sec:examples}, the algorithm is sufficient to resolve various integrals at the 1- 2- and 3-loop level in kinematic regimes of physical interest. 
We see no particular obstruction of the algorithm related to the number of loops or legs present in an integral.
With this said, we have also encountered simple examples for which this algorithm fails, one such example is provided by the 2-loop 4-point \emph{planar} double-box integral in the regime $\mathbf{s}_R = \{ 0 < s_{12} < \infty, -s_{12} < s_{23} < 0 \}$, which is not univariate bisectable, unlike its non-planar cousin (called BNP7 in Section~\ref{sec:examples}) which is UB.
One immediate generalisation of Algorithm~\ref{alg:univariate}, which we have verified to work on several examples, is to iterate the bisection using several variables until the integral is resolved. 
The primary complication of this procedure is ensuring that the entire original integration domain is covered without any double-counting.
This complication motivates the general decomposition formula of Eq.~\eqref{eq:decomp} in which we do not specify the total number of positive and negative resolutions.

In Section~\ref{sec:examples}, we will examine Feynman integrals involving internal massive propagators.
In this case, the $\mathcal{F}(\mathbf{x};\mathbf{s})$ polynomial becomes quadratic in the Feynman parameters associated with the massive propagators, and it may not be possible to \emph{bisect} the Feynman integral in such parameters (although we are aware of massive examples which are UB with respect to a massive parameter).
We will focus on the study of individual integrals involving internal masses, including integrals known to be elliptic and hyperelliptic and we will show that despite the failure of the UB algorithm in these cases, the integrals can nevertheless be resolved at the expense of introducing, in general, \emph{algebraic} transformations related to the roots of a quadratic equation with coefficients generally depending on Feynman parameters and kinematic invariants.

The goal of the present work is to propose and demonstrate that a resolution of the form Eq.~\eqref{eq:decomp} is a desirable representation of a Feynman integral in some contexts and that such a resolution can be obtained for some non-trivial integrals (massless/massive/non-planar and with various underlying integral geometries).
A general solution to the problem of finding a resolution is given by a generic (or open) cylindrical algebraic decomposition~\cite{10.1007/3-540-07407-4_17}, for which constructive algorithms exist\footnote{We thank in particular Bakar Chargeishvili for his input to this discussion in the context of our ongoing collaboration focused on generalising/automating the resolution procedure.}.
In future work, we intend to investigate such general solutions as well as other, simpler algorithms which are sufficient for various classes of Feynman integrals.

\section{Examples}
\label{sec:examples}
\subsection{Massless Examples}
\label{ssec:massless}
In this section, we provide examples of massless integrals which are resolved by the algorithm presented in Section~\ref{ssec:method_ub}. 
We show that this procedure can be successfully applied to several integrals with up to 3 loops (including non-planar integrals) and up to 5 legs. 
For pedagogical reasons, we apply each step of the algorithm in detail to a simple massless box with on-shell legs before presenting the remaining examples more succinctly.

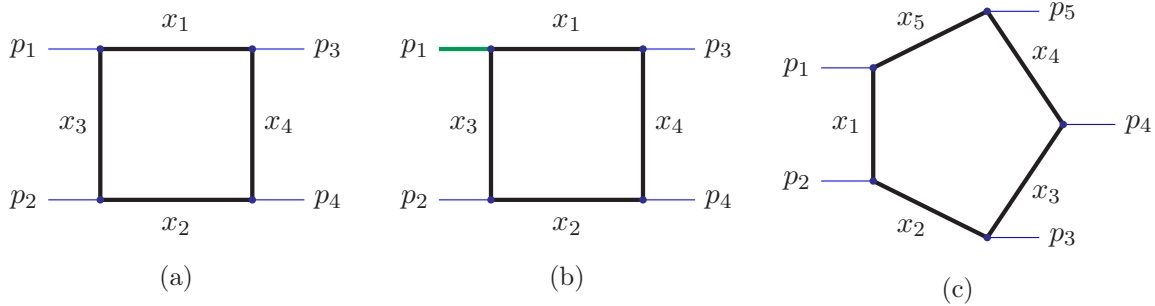
\begin{figure}[t]
    \centering
    \begin{subfigure}[t]{0.32\textwidth}
        \centering
        \begin{tikzpicture}[baseline=13ex,scale=1.0]
            \coordinate (x1) at (1, 1) ;
            \coordinate (x2) at (3, 1) ;
            \coordinate (x3) at (3, -1) ;
            \coordinate (x4) at (1,-1) ;
            \node (p1) at (0, 1) {$p_1$};
            \node (p2) at (0, -1) {$p_2$};
            \node (p3) at (4, 1) {$p_3$};
            \node (p4) at (4, -1) {$p_4$};
            \draw[color=blue] (x1) -- (p1);
            \draw[color=blue] (x2) -- (p3);
            \draw[color=blue] (x3) -- (p4);
            \draw[color=blue] (x4) -- (p2);

            \draw[ultra thick,color=Black] (x1) -- (x2) node [midway,yshift=+10pt,color=Black] {$x_1$};
            \draw[ultra thick,color=Black] (x2) -- (x3) node [midway,xshift=+10pt,color=Black] {$x_4$};
            \draw[ultra thick,color=Black] (x3) -- (x4) node [midway,yshift=-10pt,color=Black] {$x_2$};
            \draw[ultra thick,color=Black] (x4) -- (x1) node [midway,xshift=-10pt,color=Black] {$x_3$};

            \draw[fill,thick,color=Blue] (x1) circle (1pt);
            \draw[fill,thick,color=Blue] (x2) circle (1pt);
            \draw[fill,thick,color=Blue] (x3) circle (1pt);
            \draw[fill,thick,color=Blue] (x4) circle (1pt);
        \end{tikzpicture}
       \caption{}\label{massless_box_onshell}
    \end{subfigure}
    \hfill
    \begin{subfigure}[t]{0.32\textwidth}
        \centering
        \begin{tikzpicture}[baseline=13ex,scale=1.0]
            \coordinate (x1) at (1, 1) ;
            \coordinate (x2) at (3, 1) ;
            \coordinate (x3) at (3, -1) ;
            \coordinate (x4) at (1,-1) ;
            \node (p1) at (0, 1) {$p_1$};
            \node (p2) at (0, -1) {$p_2$};
            \node (p3) at (4, 1) {$p_3$};
            \node (p4) at (4, -1) {$p_4$};
            \draw[ultra thick,color=ForestGreen] (x1) -- (p1);
            \draw[color=blue] (x2) -- (p3);
            \draw[color=blue] (x3) -- (p4);
            \draw[color=blue] (x4) -- (p2);

            \draw[ultra thick,color=Black] (x1) -- (x2) node [midway,yshift=+10pt,color=Black] {$x_1$};
            \draw[ultra thick,color=Black] (x2) -- (x3) node [midway,xshift=+10pt,color=Black] {$x_4$};
            \draw[ultra thick,color=Black] (x3) -- (x4) node [midway,yshift=-10pt,color=Black] {$x_2$};
            \draw[ultra thick,color=Black] (x4) -- (x1) node [midway,xshift=-10pt,color=Black] {$x_3$};

            \draw[fill,thick,color=Blue] (x1) circle (1pt);
            \draw[fill,thick,color=Blue] (x2) circle (1pt);
            \draw[fill,thick,color=Blue] (x3) circle (1pt);
            \draw[fill,thick,color=Blue] (x4) circle (1pt);
        \end{tikzpicture}
       \caption{}\label{massless_box_offshell}
    \end{subfigure}
    \hfill
    \begin{subfigure}[t]{0.32\textwidth}
        \centering
        \begin{tikzpicture}[baseline=13ex,scale=1.0]
            \coordinate (x1) at (1, 0.75) ;
            \coordinate (x2) at (2.5, 1.5) ;
            \coordinate (x3) at (3.5, 0) ;
            \coordinate (x4) at (2.5,-1.5) ;
            \coordinate (x5) at (1, -0.75) ;
            \node (p1) at (0, 0.75) {$p_1$};
            \node (p2) at (0, -0.75) {$p_2$};
            \node (p5) at (3.5, 1.5) {$p_5$};
            \node (p4) at (4.5, 0) {$p_4$};
            \node (p3) at (3.5, -1.5) {$p_3$};
            \draw[color=blue] (x1) -- (p1);
            \draw[color=blue] (x5) -- (p2);
            \draw[color=blue] (x2) -- (p5);
            \draw[color=blue] (x3) -- (p4);
            \draw[color=blue] (x4) -- (p3);

            \draw[ultra thick,color=Black] (x1) -- (x2) node [midway,,xshift=-7pt,yshift=+7pt,,color=Black] {$x_5$};
            \draw[ultra thick,color=Black] (x2) -- (x3) node [midway,xshift=+8pt,yshift=+5pt,color=Black] {$x_4$};
            \draw[ultra thick,color=Black] (x3) -- (x4) node [midway,xshift=+8pt,yshift=-5pt,color=Black] {$x_3$};
            \draw[ultra thick,color=Black] (x4) -- (x5) node [midway,,xshift=-7pt,yshift=-7pt,,color=Black] {$x_2$};
            \draw[ultra thick,color=Black] (x5) -- (x1) node [midway,xshift=-10pt,color=Black] {$x_1$};

            \draw[fill,thick,color=Blue] (x1) circle (1pt);
            \draw[fill,thick,color=Blue] (x2) circle (1pt);
            \draw[fill,thick,color=Blue] (x3) circle (1pt);
            \draw[fill,thick,color=Blue] (x4) circle (1pt);
            \draw[fill,thick,color=Blue] (x5) circle (1pt);
        \end{tikzpicture}
        \caption{}\label{massless_pentagon}
    \end{subfigure}
    \caption{The massless box with all on-shell legs (\ref{massless_box_onshell}), an off-shell leg ($p_1$) (\ref{massless_box_offshell}) and the massless pentagon (\ref{massless_pentagon}).
    }
    \label{fig:massless_1L}
\end{figure}
\subsubsection{1-Loop Box with On-Shell Legs}
\label{sec:box-on}
To clarify the application of the algorithm presented in Section~\ref{ssec:method_ub}, let us begin by analysing the simple case of a 1-loop massless box with all external legs on-shell, shown in Fig.~\ref{massless_box_onshell}. 
Each step will be carried out in detail for this simple example in the hope that this illuminates the abstract procedure.

The integral we wish to resolve can be written in Feynman parameter space as,
\begin{align}
J_{\mathrm{box}}(\mathbf{s})&= \Gamma\left(2+\epsilon\right) \lim_{\delta\to0^+} I_{\mathrm{box}}(\mathbf{s};\delta), \\
I_{\mathrm{box}}(\mathbf{s};\delta)& = \int_{\mathbb{R}^4_{\geq0}}\prod_{i=1}^4\mathrm{d}x_i
\, \frac{\mathcal{U}(\mathbf{x})^{2\epsilon}}{\left(\mathcal{F}(\mathbf{x}; \mathbf{s}) - i\delta\right)^{2+\epsilon}}\delta\left(1-\alpha(\mathbf{x})\right),
\label{onshell-box-integral}
\end{align}
where the $\mathcal{U}(\mathbf{x})$ and $\mathcal{F}(\mathbf{x};\mathbf{s})$ polynomials are given by
\begin{align}
\mathcal{U}(\mathbf{x}) &= x_1+x_2+x_3+x_4,\\
\mathcal{F}(\mathbf{x}; \mathbf{s}) &= -s_{12} x_1x_2- s_{13} x_3x_4, 
\end{align}
with $s_{ij}=(p_i+p_j)^2$.
Suppose that we are interested in evaluating this integral for $2 \rightarrow 2$ physical scattering kinematics $\mathbf{s}_\mathrm{phys}=\{0<s_{12}<\infty,\ -s_{12}<s_{13}<0\}$. 
Usually we would need a contour deformation as this regime is above the threshold at $s_{12}=0$ and the $\mathcal{F}$ polynomial contains monomials of different sign and can itself be both positive or negative inside the domain of integration.
Recall that in this kinematic regime $s_{12} >0$ and $(-s_{13}) >0$.

Applying Algorithm~\ref{alg:univariate}, we begin with the choice $x_i = x_1$.
Attempting to reduce with respect to $x_1$ the set of inequalities,
\begin{align}
\{-s_{12} x_1x_2- s_{13}x_3x_4<0\} 
\cup \{0<x_1,\ 0<x_2,\ 0<x_3,\ 0<x_4\} \cup \,\mathbf{s}_\mathrm{phys},
\end{align}
we obtain the solution,
\begin{align}
r = \{\frac{-s_{13} x_3x_4}{s_{12} x_2}<x_1\} \cup \{0<x_2,\ 0<x_3,\ 0<x_4\}  \cup \,\mathbf{s}_\mathrm{phys}.
\end{align}
We observe that $r$ is of the form given in Eq.~\eqref{eq:ub_b} with
\begin{align}
f(\mathbf{x}_{\neq1})=\frac{(-s_{13})x_3x_4}{s_{12}x_2},
\end{align}
therefore, $x_1$ is a valid bisection parameter.
As dictated by the algorithm, we can now construct the positive and negative contributions by transforming the variable $x_1$.
Applying the transformation given in Eq.~\eqref{eq:ub_b},
\begin{align}
x_1 \rightarrow y_1^\prime = x_1 + \frac{(-s_{13})x_3x_4}{s_{12}x_2},
\end{align}
we map the variety $\mathcal{F}(\mathbf{x};\mathbf{s}) = 0$ to $x_1=0$ while keeping the boundary at $x_1 \rightarrow \infty$ fixed.
The resulting integrand is given by
\begin{align}
\mathcal{J}^-(\mathbf{x}) &= 1,
\quad \mathcal{U}^-(\mathbf{x}) = x_1+x_2+x_3+x_4+\frac{-s_{13} x_3 x_4}{s_{12} x_2},\quad
\mathcal{F}^-(\mathbf{x};\mathbf{s}) = s_{12} x_1x_2,& \\
\mathcal{I}^-_{\mathrm{box}}(\mathbf{x};\mathbf{s}) &= \mathcal{J}^-(\mathbf{x}) \frac{\mathcal{U}^-(\mathbf{x})^{2 \epsilon}}{\mathcal{F}^-(\mathbf{x};\mathbf{s})^{2+\epsilon}}\nonumber\\
&= x_1^{-2-\epsilon} (s_{12} x_2)^{-2-3\epsilon} \left(s_{12} x_2\left(x_1+x_2+x_3+x_4\right)-s_{13} x_3x_4\right)^{2\epsilon}.
\end{align}
In this example we are not considering an integral with dots (propagators raised to a higher power) or a numerator, in general one would apply the transformation also to any Feynman parameters appearing in the numerator.

The positive contribution is given by transforming the variable $x_1$ according to Eq.~\eqref{eq:sub_a} with the arbitrary choice $x_j = x_N = x_4$,
\begin{align}
x_1 \rightarrow y_1 = \frac{x_1}{x_1+x_4} \frac{(-s_{13}) x_3 x_4}{s_{12} x_2}
\end{align}
which maps the variety $\mathcal{F}(\mathbf{x};\mathbf{s}) = 0$ to $x_1 \rightarrow \infty$ keeping $x_1 =0$ fixed.
The resulting integrand is given by
\begin{align}
\mathcal{J}^+(\mathbf{x}) &= \frac{(-s_{13}) x_3 x_4^2}{s_{12} x_2 (x_1+x_4)^2},\quad
\mathcal{U}^+(\mathbf{x}) = \frac{x_1}{x_1+x_4} \frac{(-s_{13}) x_3 x_4}{s_{12} x_2} + x_2 +x_3 + x_4, \\
\mathcal{F}^+(\mathbf{x};\mathbf{s}) &= \frac{(-s_{13}) x_3 x_4^2}{x_1+x_4}, \\
\mathcal{I}^+_{\mathrm{box}}(\mathbf{x};\mathbf{s}) &= \mathcal{J}^+(\mathbf{x}) \frac{\mathcal{U}^+(\mathbf{x})^{2 \epsilon}}{\mathcal{F}^+(\mathbf{x};\mathbf{s})^{2+\epsilon}} \nonumber\\
&= \left(x_1+x_4\right)^{-\epsilon}\left(s_{12}x_2\right)^{-1-2\epsilon}\left(-s_{13}x_3x_4^2\right)^{-1-\epsilon} \nonumber\\
& \qquad\!\!\!\!\!\left(s_{12} x_2\left(x_1+x_4\right)\left(x_2+x_3+x_4\right)-s_{13} x_1x_3x_4\right)^{2\epsilon}.
\end{align}

Sewing the positive and negative contributions together, the final result for the on-shell box integral is given by,
\begin{align}
J_{\mathrm{box}}(\mathbf{s})&=\Gamma\left(2+\epsilon\right) \lim_{\delta\to0^+} I_{\mathrm{box}}(\mathbf{s};\delta)\\
I_{\mathrm{box}}(\mathbf{s};\delta)&=I^{+}_{\mathrm{box}}(\mathbf{s})+\left(-1-i\delta\right)^{-2-\epsilon}I^{-}_{\mathrm{box}}(\mathbf{s}).
\label{eq:decompboxonshell}
\end{align}

Neglecting the $\delta$-functions, the integrands of both $I^+_{\mathrm{box}}$ and $I^-_{\mathrm{box}}$ pick up a factor of $\lambda^{-N}=\lambda^{-4}$ under the scaling transformation ${\{x_1,\dots,x_4\}\rightarrow\{\lambda x_1,\dots,\lambda x_4\}}$.
This stems from the homogeneity of the original $\mathcal{U}$ and $\mathcal{F}$ polynomials and the fact that the algorithm preserves the homogeneity.

\subsubsection{1-Loop Box with One Off-Shell Leg}
\label{sec:box-off}
Here, we extend the previous example to a massless box with one off-shell leg, $p_1^2>0$, shown in Fig.~\ref{massless_box_offshell}. 
For this integral and the remaining massless examples, we present the resolution procedure more concisely. 
The integral we wish to consider is
\begin{align}
J_{\mathrm{box},\  \!\!p_1^2>0}(\mathbf{s})&=\Gamma\left(2+\epsilon\right) \lim_{\delta\to0^+} I_{\mathrm{box},\  \!\!p_1^2>0}(\mathbf{s};\delta)\\
I_{\mathrm{box},\  \!\!p_1^2>0}(\mathbf{s};\delta)&=\int_{\mathbb{R}^4_{\geq0}}\prod_{i=1}^4\mathrm{d}x_i
\, \frac{\mathcal{U}(\mathbf{x})^{2\epsilon}}{\left(\mathcal{F}(\mathbf{x}; \mathbf{s}) - i\delta\right)^{2+\epsilon}}\delta\left(1-\alpha(\mathbf{x})\right)
\label{offshell-box-integral}
\end{align}
where the $\mathcal{U}$ polynomial remains unchanged and the $\mathcal{F}$ polynomial gets modified by an extra term proportional to the off-shellness, $p_1^2>0$:
\begin{align}
       \mathcal{U}(\mathbf{x}) &= x_1+x_2+x_3+x_4\\
\mathcal{F}(\mathbf{x}; \mathbf{s}) &= -s_{12} x_1x_2- s_{13}x_3x_4-p_1^2x_1x_3. 
\end{align}
We resolve this integral in the regime $\mathbf{s}_{p_1^2>0}= \{ 0 < p_1^2 < \infty, 0 <s_{12} < \infty, -s_{12}<s_{13}<0 \}$.
According to the algorithm, we can bisect the integral for $x_i = x_1$, choosing $x_j = x_4$, we obtain,
\begin{align}
&x_1 \rightarrow y_1^\prime = x_1+\left(\frac{- s_{13}x_3x_4}{s_{12} x_2+p_1^2x_3}\right),& &\mathcal{F}^-(\mathbf{x};\mathbf{s})= x_1 (s_{12} x_2 + p_1^2 x_3), &\\
&x_1 \rightarrow y_1 = \frac{x_1}{x_1+x_4}\left(\frac{-s_{13} x_3 x_4}{s_{12}x_2+p_1^2x_3}\right),& &\mathcal{F}^+(\mathbf{x};\mathbf{s})= \frac{-s_{13}x_3 x_4^2}{x_1+x_4}.&
\end{align}
In this instance, we can see that $\mathcal{F}^-(\mathbf{x};\mathbf{s})$ and $\mathcal{F}^+(\mathbf{x};\mathbf{s})$ are manifestly non-negative in $\mathbf{s}_{p_1^2>0}$.
Including also the transformed $\mathcal{U}(\mathbf{x})$ and Jacobian we obtain the resolved integrands,
\begin{align}
\mathcal{I}_{\mathrm{box},\ \!\!p_1^2>0}^-(\mathbf{x};\mathbf{s})&= \frac{\left[\left(s_{12} x_2+p_1^2x_3\right)\left(x_1+x_2+x_3+x_4\right)-s_{13} x_3x_4\right]^{2\epsilon}}{x_1^{2+\epsilon}\left(s_{12} x_2+p_1^2x_3\right)^{2+3\epsilon}}, \\
\mathcal{I}_{\mathrm{box},\ \!\!p_1^2>0}^+(\mathbf{x};\mathbf{s})&= \frac{\left[\left(s_{12} x_2+p_1^2x_3\right)\left(x_1+x_4\right)\left(x_2+x_3+x_4\right)-s_{13}x_1x_3x_4\right]^{2\epsilon}}{x_4^{2+2\epsilon}\left(x_1+x_4\right)^{\epsilon}\left(-s_{13} x_3\right)^{1+\epsilon}\left(s_{12}x_2+p_1^2x_3\right)^{1+2\epsilon}}.
\end{align}
The final result for the off-shell box integral is given by,
\begin{align}
J_{{\mathrm{box},\  \!\!p_1^2>0}}(\mathbf{s})&= \Gamma\left(2+\epsilon\right) \lim_{\delta\to0^+}I_{{\mathrm{box},\  \!\!p_1^2>0}}(\mathbf{s};\delta),\\
I_{{\mathrm{box},\  \!\!p_1^2>0}}(\mathbf{s};\delta)&= I^{+}_{{\mathrm{box},\  \!\!p_1^2>0}}(\mathbf{s})+\left(-1-i\delta\right)^{-2-\epsilon}I^{-}_{{\mathrm{box},\  \!\!p_1^2>0}}(\mathbf{s}).
    \label{decompboxoffshell}
\end{align}

In Section~\ref{subsec:1-loopboxtimings} we present a study of the numerical performance before and after resolution using sector decomposition.

\subsubsection{1-Loop On-Shell Pentagon}
\label{sec:pentagon}
A massless pentagon (see Fig.~\ref{massless_pentagon}) is minimally parameterised with five kinematic invariants, for example with
the set of cyclic scalar products $(s_{12}, s_{23}, s_{34}, s_{45}, s_{51})$, where ${s_{ij}=(p_i+p_j)^2}$. In Feynman parameterised form, the integral can be written as,
\begin{align}
J_{\mathrm{pen}}(\mathbf{s})&=-\Gamma\left(3+\epsilon\right) \lim_{\delta\to0^+}I_{\mathrm{pen}}(\mathbf{s};\delta)\\
I_{\mathrm{pen}}(\mathbf{s};\delta)&=\int_{\mathbb{R}^5_{\geq0}}\prod_{i=1}^5\mathrm{d}x_i\, \frac{\mathcal{U}(\mathbf{x})^{1+2\epsilon}}{\left(\mathcal{F}(\mathbf{x}; \mathbf{s}) - i\delta\right)^{3+\epsilon}}\delta\left(1-\sum\nolimits_{i=1}^5\alpha_ix_i\right)
\label{pentagon-integral}
\end{align}
where the pentagon $\mathcal{U}(\mathbf{x})$ and $\mathcal{F}(\mathbf{x};\mathbf{s})$ polynomials are given by,
\begin{align}
          \mathcal{U}(\mathbf{x}) &= x_1+x_2+x_3+x_4+x_5\\
\mathcal{F}(\mathbf{x}; \mathbf{s}) &= -s_{12}x_2x_5 -s_{23}x_1x_3 -s_{34}x_2x_4 -s_{45}x_3x_5 -s_{51}x_1x_4 .  
\end{align}
We arbitrarily consider the kinematic regime ${\mathbf{s}_R = \{ 0< s_{12},s_{34},s_{51} < \infty, -\infty< s_{23},s_{45} <0 \}}$.
Imposing these constraints, the algorithm finds a valid bisection in parameter $x_i=x_3$, giving,
\begin{align}
&f(\mathbf{x}_{\neq3}) =\frac{s_{51}x_1x_4+s_{34}x_2x_4+s_{12}x_2x_5}{-s_{23}x_1-s_{45}x_5}.
\end{align}
Choosing $x_j = x_5$, we obtain,
\begin{align}
&x_3\rightarrow y_3 = \frac{x_3}{x_3+x_5}f(\mathbf{x}_{\neq3}),& 
&\mathcal{F}^-(\mathbf{x};\mathbf{s})=\frac{x_5 (s_{51} x_1 x_4 + s_{34} x_2 x_4 + s_{12} x_2 x_5)}{x_3 + x_5}, &
\\
&x_3\rightarrow y_3^\prime = x_3+f(\mathbf{x}_{\neq3}),& &\mathcal{F}^+(\mathbf{x};\mathbf{s}) = -x_3(s_{23} x_1 + s_{45} x_5).& 
\end{align}
The resulting integrands are given by,
\begin{align}
\mathcal{I}_{\mathrm{pen}}^-=\;&x_5^{-2-\epsilon}\left(x_3+x_5\right)^{-\epsilon}\left(-s_{23}x_1-s_{45}x_5\right)^{-2-2\epsilon}\left(s_{51}x_1x_4+s_{34}x_2x_4+s_{12}x_2x_5\right)^{-2-\epsilon}\times\nonumber\\&[\left(x_3\!+\!x_5\right)\!\left(-s_{23}x_1\!-\!s_{45}x_5\right)\!\left(x_1\!+\!x_2\!+\!x_3\!+\!x_4\!+\!x_5\right)\!+\nonumber\\&\!x_3\left(s_{51}x_1x_4\!+\!s_{34}x_2x_4\!+\!s_{12}x_2x_5\right)]^{1+2\epsilon}, \\
\mathcal{I}_{\mathrm{pen}}^+=\;&{x_3^{-3-\epsilon}\left(-s_{23}x_1-s_{45}x_5\right)^{-4-3\epsilon}}\times\nonumber\\&\left[\left(-s_{23}x_1-s_{45}x_5\right)\left(x_1+x_2+x_3+x_4+x_5\right)+s_{51}x_1x_4+s_{34}x_2x_4+s_{12}x_2x_5\right]^{1+2\epsilon}.
\end{align}
Summing the positive and negative contributions yields the resolved pentagon integral,
\begin{align}
J_{\mathrm{pen}}(\mathbf{s})=& -\Gamma\left(3+\epsilon\right) \lim_{\delta\to0^+}I_{\mathrm{pen}}(\mathbf{s};\delta),\\
I_{\text{pen}}(\mathbf{s};\delta)=& I^{+}_{\text{pen}}(\mathbf{s})+\left(-1-i\delta\right)^{-3-\epsilon}I^{-}_{\text{pen}}(\mathbf{s}).
    \label{decomppentagon}
\end{align} 
In Section.~\ref{subsec:1-looppenttimings} we present a performance study of the numerical integration of the original and resolved pentagon integral.  

The resolution described above is valid for one particular sign combination of the kinematics, $\mathbf{s}_R$. 
To evaluate the pentagon for any kinematic point, one would naively have to repeat the procedure above $2^5=32$ times. 
Due to reparametrisation invariance, however, only 4 resolution procedures are required, corresponding to the presence of 0, 1 or 2 negative invariants, where the case with 2 negative invariants requires different resolutions depending on whether the legs with negative invariants are adjacent or not. 
For example, the regime where $-\infty < s_{12},s_{34},s_{51} < 0 < s_{23},s_{45} < \infty$ can be directly related to the case described above by factoring out a minus sign from $\mathcal{F}$. 
For completeness, we highlight that the UB algorithm is sufficient to resolve kinematic regions where either 2 or 3 invariants are negative, and where the propagators that connect the legs corresponding to each of the same sign invariants are not adjacent to each other.
In other configurations, the space is only dissected into positive and negative parts with constraints on more than one Feynman parameter.

If we wish to compute 1-loop integrals with more than 5 legs, we can continue to derive resolutions using the techniques described above.
However, all dimensionally regulated 1-loop $N$-point integrals can be reduced at all orders in $\epsilon$ to integrals with at most 5 legs~\cite{Bern:1993kr,Binoth:1999sp,Duplancic:2003tv,Giele:2004iy,Binoth:2005ff}, therefore, it is sufficient to resolve all 1-loop integrals up to and including pentagons.
Furthermore, if we wish to evaluate only up to and including the finite order in $\epsilon$, we can reduce all 1-loop integrals with 5 or more legs to box integrals with off-shell legs and resolve these instead.
For example, the 1-loop pentagon integral can be written in terms of boxes as \cite{Bern:1993kr,Binoth:1999sp},
\begin{equation}
I_{\mathrm{pen}} = -\frac{1}{2} \sum_{l,k=1}^5 S_{lk}^{-1} I_{\mathrm{box},l} + \mathcal{O}(\varepsilon),
\label{eq:pentagon-reduction}
\end{equation}
where $S_{lk}$ are elements of the S-matrix
\begin{equation}
S = -\frac{1}{2}\begin{pmatrix}
0 \quad \quad & p_2^2 \quad \quad & s_{23} \quad \quad & s_{51} \quad \quad & p_1^2 \\
p_2^2 \quad \quad & 0 \quad \quad & p_3^2 \quad \quad & s_{34} \quad \quad & s_{12} \\
s_{23} \quad \quad & p_3^2 \quad \quad & 0 \quad \quad & p_4^2 \quad \quad & s_{45} \\
s_{51} \quad \quad & s_{34} \quad \quad & p_4^2 \quad \quad & 0 \quad \quad & p_5^2 \\
p_1^2 \quad \quad & s_{12} \quad \quad & s_{45} \quad \quad & p_5^2 \quad \quad & 0
\end{pmatrix}.
\label{S-matrix}
\end{equation}
In Eq.~\eqref{eq:pentagon-reduction}, each box integral has one off-shell leg corresponding to the pinching of a propagator of the pentagon.

\begin{figure}[t]
    \centering
    \begin{subfigure}[t]{0.32\textwidth} 
    \centering
    \begin{tikzpicture}[baseline=13ex,scale=1.0]
        \coordinate (x1) at (1, 3) ;
        \coordinate (x2) at (1, 1) ;
        \coordinate (x3) at (2, 2) ;
        \coordinate (x5) at (3, 1) ;
        \coordinate (x4) at (3, 3) ;
        \node (p1) at (0, 3) {$p_1$};
        \node (p2) at (0, 1) {$p_2$};
        \node (p3) at (1.33, 1.33) {$p_3$};
        \node (p4) at (4, 3) {$p_4$};
        \draw[color=blue] (x1) -- (p1);
        \draw[color=blue] (x2) -- (p2);
        \draw[color=blue] (x3) -- (p3);
        \draw[color=blue] (x4) -- (p4);
        \draw[ultra thick,color=Black] (x1) -- (x2) node [midway,,xshift=-7pt,yshift=0,color=Black] {$x_1$};
        \draw[ultra thick,color=Black] (x1) -- (x4) node [midway,,xshift=0pt,yshift=7pt,color=Black] {$x_3$};
        \draw[ultra thick,color=Black] (x2) -- (x5) node [midway,,xshift=0pt,yshift=-7pt,color=Black] {$x_6$};
        \draw[ultra thick,color=Black] (x3) -- (x1) node [midway,,xshift=+9pt,yshift=+1.6pt,color=Black] {$x_2$};
        \draw[ultra thick,color=Black] (x3) -- (x5) node [midway,,xshift=-6pt,yshift=-5pt,color=Black] {$x_5$};
        \draw[ultra thick,color=Black] (x4) -- (x5) node [midway,,xshift=7.75pt,yshift=0pt,color=Black] {$x_4$};
        \draw[fill,thick,color=Blue] (x1) circle (1pt);
        \draw[fill,thick,color=Blue] (x2) circle (1pt);
        \draw[fill,thick,color=Blue] (x3) circle (1pt);
        \draw[fill,thick,color=Blue] (x4) circle (1pt);
        \draw[fill,thick,color=Blue] (x5) circle (1pt);
    \end{tikzpicture}
    \caption{}\label{massless_BNP6}
    \end{subfigure}
    \hspace{1cm}    
    \begin{subfigure}[t]{0.32\textwidth} 
    \centering
    \begin{tikzpicture}[baseline=13ex,scale=1.0]
        \coordinate (x1) at (1, 3) ;
        \coordinate (x2) at (3, 3) ;
        \coordinate (x3) at (3.9, 2) ;
        \coordinate (x4) at (2.1, 2) ;
        \coordinate (x5) at (3, 1) ;
        \coordinate (x6) at (1, 1) ;
        \node (p1) at (0, 3) {$p_1$};
        \node (p2) at (0, 1) {$p_2$};
        \node (p3) at (1.35, 2) {$p_3$};
        \node (p4) at (5, 2) {$p_4$};
        \draw[color=blue] (x1) -- (p1);
        \draw[color=blue] (x6) -- (p2);
        \draw[color=blue] (x4) -- (p3);
        \draw[color=blue] (x3) -- (p4);
        \draw[ultra thick,color=Black] (x1) -- (x2) node [midway,,xshift=0pt,yshift=7pt,color=Black] {$x_3$};
        \draw[ultra thick,color=Black] (x2) -- (x3) node [midway,,xshift=-9.5pt,yshift=0,color=Black] {$x_7$};
        \draw[ultra thick,color=Black] (x2) -- (x4) node [midway,,xshift=-10pt,yshift=0,color=Black] {$x_5$};
        \draw[ultra thick,color=Black] (x3) -- (x5)  node [midway,,xshift=-9.5pt,yshift=0,color=Black] {$x_6$};
        \draw[ultra thick,color=Black] (x4) -- (x5) node [midway,,xshift=-10pt,yshift=0,color=Black] {$x_4$};
        \draw[ultra thick,color=Black] (x5) -- (x6) node [midway,,xshift=0pt,yshift=-7pt,color=Black] {$x_2$};
        \draw[ultra thick,color=Black] (x6) -- (x1) node [midway,,xshift=-7pt,yshift=0,color=Black] {$x_1$};
        \draw[fill,thick,color=Blue] (x1) circle (1pt);
        \draw[fill,thick,color=Blue] (x2) circle (1pt);
        \draw[fill,thick,color=Blue] (x3) circle (1pt);
        \draw[fill,thick,color=Blue] (x4) circle (1pt);
        \draw[fill,thick,color=Blue] (x5) circle (1pt);
        \draw[fill,thick,color=Blue] (x6) circle (1pt);
    \end{tikzpicture}
    \caption{}\label{massless_BNP7}
    \end{subfigure}
    \caption{Massless non-planar 2-loop boxes with 6 (\ref{massless_BNP6}: BNP6) and 7 (\ref{massless_BNP7}: BNP7) propagators respectively.}
    \label{fig:massless_BNP}
\end{figure}
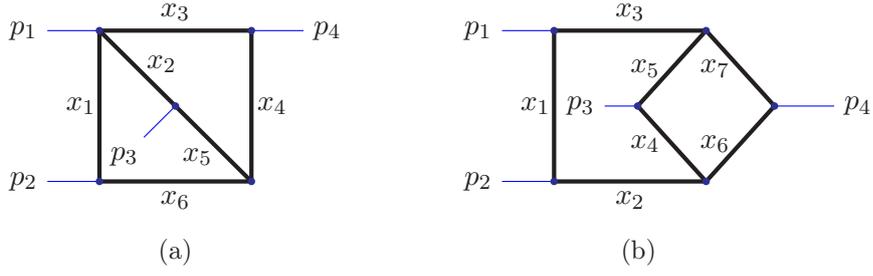
\subsubsection{2-Loop Non-Planar 6 Propagator Box}
\label{sec:BNP6}
The non-planar box with six propagators (BNP6, see Fig.~\ref{massless_BNP6}) can be parameterised by the Mandelstam invariants $s_{12}=(p_1+p_2)^2$ and $s_{23} = (p_2+p_3)^2$ after applying the momentum conservation rule $s_{12}+s_{23}+s_{13}=0$ to eliminate $s_{13}$. 
The integral may be written as,
\begin{align}
    J_{\mathrm{BNP6}}(\mathbf{s})&=\Gamma\left(2+2\epsilon\right)\lim_{\delta\to0^+}I_{\mathrm{BNP6}}(\mathbf{s};\delta)\\I_{\mathrm{BNP6}}(\mathbf{s};\delta)&=\int_{\mathbb{R}^6_{\geq0}}\prod_{i=1}^6\mathrm{d}x_i\, \frac{\mathcal{U}(\mathbf{x})^{3\epsilon}}{\left(\mathcal{F}(\mathbf{x}; \mathbf{s}) - i\delta\right)^{2+2\epsilon}}\delta\left(1-\alpha(\mathbf{x})\right)
\label{BNP6-integral}
\end{align}
with the $\mathcal{U}$ and $\mathcal{F}$ polynomials,
\begin{align}
\mathcal{U}(\mathbf{x}) = \, &x_1x_2 + x_1x_3 + x_1x_4 + x_1x_5 + x_2x_3 + x_2x_4 + x_2x_6 + \nonumber \\ 
&x_3x_5 + x_3x_6 + x_4x_5 + x_4x_6 + x_5x_6,\\
\mathcal{F}(\mathbf{x}; \mathbf{s}) = &-s_{12} x_2 x_3 x_6 -s_{23} x_1 x_2 x_4+(s_{12}+s_{23}) x_1 x_3 x_5.
\end{align}
We restrict to the physical kinematic regime for massless $2\rightarrow2$ scattering, $\mathbf{s}_\mathrm{phys}=\{0<s_{12}<\infty,\ -s_{12}<s_{23}<0\}$. Applying the algorithm, we find that we can bisect the integral with the parameter $x_i=x_1$, choosing $x_j=x_6$:
\begin{align}
&x_1\rightarrow y_1=\frac{x_1}{x_1+x_6} f(\mathbf{x}_{\neq1}),& &\mathcal{F}^-(\mathbf{x};\mathbf{s})=\frac{s_{12}x_2x_3x_6^2}{x_1+x_6} , &\\
&x_1 \rightarrow y_1' = x_1 +  f(\mathbf{x}_{\neq1}),& &\mathcal{F}^+(\mathbf{x};\mathbf{s})= x_1\left[(s_{12}+s_{23})x_3x_5-s_{23}x_2x_4\right],&
\end{align}
with
\begin{equation} f(\mathbf{x}_{\neq1})=\frac{s_{12}x_2x_3x_6}{\left(s_{12}+s_{23}\right)x_3x_5-s_{23}x_2x_4},
\end{equation}
and we emphasise that $(s_{12}+s_{23})>0$ and $(-s_{23})>0$ in our restricted kinematic regime. These transformations result in the resolved integrands,
\begin{align}
            \mathcal{I}_{\mathrm{BNP6}}^-=&\left(s_{12}x_2x_3x_6^2\right)^{-1-2\epsilon}\left(x_1+x_6\right)^{-\epsilon}\left[\left(s_{12}+s_{23}\right)x_3x_5-s_{23}x_2x_4\right]^{-1-3\epsilon}\times\nonumber\\&\Bigl[\left[\left(s_{12}+s_{23}\right)x_3 x_5 -s_{23} x_2 x_4\right]\left(x_1+x_6\right) [\left(x_3+x_4\right) \left(x_2+x_5\right)+\nonumber \\&\left(x_2+x_3+x_4+x_5\right) x_6] + s_{12} x_1 x_2 x_3 x_6\left(x_2+x_3+x_4+x_5\right)\Bigr]^{3\epsilon},\\
             \mathcal{I}_{\mathrm{BNP6}}^+=&\ x_1^{-2-2\epsilon}\left[\left(s_{12}+s_{23}\right)x_3x_5-s_{23}x_2x_4\right]^{-2-5\epsilon}\times\nonumber\\&\Bigl[\left[ \left(s_{12}+s_{23}\right)x_3 x_5-s_{23} x_2 x_4\right]\bigl[\left(x_3+x_4\right) \left(x_2+x_5\right)+\nonumber\\&\left(x_1+x_6\right) \left(x_2+x_3+x_4+x_5\right)\bigr] +s_{12}x_2x_3x_6\left(x_2+x_3+x_4+x_5\right)\Bigr]^{3\epsilon}.
\end{align}
Using the decomposition formula Eq.~\eqref{eq:decomp}, we may therefore write BNP6 as,
\begin{align}
J_{\mathrm{BNP6}}(\mathbf{s})&=\Gamma\left(2+2\epsilon\right)\lim_{\delta\to0^+}I_{\mathrm{BNP6}}(\mathbf{s};\delta)\\
    I_{\text{BNP6}}(\mathbf{s};\delta)&=I^{+}_{\text{BNP6}}(\mathbf{s})+\left(-1-i\delta\right)^{-2-2\epsilon}I^{-}_{\text{BNP6}}(\mathbf{s}).
    \label{decompBNP6}
\end{align}   
\subsubsection{2-Loop Non-Planar 7 Propagator Box}
\label{sec:BNP7}
The non planar box with seven propagators (BNP7, see Fig.~\ref{massless_BNP7}) depends on the two kinematic invariants $s_{12}=(p_1+p_2)^2$ and $s_{23} = (p_2+p_3)^2$ and is given by
\begin{align}
J_{\mathrm{BNP7}}&=-\Gamma\left(3+2\epsilon\right)\lim_{\delta\to0^+}I_{\mathrm{BNP7}}\\I_{\mathrm{BNP7}}&=\int_{\mathbb{R}^7_{\geq0}}\prod_{i=1}^7\mathrm{d}x_i \,\frac{\mathcal{U}(\mathbf{x})^{1+3\epsilon} }{\left(\mathcal{F}(\mathbf{x}; \mathbf{s})- i\delta\right)^{3+2\epsilon}}\delta\left(1-\alpha(\mathbf{x})\right)
\label{BNP7-integral}
\end{align}
with the $\mathcal{U}$ and $\mathcal{F}$ polynomials,
    \begin{align}
        \mathcal{U}(\mathbf{x}) = \, &x_1x_4 + x_1x_5 + x_1x_6 + x_1x_7 + x_2x_4 + x_2x_5 + x_2x_6 + x_2x_7 + \nonumber\\&x_3x_4 + x_3x_5 + x_3x_6 + x_3x_7 + x_4x_6 + x_4x_7 + x_5x_6 + x_5x_7 \\ 
\mathcal{F}(\mathbf{x}, \mathbf{s}) = &-s_{12}(x_3x_4x_6 + x_2x_5x_7 + x_2x_3x_7 + x_2x_3x_6 + x_2x_3x_5 + x_2x_3x_4) \nonumber\\& -s_{23}x_1x_5x_6 + (s_{12} + s_{23})x_1x_4x_7.
    \end{align}
We consider the kinematic regime $\mathbf{s}_\mathrm{phys}=\{0<s_{12}<\infty,\ -s_{12}<s_{23}<0\}$ and find that the algorithm gives us the following transformations for $x_i=x_1$, choosing $x_j=x_7$:
\begin{align}
&        x_1\rightarrow y_1=\frac{x_1}{x_1+x_7}f(\mathbf{x}_{\neq1}),& &\mathcal{F}^-(\mathbf{x};\mathbf{s})=\frac{s_{12}x_7\left[x_3x_4x_6+x_2x_5x_7+x_2x_3\left(x_4+x_5+x_6+x_7\right)\right]}{x_1+x_7} , &\\
&x_1 \rightarrow y_1' = x_1 +  f(\mathbf{x}_{\neq1}),& &\mathcal{F}^+(\mathbf{x};\mathbf{s})= x_1\left[\left(s_{12}+s_{23}\right)x_4x_7-s_{23}x_5x_6\right]&
\end{align}
where
\begin{align}
f(\mathbf{x}_{\neq1})=\frac{s_{12}\left[x_3x_4x_6+x_2x_5x_7+x_2x_3\left(x_4+x_5+x_6+x_7\right)\right]}{\left(s_{12}+s_{23}\right)x_4x_7-s_{23}x_5x_6}
\end{align}
with $(s_{12}+s_{23})>0$ and $(-s_{23})>0$ in the regime considered. 
We may therefore generate,
\begin{align}
        \mathcal{I}_{\mathrm{BNP7}}^-=&\left(s_{12}x_7\right)^{-2-2\epsilon}(x_1+x_7)^{-\epsilon}\left[\left(s_{12}+s_{23}\right)x_4x_7-s_{23}x_5x_6\right]^{-2-3\epsilon}\times\label{BNP7-neg}\nonumber\\&\left[x_3x_4x_6+x_2x_5x_7+x_2x_3\left(x_4+x_5+x_6+x_7\right)\right]^{-2-2\epsilon}\times\nonumber\\&\Bigl[s_{12} x_1 \left(x_4+x_5+x_6+x_7\right) \left[x_3 x_4 x_6+x_2 x_5 x_7+x_2 x_3 \left(x_4+x_5+x_6+x_7\right)\right]+\nonumber\\&\left[(s_{12}+s_{23})x_4 x_7 -s_{23} x_5 x_6\right]\left(x_1+x_7\right) [\left(x_4+x_5\right) \left(x_6+x_7\right)+\nonumber\\&\left(x_2+x_3\right) \left(x_4+x_5+x_6+x_7\right)] \Bigr]^{1+3\epsilon},\\
        \mathcal{I}_{\mathrm{BNP7}}^+=&\ x_1^{-3-2\epsilon}\left[\left(s_{12}+s_{23}\right)x_4x_7-s_{23}x_5x_6\right]^{-4-5\epsilon}\times\label{BNP7-pos}\nonumber\\&\Bigl[s_{12} \left(x_4+x_5+x_6+x_7\right) \left[x_3 x_4 x_6+x_2 x_5 x_7+x_2 x_3 \left(x_4+x_5+x_6+x_7\right)\right]+\nonumber\\&\left[(s_{12}+s_{23})x_4 x_7 -s_{23} x_5 x_6\right]\bigl[\left(x_4+x_5\right) x_6+x_3 \left(x_4+x_5+x_6\right)+\nonumber\\&\left(x_3+x_4+x_5\right) x_7+\left(x_1+x_2\right) \left(x_4+x_5+x_6+x_7\right)\bigr] \Bigr]^{1+3\epsilon}.       
\end{align}
All factors in the integrands above are positive in this kinematic regime within the integration domain (with zeroes only on the boundary); however, this is not immediately apparent, as it may only become manifest after simplification of the transformed integrands.

Combining the positive and negative contributions appropriately gives us the resolution of BNP7,
\begin{align}
J_{\mathrm{BNP7}}(\mathbf{s})&=-\Gamma\left(3+2\epsilon\right)\lim_{\delta\to0^+}I_{\mathrm{BNP7}}(\mathbf{s};\delta)\\
I_{\text{BNP7}}(\mathbf{s};\delta)&=I^{+}_{\text{BNP7}}(\mathbf{s})+\left(-1-i\delta\right)^{-3-2\epsilon}I^{-}_{\text{BNP7}}(\mathbf{s}).
    \label{decompBNP7}
\end{align}   

\subsubsection{3-Loop Non-Planar Box}
\label{sec:crown}
In this section, we consider a massless 3-loop 4-point example ($G_{\bullet\bullet}$ in the notation of Ref.~\cite{Gardi:2024axt}) where a naive contour deformation in Feynman parameter space, as described in Section~\ref{ssec:contourdef}, fails due to the presence of a leading Landau singularity within the domain of integration. 
We show that, after first dissecting the integral on the parameter space hypersurface associated with the Landau singularity, we can apply a combination of shifts, rescalings, and rational transformations of the Feynman parameters to resolve the mixed-sign integrals.

\begin{figure}[t]
\centering
\begin{tikzpicture}[baseline=12ex,scale=0.35]
\draw (0.5,1) edge [color=blue] (2,2) node [] {};
\draw (9.5,1) edge [color=blue] (8,2) node [] {};
\draw (0.5,9) edge [color=blue] (2,8) node [] {};
\draw (9.5,9) edge [color=blue] (8,8) node [] {};
\draw (5,7.5) edge [ultra thick, bend right = 10] (2,2) node [] {};
\draw (5,7.5) edge [ultra thick, bend left = 10] (8,2) node [] {};
\draw (5,7.5) edge [ultra thick, bend right = 10] (2,8) node [] {};
\draw (5,7.5) edge [ultra thick, bend left = 10] (8,8); 
\draw (5,2.5) edge [ultra thick, draw=white, double=white, double distance=0pt, bend left = 10] (2,2) node [] {};\draw (5,2.5) edge [ultra thick, bend left = 10] (2,2) node [] {};
\draw (5,2.5) edge [ultra thick, draw=white, double=white, double distance=0pt, bend right = 10] (8,2) node [] {};\draw (5,2.5) edge [ultra thick, bend right = 10] (8,2) node [] {};
\draw (5,2.5) edge [ultra thick, draw=white, double=white, double distance=0.5pt, bend left = 10] (2,8) node [] {};\draw (5,2.5) edge [ultra thick, bend left = 10] (2,8) node [] {};
\draw (5,2.5) edge [ultra thick, draw=white, double=white, double distance=0.5pt, bend right = 10] (8,8) node [] {};\draw (5,2.5) edge [ultra thick, bend right = 10] (8,8) node [] {};

\node () at (0.3,0.5) {$p_1$};
\node () at (9.7,0.5) {$p_3$};
\node () at (0.3,9.5) {$p_2$};
\node () at (9.7,9.5) {$p_4$};

\node () at (3.6,1.3) {$x_2$};
\node () at (6.5,1.3) {$x_6$};
\node () at (1.5,3.6) {$x_1$};
\node () at (1.5,6.3) {$x_4$};
\node () at (8.5,3.6) {$x_5$};
\node () at (8.5,6.3) {$x_8$};
\node () at (3.6,8.7) {$x_3$};
\node () at (6.5,8.7) {$x_7$};

\draw[fill, thick, color=Black] (2,2) circle (5pt);
\draw[fill, thick, color=Black] (8,2) circle (5pt);
\draw[fill, thick, color=Black] (2,8) circle (5pt);
\draw[fill, thick, color=Black] (8,8) circle (5pt);
\draw[fill, thick, color=Black] (5,7.5) circle (5pt);
\draw[fill, thick, color=Black] (5,2.5) circle (5pt);
\end{tikzpicture}
\caption{Massless non-planar 3-loop box (the crown graph, $G_{\bullet\bullet}$)}
\label{fig:3-loop-diagram}
\end{figure}
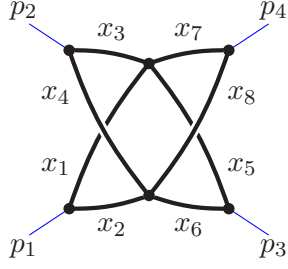

The non-planar crown graph, $G_{\bullet\bullet}$, see Fig.~\ref{fig:3-loop-diagram}, depends on the Mandelstam invariants $s_{12}$, $s_{13}$ and $s_{23}$.
We can use momentum conservation to eliminate $s_{23}=-s_{12}-s_{13}$.
The integral can be written in parameter space as,
\begin{align}
J_{G_{\bullet\bullet}}(\mathbf{s})&=\Gamma\left(2+3\epsilon\right) \lim_{\delta\to0^+}I_{G_{\bullet\bullet}}(\mathbf{s};\delta)\\
I_{G_{\bullet\bullet}}(\mathbf{s};\delta)&=\int_{\mathbb{R}^8_{\geq0}}\prod_{i=1}^8\mathrm{d}x_i \,\frac{\mathcal{U}(\mathbf{x})^{4\epsilon} }{\left(\mathcal{F}(\mathbf{x}, \mathbf{s})- i\delta\right)^{2+3\epsilon}}\delta\left(1-\alpha(\mathbf{x})\right)
\label{crown-integral}
\end{align}
where the $\mathcal{U}$ and $\mathcal{F}$ polynomials for $G_{\bullet\bullet}$ are given by
\begin{align}
\mathcal{U}(\mathbf{x}) = \, &\left(x_1+x_2\right)\left(x_3+x_4\right)\left(x_5+x_6\right)+\left(x_1+x_2\right)\left(x_3+x_4\right)\left(x_7+x_8\right)+ \nonumber \\ &\left(x_1+x_2\right)\left(x_5+x_6\right)\left(x_7+x_8\right)+\left(x_3+x_4\right)\left(x_5+x_6\right)\left(x_7+x_8\right),\\
\mathcal{F}(\mathbf{x}, \mathbf{s}) = &-s_{12}\left(x_2x_5-x_1x_6\right)\left(x_4x_7-x_3x_8\right)-s_{13}\left(x_2x_3-x_1x_4\right)\left(x_6x_7-x_5x_8\right).
\end{align}
We restrict to the massless $2\rightarrow2$ physical scattering regime,
\begin{align}
\mathbf{s}_\mathrm{phys} = \{ 0 < s_{12} < \infty, -s_{12} < s_{13} < 0 \}.
\end{align}

In Ref.~\cite{Gardi:2024axt} it was shown that this integral has a solution of the Landau equations within the integration domain for generic physical kinematics, preventing evaluation with the usual contour deformation procedure.
The solution proposed by the authors is to transform the Feynman parameters such that each factor of $\mathcal{F}(\mathbf{x};\mathbf{s})$ is linear.
This can be achieved by the transformations
\begin{align}
x_1 \rightarrow x_1\frac{x_2}{x_8},\quad x_3 \rightarrow x_3\frac{x_4}{x_8},\quad x_5 \rightarrow x_5\frac{x_6}{x_8},\quad x_7 \rightarrow x_7\frac{x_8}{x_8}\quad,
\end{align}
with Jacobian,
\begin{align}
\mathcal{J} = \frac{x_2x_4x_6}{x_8^3}.
\end{align}
The resulting $\mathcal{F}$-polynomial is given by,
\begin{align}
\mathcal{F}(\mathbf{x};\mathbf{s}) = \frac{x_2 x_4 x_6}{x_8} \left( -s_{12} (x_1-x_5)(x_3-x_7) -s_{13} (x_1-x_3)(x_5-x_7)\right).
\end{align}
The integral can then be dissected into  $4!=24$ regions each defined by a strict ordering of $x_i \ge x_j \ge x_k \ge x_l$ with $\{i,j,k,l\}$ all permutations of $\{1,3,5,7\}$.
Each of the resulting integrals will have polynomials of definite sign multiplying each invariant $s_{12}$ or $s_{13}$ and can in principle be evaluated using contour deformation.
Taking into account the symmetry of the integral, $I_{G_{\bullet\bullet}}$ can now be expressed as a sum over six integrals,
\begin{align}
        I_{G_{\bullet\bullet}}(\mathbf{s};\delta)&=4\sum_{K}I_K(\mathbf{s};\delta),\\[10pt]I_K(\mathbf{s};\delta)&=\int_{\mathbb{R}^8_{\geq0}}\prod_{i=1}^8\mathrm{d}x_i \,\frac{\mathcal{U}_K(\mathbf{x})^{4\epsilon} }{\left(x_2x_4x_6\right)^{1+3\epsilon}x_8^{1+9\epsilon}\left(\mathcal{F}_K\left(\mathbf{x},\mathbf{s}\right)-i\delta\right)^{2+3\epsilon}}\delta\left(1-\alpha(\mathbf{x})\right),
    \label{Landau_sum}
\end{align}
where $K \in \{A,B,C,D,E,F\}$, denoting each of the six integrals resulting from the resolution of the Landau singularity. 
The procedure protects the homogeneity of $\mathcal{U}_K$ and $\mathcal{F}_K$ as well as retaining the positive definiteness of the former.

Strictly, not all terms in the sum of \eqref{Landau_sum} require an $i\delta$-prescription, this is clear from examining the resultant $\mathcal{F}_K$ polynomials themselves\footnote{We use the symmetry to reduce to the six where $\mathcal{F}_K$ depends only on $\{x_1,x_3,x_5\}$ and not $x_7$.}:
    \begin{align}  \mathcal{F}_A\left(\mathbf{x},\mathbf{s}\right)&=-\left[s_{12}x_3\left(x_1+x_3+x_5\right)+\left(s_{12}+s_{13}\right)x_1x_5\right]\\
    \mathcal{F}_B\left(\mathbf{x},\mathbf{s}\right)&=-\left[\left(s_{12}+s_{13}\right)x_1x_3+s_{13} x_5\left(x_1+x_3+x_5\right)\right]\label{eq:fb}\\
    \mathcal{F}_C\left(\mathbf{x},\mathbf{s}\right)&=-\left[s_{12} x_1\left(x_1+x_3+x_5\right)-s_{13} x_3x_5\right]\\
    \mathcal{F}_D\left(\mathbf{x},\mathbf{s}\right)&=+\left[\left(s_{12}+s_{13}\right)x_5\left(x_1+x_3+x_5\right)+s_{13}x_1x_3\right]\label{eq:fd}\\
    \mathcal{F}_E\left(\mathbf{x},\mathbf{s}\right)&=+\left[s_{12}x_3x_5-s_{13}x_1\left(x_1+x_3+x_5\right)\right]\\
    \mathcal{F}_F\left(\mathbf{x},\mathbf{s}\right)&=+\left[s_{12}x_1x_5+\left(s_{12}+s_{13}\right)x_3\left(x_1+x_3+x_5\right)\right].
    \end{align}
In the kinematic regime $\mathbf{s}_\mathrm{phys}$, we have $s_{12}+s_{13}>0$ and $-s_{13}>0$, using this, we can immediately make a number of remarks.
Firstly, it is clear that integrals $E$ and $F$ will have manifestly positive integrands, hence, they do not require an $i\delta$-prescription. 
Secondly, integrals $A$ and $C$ can be trivially brought into Euclidean form by factoring out $(-1-i\delta)^{-2-3\epsilon}$ from the integrals, similarly to the example in Eq.~\eqref{negfactoring}, since $\mathcal{F}_A$ and $\mathcal{F}_C$ are manifestly negative within the domain of integration for the assumed kinematic regime.
Finally, we remark that integrals $B$ and $D$ would naively require a contour deformation due to zeroes of the $\mathcal{F}_K$ polynomials within the domain of integration (strictly away from the boundary); these can be traced to the appearance of $+s_{13}<0$ appearing inside the brackets of $\mathcal{F}_B$ and $\mathcal{F}_D$ in Eq.~\eqref{eq:fb} and Eq.~\eqref{eq:fd}. 
We observe that, had we assumed a different kinematic regime, a different subset of the six integrals would be automatically Euclidean, but we stress that there is no regime where all six integrals are same-sign.
\begin{figure}
\centering
\usetikzlibrary{shapes.geometric, arrows}

\tikzstyle{startstop} = [rectangle, rounded corners, 
minimum width=3cm, 
minimum height=1cm,
text centered, 
draw=black, 
fill=red!5]

\tikzstyle{io} = [rectangle, 
minimum width=3cm, 
minimum height=1cm, text centered, 
draw=black, fill=blue!5]

\tikzstyle{process} = [rectangle, 
minimum width=3cm, 
minimum height=1cm, 
text centered, 
draw=black, 
fill=orange!5]

\tikzstyle{decision} = [rectangle, rounded corners, 
minimum width=3cm, 
minimum height=1cm, 
text centered, 
draw=black, 
fill=green!5]
\tikzstyle{arrow} = [thick,->,>=stealth]

\begin{tikzpicture}[node distance=2cm,thick,scale=1, every node/.style={scale=0.8}]

\node (start) [startstop] {$\mathcal{F}_B\left(\mathbf{x};\mathbf{s}\right)=-\left[\left(s_{12}+s_{13}\right)x_1x_3+s_{13}x_5\left(x_1+x_3+x_5\right)\right]$};

\node (rescale1) [io, below of=start,yshift=-1cm] {$\mathcal{F}_B\rightarrow\frac{s_{12}+s_{13}}{-s_{13}}\left[s_{12} x_5^2+s_{13}\left(x_1-x_5\right)\left(x_3-x_5\right)\right]$};

\node (split1L) [process, below of=rescale1, yshift=-1cm, xshift=-6.5cm] {$\mathcal{F}_B^\prime\rightarrow\frac{s_{12}+s_{13}}{-s_{13}}\left[s_{12}x_5^2+s_{13}x_1\left(x_3-x_5\right)\right]$};

\node (split1R) [decision, right of=split1L, xshift=+7cm] {$\mathcal{F}_B^{+,1}=\frac{s_{12}+s_{13}}{-s_{13}}\left[\left(s_{12}x_1+\left(s_{12}+s_{13}\right)x_5\right)\left(x_1+x_5\right)-s_{13}x_3x_5\right]$};

\node (split2L) [process, below  of=split1L, yshift=-1cm, xshift=-0.3cm] {$\mathcal{F}_B^{\prime \prime}\rightarrow\frac{s_{12}+s_{13}}{-s_{13}}\left[s_{12}x_5^2+s_{13} x_1 x_3\right]$};

\node (split2R) [decision, right of=split2L, xshift=+5cm] {$\mathcal{F}_B^{+,2}=\frac{s_{12}+s_{13}}{-s_{13}}\left[s_{12}\left(x_3+x_5\right)^2-s_{13} x_1 x_5\right]$};

\node (split3L) [decision, below  of=split2L, yshift=-1cm, xshift=-0.3cm] {$\mathcal{F}_B^{+,3}=\frac{s_{12}+s_{13}}{-s_{13}}\left[\frac{s_{12}x_5^2x_8}{x_1+x_8}\right]$};

\node (split3R) [decision, right of=split3L, xshift=+5cm] {$\mathcal{F}_B^{-}=-\left[\left(s_{12}+s_{13}\right)x_1x_3\right]$};

\draw [arrow] (start) -- node[anchor=west,xshift=+0.1cm] {$x_5\rightarrow\frac{s_{12}+s_{13}}{-s_{13}}x_5$} (rescale1);
\draw [arrow] (rescale1) -- node[anchor=east,xshift=-0.5cm] {$x_1\rightarrow x_1+x_5$} (split1L);
\draw [arrow] (rescale1) -- node[anchor=west,xshift=+0.5cm] {$x_5\rightarrow x_1+x_5$} (split1R);
\draw [arrow] (split1L) -- node[anchor=east,xshift=-0.2cm] {$x_3\rightarrow x_3+x_5$} (split2L);
\draw [arrow] (split1L) -- node[anchor=west,xshift=+0.8cm] {$x_5\rightarrow x_3+x_5$} (split2R);
\draw [arrow] (split2L) -- node[anchor=east,xshift=-0.2cm] {$x_1\rightarrow \frac{x_1}{x_1+x_8}\left(\frac{s_{12} x_5^2}{-s_{13}x_3}\right)$} (split3L);
\draw [arrow] (split2L) -- node[anchor=west,xshift=+0.8cm] {$x_1\rightarrow x_1+\left(\frac{s_{12} x_5^2}{-s_{13}x_3}\right)$} (split3R);

\end{tikzpicture}
    \caption{
    Example transformations for resolving integral $B$.
    The initial integral (red) is mapped to four integrals (green), three of which are manifestly positive and one of which is manifestly negative.}
    \label{fig:crown_flowchart}
\end{figure}
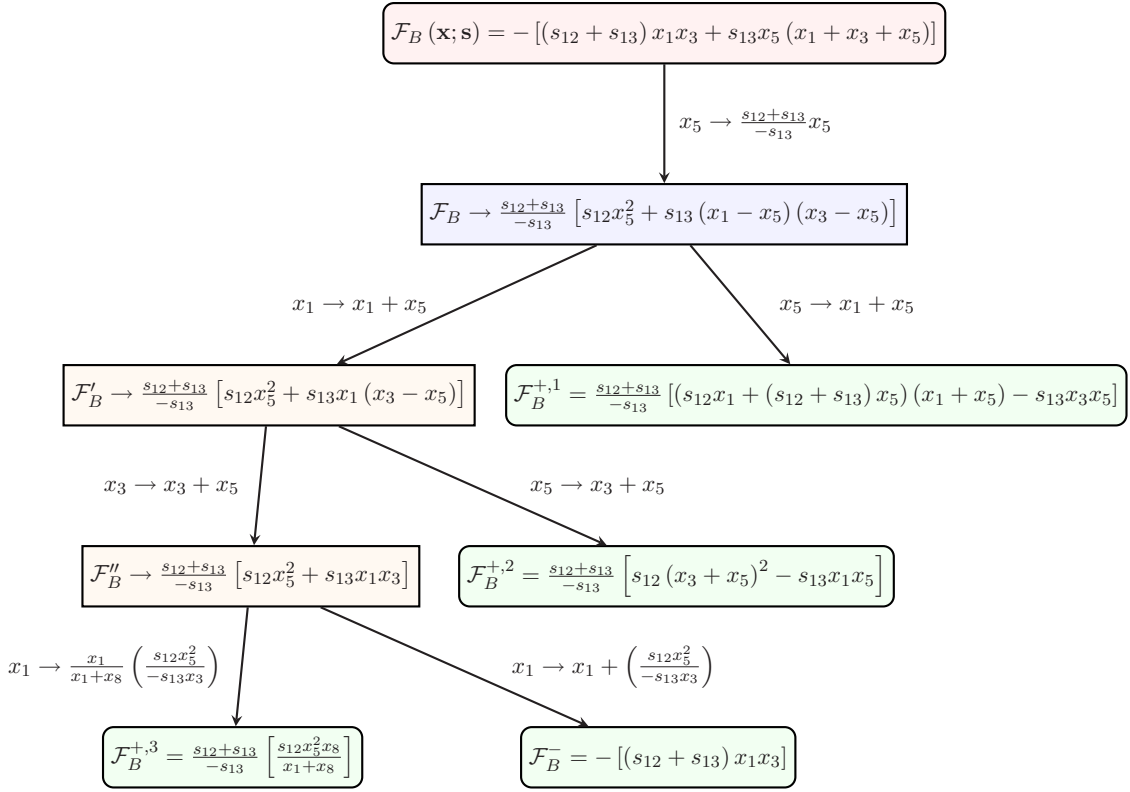

We focus on integrals $B$ and $D$ and resolve them such that we do not need an explicit contour deformation. 
The integral can be resolved both with the default application of Algorithm~\ref{alg:univariate}, at the expense of introducing algebraic transformations, as well as with an iterated generalisation of the algorithm (the detail of which we leave for future work).
Here, we demonstrate an alternative resolution procedure, first described in Ref.~\cite{Jones:2024gmw} where it was used to resolve the 1-loop box with an off-shell leg of Section~\ref{sec:box-off} in detail.
Generally, for Feynman integrals, we can perform the following changes of variables,
\begin{align}
x_i &\rightarrow \alpha^n x_i,\\
x_i &\rightarrow x_j^n x_i,\\
x_i &\rightarrow x_i + x_j \quad \& \quad x_j\rightarrow x_i + x_j,
\end{align}
where $\alpha >0$ is a positive constant possibly depending on kinematic invariants or masses.
Whenever we shift a variable such that ${x_a\rightarrow x_a+x_b}$, we must consider the converse case ${x_b\rightarrow x_a+x_b}$ to cover the entire original domain. 
This corresponds to bisecting the integral by inserting  $\theta\left(x_i-x_j\right)+\theta\left(x_j-x_i\right)=1$.
These transformations are similar to those that appear in sector decomposition, which resolves singularities on the boundary of integration.

In Fig.~\ref{fig:crown_flowchart}, we record one possible chain of transformations for resolving integral $B$. 
The integral is mapped to four different integrals, three positive contributions and one negative contribution,
\begin{align}
I_B(\mathbf{s};\delta)=\sum_{n_+=\ \!\!1}^{3}I_B^{+,n_+}(\mathbf{s})+\left(-1-i\delta\right)^{-2-3\epsilon}I_B^{-}(\mathbf{s}).
\label{I_B_decomp}
\end{align}
This approach generated more integrals than necessary; however, the transformations at each step are simple, as are the resulting integrands.
The complete integrand of the negative contribution $I_B^-$, is given by,
\begin{align}
    I_B^-(\mathbf{s})=&\ \frac{1}{\left(s_{12}+s_{13}\right)^{1+3\epsilon}\left(-s_{13}\right)^{1+8\epsilon}}\int_{\mathbb{R}^8_{\geq0}}\prod_{i=1}^8\mathrm{d}x_i\, \ \mathcal{I}_{B}^-(\mathbf{s})\ \delta\left(1-\alpha(\mathbf{x})\right)\\[10pt]
    \mathcal{I}_{B}^-(\mathbf{s})=&\ x_1^{-2-3 \epsilon} x_3^{-2-7 \epsilon}x_8^{-1-9 \epsilon} \left(x_2 x_4 x_6\right)^{-1-3 \epsilon} \times\nonumber\\& \Bigl[-s_{13} x_3 x_4 x_6 x_8 \left(x_7+x_8\right) \left(x_3+x_5+x_7+x_8\right) \left[s_{12} x_5-s_{13} \left(x_3+x_7+x_8\right)\right]+\nonumber\\&x_2 x_6 x_8 \left(x_7+x_8\right) \left[s_{12} x_5 \left(x_3+x_5\right)-s_{13} x_3 \left(x_1+x_3+x_5+x_7+x_8\right)\right]\times\nonumber\\& \left[s_{12} x_5-s_{13} \left(x_3+x_7+x_8\right)\right]+x_2 x_4 x_6 \left(x_3+x_5+x_7+x_8\right)\times \nonumber\\&\left[s_{12} x_5 \left(x_3+x_5\right)-s_{13} x_3 \left(x_1+x_3+x_5+x_7+x_8\right)\right] \left[s_{12} x_5-s_{13} \left(x_3+x_7+x_8\right)\right]+\nonumber\\&-s_{13} x_2 x_4 x_8 \left(x_7+x_8\right) \left(x_3+x_5+x_7+x_8\right) \times\nonumber\\&\left[s_{12} x_5 \left(x_3+x_5\right)-s_{13} x_3 \left(x_1+x_3+x_5+x_7+x_8\right)\right]\Bigr]^{4 \epsilon }.
\end{align}
For brevity, we do not state all of the positive resolutions, they can be obtained by applying the transformations given in Fig.~\ref{fig:crown_flowchart}.

Integral $D$ may be similarly resolved into three positive contributions and one negative contribution such that our initial integral, $I_{G_{\bullet\bullet}}$, can be expressed as sum over twelve integrals with manifestly positive integrands,
    \begin{equation}
    I_{G_{\bullet\bullet}}=4\left[\sum_{n_+=\ \!\!1}^{3}I_B^{+,n_+}+\sum_{n_+=\ \!\!1}^{3}I_D^{+,n_+}+I_E^++I_F^+\right]+4\left(-1-i\delta\right)^{-2-3\epsilon}\left[I_A^{-}+I_B^{-}+I_C^{-}+I_D^{-}\right].
    \label{gdotot_decomp}
\end{equation}
where the positive contributions $I_E^+$ and $I_F^+$ are simply $I_E$ and $I_F$ and the negative contributions $I_A^-$ and $I_C^-$ are merely $I_A$ and $I_C$ with $\left(-1-i\delta\right)^{-2-3\epsilon}$ factored out accordingly.

\subsection{Massive Examples}
\label{ssec:massive}
In this section, we present 1-, 2- and 3-loop examples of integrals with massive propagators.
Understanding the resolution of such integrals is important for applying the method to the calculation of massive phenomenologically-relevant amplitudes, for example, involving massive quarks, electroweak bosons or Higgs bosons. 
The primary complication in the massive case is that the $\mathcal{F}$ polynomial gets modified by a term proportional to $\mathcal{U}$ such that each Feynman parameter associated with a massive propagator may appear quadratically in the monomials of $\mathcal{F}$,
\begin{align}
\mathcal{F}(\mathbf{x};\mathbf{s})=\mathcal{F}_0(\mathbf{x};\mathbf{s})+\mathcal{U}(\mathbf{x};\mathbf{s})\sum_{j=1}^{N}m_j^2x_j,
\end{align}
where $\mathcal{F}_0$ is the polynomial corresponding to the massless version of the integral. 
We analyse the 1-loop massive bubble and triangle initially in Sections~\ref{sec:bubble-eq}, \ref{sec:bubble-uneq} and \ref{sec:triangle} before applying the method to 2-loop elliptic and 3-loop hyperelliptic examples in Sections~\ref{sec:sunrise} and \ref{sec:banana}, respectively.


\begin{figure}[t]
    \centering
    \begin{subfigure}[t]{0.235\textwidth}
        \centering
                \begin{tikzpicture}[baseline=13ex,scale=0.65]
            \coordinate (x1) at (1, 2) ;
            \coordinate (x2) at (3, 2) ;
            \node (p1) at (0, 2) {$s$};
            \node (p2) at (4, 2) {};
            \draw[ultra thick,color=ForestGreen] (x1) -- (p1);
            \draw[ultra thick,color=ForestGreen] (x2) -- (p2);

            \draw[ultra thick,color=Purple] (x2) arc (0:180:1) node [midway,yshift=+6.5pt,color=Black] {$m$};
            \draw[ultra thick,color=Purple] (x2) arc (0:-180:1) node [midway,yshift=+6.5pt,color=Black] {$m$};

            \draw[fill,thick,color=Blue] (x1) circle (1pt);
            \draw[fill,thick,color=Blue] (x2) circle (1pt);
        \end{tikzpicture}
       \caption{}\label{eqmassbub}
    \end{subfigure}
    \hspace{0.01cm}
    \begin{subfigure}[t]{0.235\textwidth}
        \centering
                \begin{tikzpicture}[baseline=13ex,scale=0.65]
            \coordinate (x1) at (1, 2) ;
            \coordinate (x2) at (3, 2) ;
            \node (p1) at (0, 2) {$s$};
            \node (p2) at (4, 2) {};
            \draw[ultra thick,color=ForestGreen] (x1) -- (p1);
            \draw[ultra thick,color=ForestGreen] (x2) -- (p2);

            \draw[ultra thick,color=Purple] (x2) arc (0:180:1) node [midway,yshift=+6.5pt,color=Black] { $m_1$};
            \draw[ultra thick,color=Purple] (x2) arc (0:-180:1) node [midway,yshift=+6.5pt,color=Black] {$m_2$};

            \draw[fill,thick,color=Blue] (x1) circle (1pt);
            \draw[fill,thick,color=Blue] (x2) circle (1pt);
        \end{tikzpicture}
        \caption{}\label{uneqmassbub}
    \end{subfigure}
    \hspace{0.01cm}
    \begin{subfigure}[t]{0.235\textwidth}
        \centering
        \begin{tikzpicture}[baseline=13ex,scale=0.65]
            \coordinate (x1) at (1, 2) ;
            \coordinate (x2) at (3, 2) ;
            \node (p1) at (0, 2) {$s$};
            \node (p2) at (4, 2) {};
            \draw[ultra thick,color=ForestGreen] (x1) -- (p1);
            \draw[ultra thick,color=ForestGreen] (x2) -- (p2);

            \draw[ultra thick,color=Purple] (x2) arc (0:180:1) node [midway,yshift=+6.5pt,color=Black] { $m$};
            \draw[ultra thick,color=Purple] (x1)--(x2) node [midway,yshift=+6.5pt,color=Black] {$m$};
            \draw[ultra thick,color=Purple] (x2) arc (0:-180:1) node [midway,yshift=+6.5pt,color=Black] {$m$};

            \draw[fill,thick,color=Blue] (x1) circle (1pt);
            \draw[fill,thick,color=Blue] (x2) circle (1pt);
        \end{tikzpicture}%
        \caption{}\label{eqmasssun}
    \end{subfigure}
    \hspace{0.01cm}
    \begin{subfigure}[t]{0.235\textwidth}
        \centering
        \begin{tikzpicture}[baseline=13ex,scale=0.65]
    \coordinate (x1) at (1, 2);
    \coordinate (x2) at (3, 2);

    \node (p1) at (0, 2) { $s$};
    \node (p2) at (4, 2) {};
    \node (m3) at (2, 2) {$m$};
    \node (m4) at (2, 2.65) {$m$};

    \draw[ultra thick, color=ForestGreen] (x1) -- (p1);
    \draw[ultra thick, color=ForestGreen] (x2) -- (p2);

    \draw[ultra thick, color=Purple] (x2) arc (0:180:1) node [midway, yshift=+6.5pt, color=Black] {$m$}; 
    \draw[ultra thick, color=Purple] (x2) arc (0:-180:1) node [midway, yshift=+6.5pt, color=Black] { $m$}; 

     \draw[ultra thick, color=Purple] (x1) to[out=40, in=140] (x2);
     \draw[ultra thick, color=Purple] (x1) to[out=-40, in=-140] (x2);

    \draw[fill, thick, color=Blue] (x1) circle (1pt);
    \draw[fill, thick, color=Blue] (x2) circle (1pt);
\end{tikzpicture}
        \caption{}\label{banana-diag}
    \end{subfigure}
    \caption{The $L$-loop banana integrals resolved in this section ($L\in\{1,2,3\}$): the equal mass (\ref{eqmassbub}) and unequal mass (\ref{uneqmassbub}) bubble integrals, the equal mass elliptic sunrise integral (\ref{eqmasssun}), and the 3-loop equal mass banana integral (\ref{banana-diag}).}
    \label{fig:massivediags}
\end{figure}
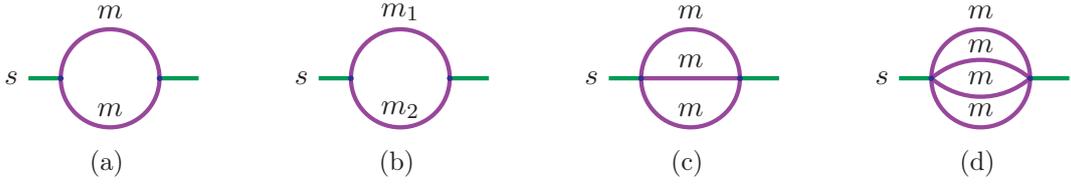
\subsubsection{Equal Mass Bubble}
\label{sec:bubble-eq}
To investigate how massive integrals can be resolved, we begin by considering the bubble integral in Fig.~\ref{eqmassbub} with internal propagators of equal (positive) mass, $m^2>0$.
In Feynman parameter space, the integral can be written as,
\begin{align}
J_{\mathrm{bub}}(\mathbf{s})&=\Gamma\left(\epsilon\right)\lim_{\delta\to0^+}I_{\mathrm{bub}}(\mathbf{s};\delta),\\
I_{\mathrm{bub}}(\mathbf{s};\delta)&=\int_{\mathbb{R}^2_{\geq0}}\!\!\mathrm{d}x_1\mathrm{d}x_2\frac{\mathcal{U}(\mathbf{x})^{-2+2\epsilon}}{\left(\mathcal{F}(\mathbf{x};\mathbf{s})-i\delta\right)^{\epsilon}}\delta\left(1-\alpha(\mathbf{x})\right).
\end{align}
with the Symanzik polynomials,
\begin{align}
\mathcal{U}(\mathbf{x}) &=  x_1+x_2,\\
\mathcal{F}(\mathbf{x};\mathbf{s}) &= -sx_1x_2+m^2\left(x_1+x_2\right)^2.
\end{align}
Making the choice $\alpha(\mathbf{x})=x_1+x_2$, we can perform the $x_2$ integral to obtain,
\begin{equation}
I_{\mathrm{bub}}(\mathbf{s};\delta)=\int_{0}^{1}\!\!\mathrm{d}x_1\frac{1}{\left(-sx_1\left(1-x_1\right)+m^2-i\delta\right)^{\epsilon}}.
\label{ibubkin}
\end{equation}
The denominator of the integrand in \eqref{ibubkin} can develop a zero within the domain of integration above the threshold for the on-shell production of the intermediate pair of massive particles, $s>4m^2$.
We wish to consider the integral within the mixed-sign, above-threshold, kinematic regime.
We find it convenient to introduce the abbreviation,
\begin{align}
\beta^2=\frac{s-4m^2}{s}\in\left(0,1\right),
\label{eqmassbeta}
\end{align}
and eliminate $s>0$ for $\beta\in\left(0,1\right)$. 
After exploiting the symmetry about $x_1=1/2$ the integral may be written as,
\begin{align}
I_{\mathrm{bub}}(\mathbf{s};\delta) &= \left(\frac{1-\beta^2}{m^2}\right)^{\epsilon} \int_{0}^{1}\!\!\mathrm{d}x_1\frac{1}{\left(\left(1-x_1\right)^2-\beta^2-i \delta\right)^{\epsilon}} \nonumber\\&= \left(\frac{1-\beta^2}{m^2}\right)^{\epsilon} \frac{(\beta^2)^{\frac{1}{2}-\epsilon}}{2}\tilde{I}_{\mathrm{bub}}(\mathbf{s};\delta), \\
\tilde{I}_{\mathrm{bub}}(\mathbf{s};\delta) &=   \int_{0}^{\frac{1}{\beta^2}}\!\!\mathrm{d}z_1\, z_1^{-\frac{1}{2}} \left(z_1-1-i \delta\right)^{-\epsilon}. \label{eq:bubble_sign}
\end{align}
In the final step, the causal prescription $i\delta$ can be absorbed into the choice of contour.
For this integral, it can be viewed as an analytic continuation prescription of the kinematic invariant $\beta^2 \rightarrow \beta^2+i\delta$.
The second factor in Eq.~\eqref{eq:bubble_sign} is negative for $z_1<1$ and non-negative for $z_1\ge 1$, the integral is therefore not of the form given in Eq.~\eqref{eq:decomp}. 
Splitting the integral at the point at which the denominator vanishes within the integration domain, $z_1=1$, and remapping the resulting integrals to the original positive unit interval gives,
\begin{align}
\tilde{I}_{\mathrm{bub}}(\mathbf{s};\delta) = &  \int_{0}^{1}\!\!\mathrm{d}z_1\, z_1^{-\frac{1}{2}} \left(z_1-1-i \delta\right)^{-\epsilon} + \int_{1}^{\frac{1}{\beta^2}}\!\!\mathrm{d}z_1\, z_1^{-\frac{1}{2}} \left(z_1-1-i \delta\right)^{-\epsilon}, \nonumber\\
= & \int_{0}^{1}\!\!\mathrm{d}z_1\, z_1^{-\frac{1}{2}} \left(z_1-1-i \delta\right)^{-\epsilon} + \int_0^1 \mathrm{d}z \, \gamma (\gamma z + 1)^{-\frac{1}{2}}(\gamma z-i\delta)^{-\epsilon}, \nonumber\\
= & (-1-i\delta)^{-\epsilon}\int_{0}^{1}\!\!\mathrm{d}z_1\, z_1^{-\frac{1}{2}} \left(1-z_1\right)^{-\epsilon} + \int_0^1 \mathrm{d}z \, \gamma (\gamma z + 1)^{-\frac{1}{2}}(\gamma z)^{-\epsilon}, \nonumber\\
= & (-1-i\delta)^{-\epsilon} \, \tilde{I}_{\mathrm{bub}}^{-}(\mathbf{s}) + \tilde{I}_{\mathrm{bub}}^{+}(\mathbf{s}),
\label{eq:equal_mass_bub}
\end{align}
where we have introduced the abbreviation $\gamma=(1-\beta^2)/\beta^2$ and $\gamma > 0$. The integrands of $\tilde{I}_{\mathrm{bub}}^{\pm}(\mathbf{s})$ are now manifestly positive inside the domain of integration.
For this 1-loop bubble, the integrals can be straightforwardly evaluated analytically,
\begin{align}
\tilde{I}_{\mathrm{bub}}^{-}(\mathbf{s}) &= \int_{0}^{1}\!\!\mathrm{d}z_1\, z_1^{-\frac{1}{2}} \left(1-z_1\right)^{-\epsilon} = \frac{\Gamma(1-\epsilon)}{\Gamma(\frac{3}{2}-\epsilon)} \pi^\frac{1}{2},\\
\tilde{I}_{\mathrm{bub}}^{+}(\mathbf{s}) &=  \int_0^1 \mathrm{d}z \, \gamma (\gamma z + 1)^{-\frac{1}{2}}(\gamma z)^{-\epsilon} = 2\int_0^{\sqrt{1+\gamma}-1} \!\!\mathrm{d}u \, u^{-\epsilon}(2+u)^{-\epsilon}\nonumber\\&=\frac{2^{1-\epsilon}}{1-\epsilon}\,(\sqrt{1+\gamma}-1)^{1-\epsilon} \,{_2F_1}\left(1-\epsilon,\,\epsilon, \,2-\epsilon,\,-\frac{1}{2}(\sqrt{1+\gamma}-1) \right).
\label{eq:equal_mass_bub_beta}
\end{align}
In this form, the integrals $\tilde{I}_{\mathrm{bub}}^{\pm}(\mathbf{s})$ are purely real and depend only on the real kinematic invariant, $\beta^2$ (or equivalently $\gamma$), the analytic continuation of the integral is manifest in the $(-1-i\delta)^{-\epsilon}$ in Eq.~\eqref{eq:equal_mass_bub}. $\tilde{I}_{\mathrm{bub}}^{+}$ can be expressed in terms of an incomplete Beta function or a ${_2F_1}$ hypergeometric function as shown above after changing variables via $z=\frac{(2+u)u}{\gamma}$.

\subsubsection{Unequal Mass Bubble}
\label{sec:bubble-uneq}
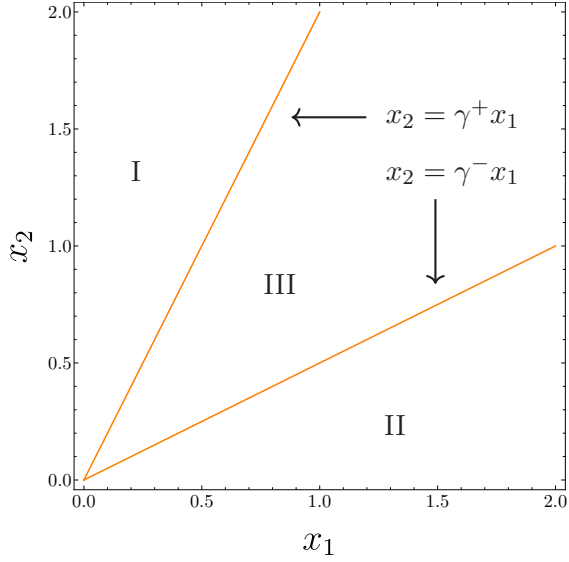
\begin{figure}[t]
    \centering
    \begin{tikzpicture}
        \node[anchor=south west,inner sep=0] (img) at (0,0) {\includegraphics[width=0.5\textwidth]{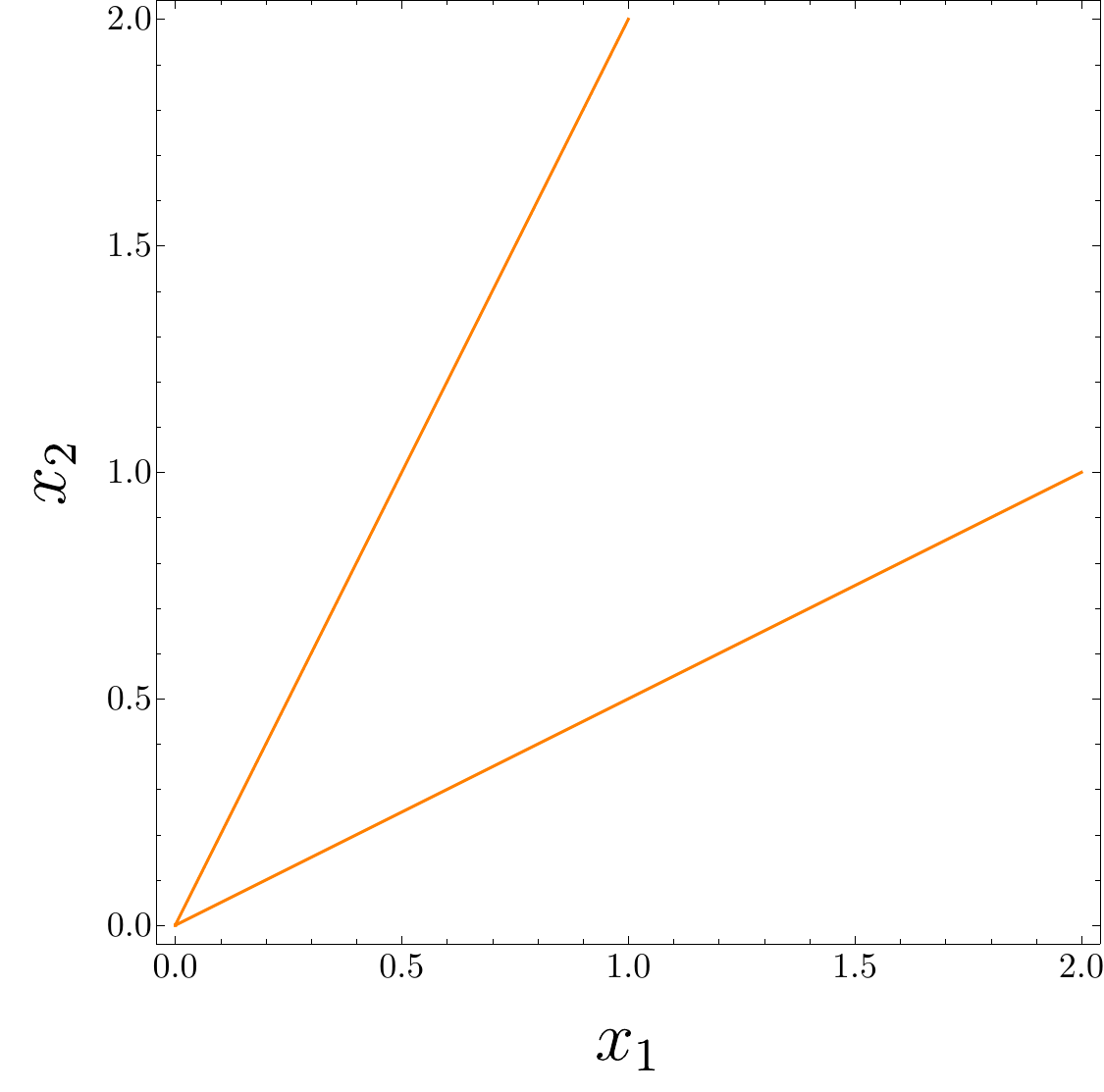}};
        \begin{scope}[x={(img.south east)}, y={(img.north west)}]
            \node at (0.8,0.8) {$x_2 = \gamma^+ x_1$};
            \node at (0.8,0.7) {$x_2 = \gamma^- x_1$};
            \draw [->, thick] (0.77,0.65) -- (0.77,0.5);
            \draw [->, thick] (0.65,0.795) -- (0.52,0.795);
            \node at (0.25,0.7) {I};
            \node at (0.7,0.25) {II};
            \node at (0.5,0.5) {III};
        \end{scope}
    \end{tikzpicture}
    \caption{The variety of $\widetilde{\mathcal{F}}$ and the three regions of the integration domain which it separates. In regions (I) and (II), $\widetilde{\mathcal{F}}>0$ whereas in region (III), $\widetilde{\mathcal{F}}<0$. The kinematic regime is given by $\beta^2\in(0,1)$ (with $m_1^2,m_2^2>0$) equivalent to $\gamma^+=\frac{1}{\gamma^-}>0$.}
    \label{fig:bubblevariety}
\end{figure}

In our analysis of the equal mass bubble integral, we first made a specific choice of $\delta$-functional and relied heavily on inspecting the analytic properties of the integrand.
This procedure can be generalised to the bubble integral with unequal masses, $m_1$ and $m_2$, depicted in Fig.~\ref{uneqmassbub}. 
However, to illustrate the general principle from a different perspective let us instead present an alternative way of resolving the massive bubble. 
In this case, we do not specify the argument of the $\delta$-functional, $\alpha(\mathbf{x})$, and focus solely on the $\mathcal{F}$ polynomial,
\begin{equation}
    \mathcal{F}=-sx_1x_2+\left(x_1+x_2\right)\left(m_1^2x_1+m_2^2x_2\right).
\end{equation}
Analogously to the equal mass bubble, we define,
\begin{equation}
    \beta^2=\frac{s-\left(m_1+m_2\right)^2}{s-\left(m_1-m_2\right)^2}\in\left(0,1\right),
\end{equation}
which reduces to Eq.~\eqref{eqmassbeta} in the limit that $m_2= m_1=m$. In the next step, we rescale the Feynman parameters $x_1$ and $x_2$ with the transformations  $x_i\rightarrow\frac{x_i}{m_i}$ such that we can perform the resolution procedure on the dimensionless polynomial,
\begin{equation}
    \widetilde{\mathcal{F}}=x_1^2+x_2^2-2\frac{1+\beta^2}{1-\beta^2}x_1 x_2.
\end{equation}
In Fig.~\ref{fig:bubblevariety}, we plot the variety of $\widetilde{\mathcal{F}}$ (that is to say, the set of points where $\widetilde{\mathcal{F}}=0$). 
We find that it separates the integration domain into three regions defined by the sign of $\widetilde{\mathcal{F}}$ -- in regions (I) and (II) of Fig.~\ref{fig:bubblevariety} we have $\widetilde{\mathcal{F}}>0$, while in region (III) we have $\widetilde{\mathcal{F}}<0$.

The essential problem is to construct transformations of the Feynman parameters such that we can convert the unequal mass bubble integral into three integrals each of which is over one of the regions in Fig.~\ref{fig:bubblevariety}. 
To do this, we first need to solve the variety $\widetilde{\mathcal{F}}=0$ for one of the Feynman parameters. 
We find that,
\begin{equation}
    x_2= \gamma^\pm x_1, \quad  \mathrm{with} \quad \gamma^\pm = \frac{1}{\gamma^\mp} = \frac{1 \pm \beta}{1 \mp \beta},
\end{equation}
defines the variety $\widetilde{\mathcal{F}}=0$. 
Once we have this solution, the construction of the transformations is straightforward,
\begin{align}
&\mathrm{I}:& &x_2 \rightarrow y_2 = x_2 + \gamma^+ x_1,& &\widetilde{\mathcal{F}}^{+,1}= x_2 ( x_2 + \frac{4 \beta}{1-\beta^2} x_1),& \\
&\mathrm{II}:& &x_1 \rightarrow y_1 = x_1 + \gamma^+ x_2,& &\widetilde{\mathcal{F}}^{+,2}=x_1( x_1 + \frac{4 \beta}{1-\beta^2} x_2),&  \\
&\mathrm{III}:& &
\begin{matrix}
x_2 \rightarrow y_2 = x_2 + \gamma^- x_1, \\
x_1 \rightarrow y_1 = x_1 + \gamma^- x_2,
\end{matrix}
& & \widetilde{\mathcal{F}}^-= -\frac{16 \beta^2}{(1-\beta)(1+\beta)^3}x_1 x_2,&
\end{align}
In region (I) we have mapped the $x_2 = \gamma^+ x_1$ hyperplane to $x_2 = 0$.
In region (II) we have mapped the $x_2 = \gamma^- x_1$ hyperplane to $x_1 = 0$. 
In region (III) we have mapped the $x_2 = \gamma^- x_1$ hyperplane to $x_2 = 0$ and, simultaneously, the $x_2 = \gamma^+ x_1$ hyperplane to $x_1 = 0$.
By construction, in each region, $\widetilde{\mathcal{F}}$ has a fixed sign (either non-negative or non-positive).
Naturally, we must also apply the corresponding transformations to the $\mathcal{U}$ polynomial as well as keeping track of the Jacobian determinant.

The original integral,
\begin{align}
J_{\mathrm{bub},m_1\neq m_2}&=\lim_{\delta\to0^+}\Gamma\left(\epsilon\right)I_{\mathrm{bub},m_1\neq m_2},\\
I_{\mathrm{bub},m_1\neq m_2}&=\int_{\mathbb{R}^2_{\geq0}}\!\!\mathrm{d}x_1\mathrm{d}x_2\frac{\left(x_1+x_2\right)^{-2+2\epsilon}}{\left(-sx_1x_2+\left(x_1+x_2\right)\left(m_1^2x_1+m_2^2x_2\right)-i\delta\right)^{\epsilon}}\delta\left(1-\alpha(\mathbf{x})\right), \label{eq:nonequal_mass_bubble}
\end{align}
in the regime $s>\left(m_1+m_2\right)^2\Leftrightarrow\beta^2\in\left(0,1\right)$ can now be written as a causally prescribed sum over three integrals with manifestly non-negative integrands:
\begin{equation}
    I_{\mathrm{bub},m_1\neq m_2}=I_{\mathrm{bub},m_1\neq m_2}^{+,1}+I_{\mathrm{bub},m_1\neq m_2}^{+,2}+\left(-1-i\delta\right)^{-\epsilon}I_{\mathrm{bub},m_1\neq m_2}^{-},
    \label{eq:bubble_unequal_sum}
\end{equation}
where
\begin{align}
    I_{\mathrm{bub},m_1\neq m_2}^{+,1}&=\nonumber(m_1 m_2)^{1-2\epsilon}(1-\beta)^{2-2\epsilon}\int_{\mathbb{R}^2_{\geq0}}\!\!\mathrm{d}x_1\mathrm{d}x_2 \,x_2^{-\epsilon}\left(x_2+\frac{4 \beta }{1-\beta^2}x_1\right)^{-\epsilon}\times\\&\left(m_1(1-\beta)x_2+[m_1(1+\beta)+m_2(1-\beta)]x_1\right)^{-2+2\epsilon}\delta\left(1-\alpha(\mathbf{x})\right),\\
    I_{\mathrm{bub},m_1\neq m_2}^{+,2}&=\nonumber(m_1 m_2)^{1-2\epsilon}(1-\beta)^{2-2\epsilon}\int_{\mathbb{R}^2_{\geq0}}\!\!\mathrm{d}x_1\mathrm{d}x_2 \,x_1^{-\epsilon}\left(x_1+\frac{4 \beta }{1-\beta^2}x_2\right)^{-\epsilon}\times\\&\left(m_2(1-\beta)x_1+[m_2(1+\beta)+m_1(1-\beta)]x_2\right)^{-2+2\epsilon}\delta\left(1-\alpha(\mathbf{x})\right),\\
    I_{\mathrm{bub},m_1\neq m_2}^{-}&=\nonumber(4m_1m_2\beta)^{1-2\epsilon}(1-\beta^2)^{\epsilon}\int_{\mathbb{R}^2_{\geq0}}\!\!\mathrm{d}x_1\mathrm{d}x_2\left(x_1 x_2\right)^{-\epsilon}\times\\&\left[(m_1+m_2)(x_1+x_2)-(m_1-m_2)(x_1-x_2)\beta\right]^{-2+2\epsilon}\delta\left(1-\alpha(\mathbf{x})\right).
\end{align}
The symmetry of the integral Eq.~\eqref{eq:nonequal_mass_bubble} is manifest in the resolved integrals. 
The integrands appearing in $I_{\mathrm{bub},m_1\neq m_2}^{+,1}$ and $I_{\mathrm{bub},m_1\neq m_2}^{+,2}$ are related by the simultaneous interchange ${x_1 \leftrightarrow x_2}$ and $m_1 \leftrightarrow m_2$, while $I_{\mathrm{bub},m_1\neq m_2}^{-}$ is invariant under this exchange.
These integrals can be analytically evaluated by direct integration and match the known result for the unequal mass bubble order-by-order in the expansion in $\epsilon$.
 Again, the analytic continuation of the result is manifest in Eq.~\eqref{eq:bubble_unequal_sum} and the expansion in $\epsilon$ of the resolved integrals and their numerical evaluation is straightforward.

\subsubsection{1-Loop Triangle with an Off-Shell Leg}
\label{sec:triangle}

\begin{figure}[t]
    \centering
    \begin{subfigure}[t]{0.19\textwidth}
        \centering
        \begin{tikzpicture}[baseline=14ex,scale=0.7]
            \coordinate (x1) at (1.1340, 1.4999) ;
            \coordinate (x2) at (2.8660, 1.5000) ;
            \coordinate (x3) at (2,3) ;
            \node (p1) at (0.2681, 0.9998) {};
            \node (p2) at (3.7320, 1.0000) {};
            \node (p3) at (2,4) [yshift=+2pt] {$p^2$};
            \draw[color=blue] (x1) -- (p1);
            \draw[ultra thick,color=ForestGreen] (x3) -- (p3);
            \draw[color=blue] (x2) -- (p2);
            \draw[ultra thick,color=Black] (x1) -- (x2);
            \draw[ultra thick,color=Black] (x2) -- (x3);
            \draw[ultra thick,color=Purple] (x3) -- (x1) node [midway,xshift=-9,color=Black] {$m$};
            \draw[fill,thick,color=Blue] (x1) circle (1pt);
            \draw[fill,thick,color=Blue] (x2) circle (1pt);
            \draw[fill,thick,color=Blue] (x3) circle (1pt);
        \end{tikzpicture}
        \caption{}
    \end{subfigure}
    \hfill
    \begin{subfigure}[t]{0.19\textwidth}
        \centering
        \begin{tikzpicture}[baseline=14ex,scale=0.7]
            \coordinate (x1) at (1.1340, 1.4999) ;
            \coordinate (x2) at (2.8660, 1.5000) ;
            \coordinate (x3) at (2,3) ;
            \node (p1) at (0.2681, 0.9998) {};
            \node (p2) at (3.7320, 1.0000) {};
            \node (p3) at (2,4) [yshift=+2pt] {$p^2$};
            \draw[color=blue] (x1) -- (p1);
            \draw[ultra thick,color=ForestGreen] (x3) -- (p3);
            \draw[color=blue] (x2) -- (p2);
            \draw[ultra thick,color=Purple] (x1) -- (x2) node [midway,yshift=+6pt,color=Black] {$m$};
            \draw[ultra thick,color=Black] (x2) -- (x3);
            \draw[ultra thick,color=Black] (x3) -- (x1);
            \draw[fill,thick,color=Blue] (x1) circle (1pt);
            \draw[fill,thick,color=Blue] (x2) circle (1pt);
            \draw[fill,thick,color=Blue] (x3) circle (1pt);
        \end{tikzpicture}
        \caption{}\label{onemasstrianglex3}
    \end{subfigure}
    \hfill
    \begin{subfigure}[t]{0.19\textwidth}
        \centering
        \begin{tikzpicture}[baseline=14ex,scale=0.7]
            \coordinate (x1) at (1.1340, 1.4999) ;
            \coordinate (x2) at (2.8660, 1.5000) ;
            \coordinate (x3) at (2,3) ;
            \node (p1) at (0.2681, 0.9998) {};
            \node (p2) at (3.7320, 1.0000) {};
            \node (p3) at (2,4) [yshift=+2pt] {$p^2$};
            \draw[color=blue] (x1) -- (p1);
            \draw[ultra thick,color=ForestGreen] (x3) -- (p3);
            \draw[color=blue] (x2) -- (p2);
            \draw[ultra thick,color=Purple] (x1) -- (x2) node [midway,yshift=+6pt,color=Black] {$m$};
            \draw[ultra thick,color=Black] (x2) -- (x3);
            \draw[ultra thick,color=Purple] (x3) -- (x1) node [midway,xshift=-9pt,color=Black] {$m$};
            \draw[fill,thick,color=Blue] (x1) circle (1pt);
            \draw[fill,thick,color=Blue] (x2) circle (1pt);
            \draw[fill,thick,color=Blue] (x3) circle (1pt);
        \end{tikzpicture}
        \caption{}
    \end{subfigure}
        \begin{subfigure}[t]{0.19\textwidth}
        \centering
        \begin{tikzpicture}[baseline=14ex,scale=0.7]
            \coordinate (x1) at (1.1340, 1.4999) ;
            \coordinate (x2) at (2.8660, 1.5000) ;
            \coordinate (x3) at (2,3) ;
            \node (p1) at (0.2681, 0.9998) {};
            \node (p2) at (3.7320, 1.0000) {};
            \node (p3) at (2,4) [yshift=+2pt] {$p^2$};
            \draw[color=blue] (x1) -- (p1);
            \draw[ultra thick,color=ForestGreen] (x3) -- (p3);
            \draw[color=blue] (x2) -- (p2);
            \draw[ultra thick,color=Black] (x1) -- (x2);
            \draw[ultra thick,color=Purple] (x2) -- (x3) node [midway,xshift=+8pt,color=Black] {$m$};
            \draw[ultra thick,color=Purple] (x3) -- (x1) node [midway,xshift=-9pt,color=Black] {$m$};
            \draw[fill,thick,color=Blue] (x1) circle (1pt);
            \draw[fill,thick,color=Blue] (x2) circle (1pt);
            \draw[fill,thick,color=Blue] (x3) circle (1pt);
        \end{tikzpicture}
        \caption{}
    \end{subfigure}
            \begin{subfigure}[t]{0.19\textwidth}
        \centering
        \begin{tikzpicture}[baseline=14ex,scale=0.7]
            \coordinate (x1) at (1.1340, 1.4999) ;
            \coordinate (x2) at (2.8660, 1.5000) ;
            \coordinate (x3) at (2,3) ;
            \node (p1) at (0.2681, 0.9998) {};
            \node (p2) at (3.7320, 1.0000) {};
            \node (p3) at (2,4) [yshift=+2pt] {$p^2$};
            \draw[color=blue] (x1) -- (p1);
            \draw[ultra thick,color=ForestGreen] (x3) -- (p3);
            \draw[color=blue] (x2) -- (p2);
            \draw[ultra thick,color=Purple] (x1) -- (x2) node [midway,yshift=+6pt,color=Black] {$m$};
            \draw[ultra thick,color=Purple] (x2) -- (x3) node [midway,xshift=+8pt,color=Black] {$m$};
            \draw[ultra thick,color=Purple] (x3) -- (x1) node [midway,xshift=-9pt,color=Black] {$m$};
            \draw[fill,thick,color=Blue] (x1) circle (1pt);
            \draw[fill,thick,color=Blue] (x2) circle (1pt);
            \draw[fill,thick,color=Blue] (x3) circle (1pt);
        \end{tikzpicture}
        \caption{}\label{fullmasstri}
    \end{subfigure}
    \caption{Independent equal mass triangles with an off-shell leg ($p^2>0$)}
    \label{triangles}
\end{figure}
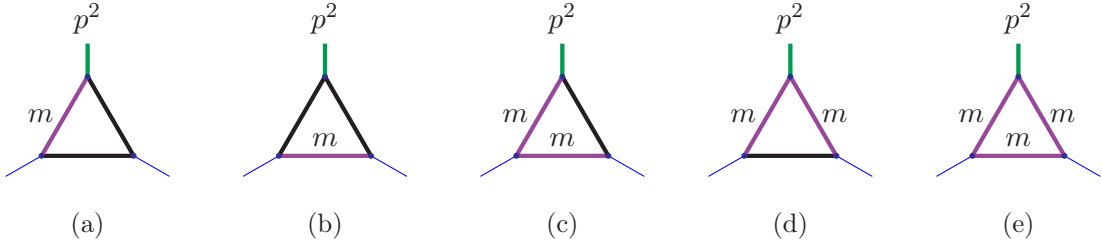

\begin{figure}[t]
    \centering
    \raisebox{-0.5\height}{\includegraphics[width=0.4\textwidth]{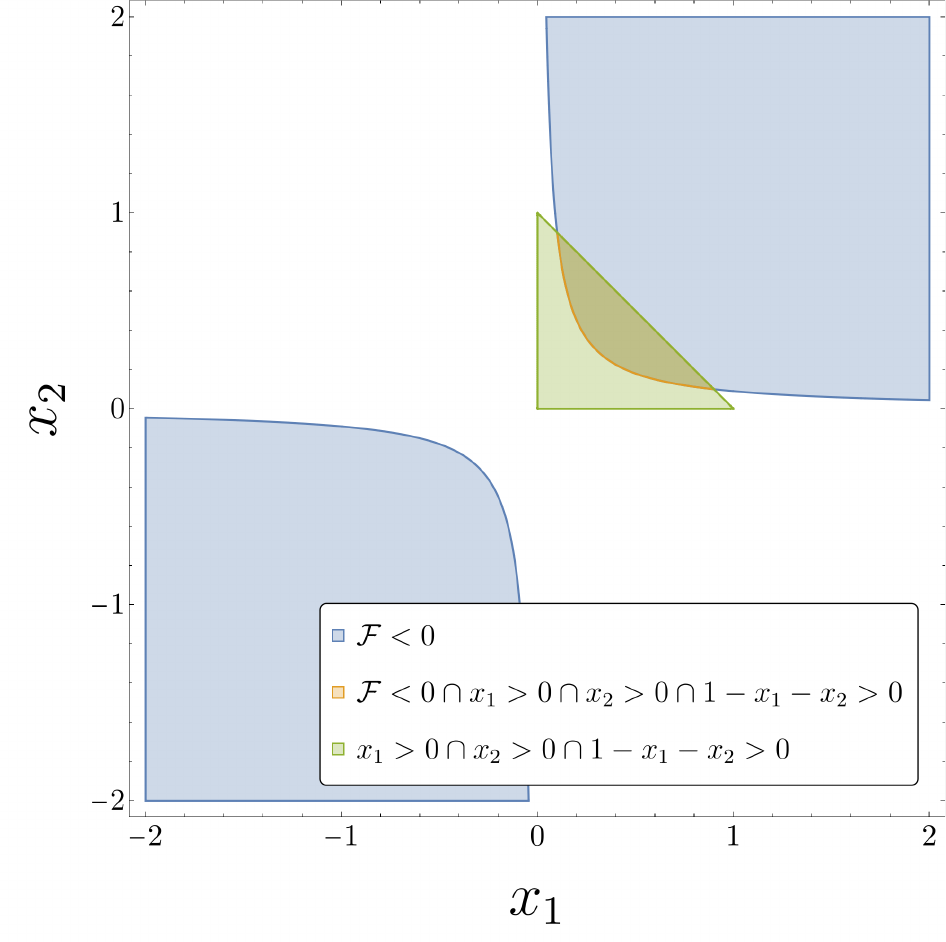}}
    \qquad $\Rightarrow$ \qquad
    \raisebox{-0.5\height}{\includegraphics[width=0.4\textwidth]{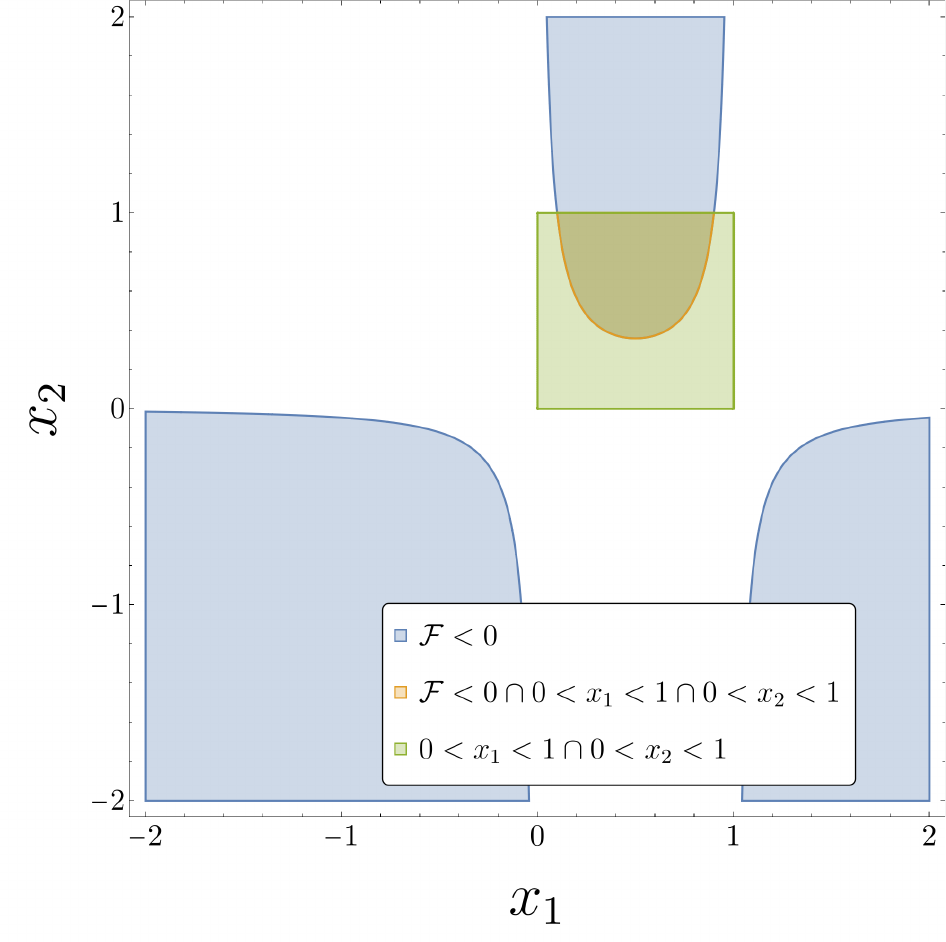}} \\
      \caption{Remapping the simplex integration region of the massive triangle (in green) to the positive unit square in $\mathbb{R}^2_{\geq0}$. Here, $\mathcal{F}$ is to be understood as $\mathcal{F}$ after the $\delta$-function has been integrated out and in the second panel, after the remapping transformation. The kinematic regime is given by $\beta^2\in(0,1)$ (with $m^2>0$).}
    \label{fig:remapping}
\end{figure}

\begin{figure}[t]
    \centering
\includegraphics[width=0.4\textwidth]{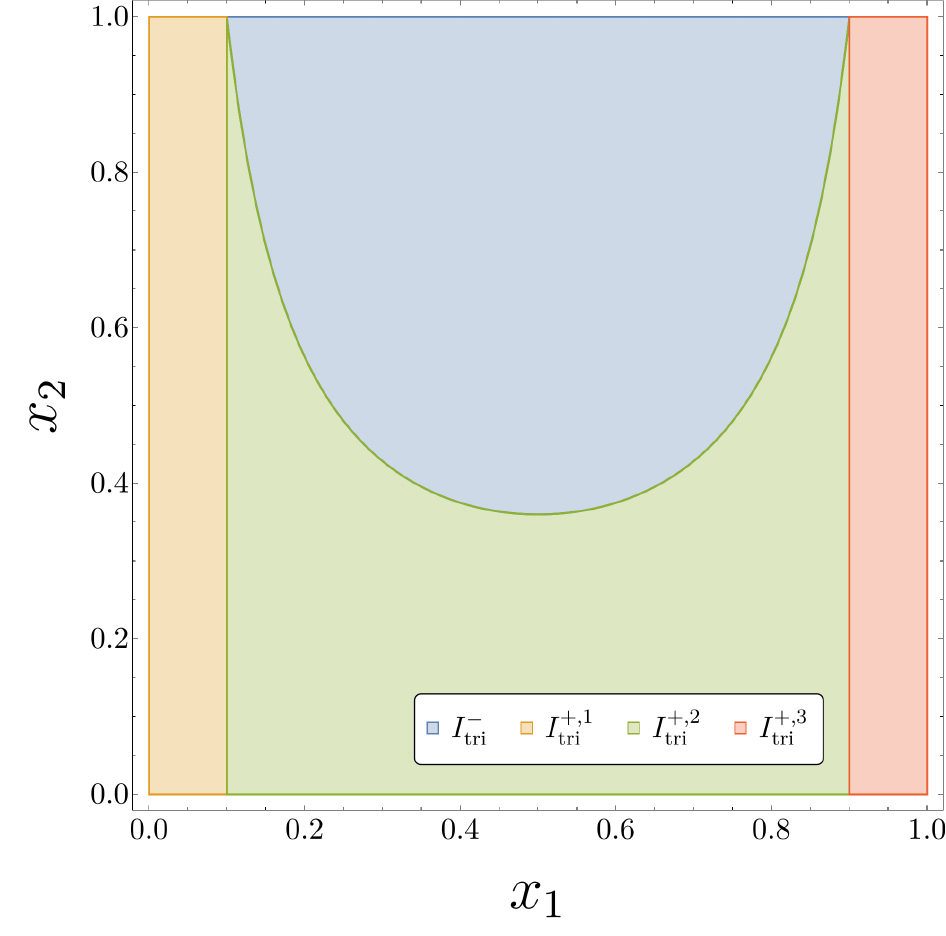}
    \caption{The integration domain of the massive triangle separated into one negative and three positive regions.}
    \label{fig:triangleregions}
\end{figure}

For our final 1-loop massive example, we consider the massive triangle with an off-shell leg (the independent equal-mass configurations of which are shown in Fig.~\ref{triangles}). For brevity, we present the resolution of the fully massive triangle (Fig.~\ref{fullmasstri}) in detail and remark that the others may be similarly resolved using the procedure outlined here. We pick the fully massive triangle for our exposition as it is the most difficult, and the analysis of the other examples is carried out almost identically.
The integral we wish to consider is
\begin{align}
J_{\mathrm{tri}}&=\lim_{\delta\to0^+}-\Gamma\left(1+\epsilon\right)I_{\mathrm{tri}}\\I_{\mathrm{tri}}&=\int_{\mathbb{R}^3_{\geq0}}\!\!\mathrm{d}x_1\mathrm{d}x_2\mathrm{d}x_3\frac{\left(x_1+x_2+x_3\right)^{-1+2\epsilon}}{\left(-p^2x_1x_2+m^2\left(x_1+x_2+x_3\right)^2-i\delta\right)^{1+\epsilon}}\delta\left(1-\alpha(\mathbf{x})\right).
\end{align}
It is possible to make multiple choices to parameterise the projective integral, making different choices of hyperplane $\alpha\!\left(\mathbf{x}\right)=\sum_i\alpha_ix_i$ can lead to different solutions of the problem. 
In our experience, being guided by the symmetry of the problem, where possible, leads to the neatest solutions, though one may choose, e.g. $\delta\left(1-x_1\right)$ and successfully avoid contour deformation (although not necessarily easily avoiding square roots involving the Feynman parameters). Following this philosophy, we make the symmetric choice (here, $\delta\left(1-x_1-x_2-x_3\right)$) which is often the optimal choice at 1-loop as it sets the $\mathcal{U}$ polynomial immediately to 1.
Defining $\beta^2=\frac{p^2-4m^2}{p^2}\in\left(0,1\right)$ and using the $\delta$-function to perform the $x_3$ integral, we have
\begin{align}
    I_{\mathrm{tri}}&=\left(\frac{1-\beta^2}{m^2}\right)^{1+\epsilon}\tilde{I}_{\mathrm{tri}}\\  \tilde{I}_{\mathrm{tri}}&=\int_{\mathbb{R}^2_{\geq0}}\!\!\mathrm{d}x_1\mathrm{d}x_2\theta\left(1-x_1-x_2\right)\left(1-\beta^2-4 x_1 x_2-i\delta\right)^{-1-\epsilon}
\end{align}
where the Heaviside function is a result of the symmetric choice of hyperplane and restricts the remaining domain of integration to a simplex. With the foresight that we will wish to integrate this numerically with a standard package like \pysecdec, we apply a coordinate transformation that remaps this domain to the positive unit square in $\mathbb{R}^2_{\geq0}$ as in Fig.~\ref{fig:remapping}. After this mapping ($x_2\rightarrow\left(1-x_1\right)x_2$), we have
\begin{equation}
    \tilde{I}_{\mathrm{tri}}=\int_{0}^{1}\!\!\mathrm{d}x_1\mathrm{d}x_2\left(1-x_1\right)\left(1-\beta^2-4 \left(1-x_1\right)x_1x_2-i\delta\right)^{-1-\epsilon}.
\end{equation}
Visualising the variety of the transformed $\mathcal{F}$ allows us to separate the integration domain into a number of regions, the choice (and number) of which is guided by what is practically simpler to deal with. We show this choice of four regions in Fig.~\ref{fig:triangleregions} where we have separated the integration domain (the positive unit square in $\mathbb{R}^2_{\geq0}$) into three positive regions where $\mathcal{F}>0$ and one negative one where $\mathcal{F}<0$ (which will generate the entire imaginary part of the full integral).
We will show the chain of transformations which maps the negative region (blue in Fig.~\ref{fig:triangleregions}) to the positive unit square and state the result for the remaining positive regions. Firstly, we need to map the orange and red positive regions in Fig.~\ref{fig:triangleregions} outside the square. To do this, we find the $x_1$-values where the variety of $\mathcal{F}$ intersects with the boundary of the domain at $x_2=1$. A trivial calculation reveals these values to be $\frac{1\pm\beta}{2}$ and we want to map the $x_1$-lines defined by these values to the boundaries of integration $x_1=0$ and $x_1=1$. We can easily construct a transformation which satisfies these demands:
\begin{equation}
    x_1'\overset{!}{=}\frac{x_1-\frac{1-\beta}{2}}{\frac{1+\beta}{2}-\frac{1-\beta}{2}}\quad\Rightarrow\quad x_1\rightarrow\frac{1}{2}+\beta\left(x_1-\frac{1}{2}\right)
\end{equation}
\begin{figure}[t]
    \centering
    \raisebox{-0.5\height}{\includegraphics[width=0.3\textwidth]{figures/triangleregions.pdf}}
    \hfill $\Rightarrow$ \hfill
    \raisebox{-0.5\height}{\includegraphics[width=0.3\textwidth]{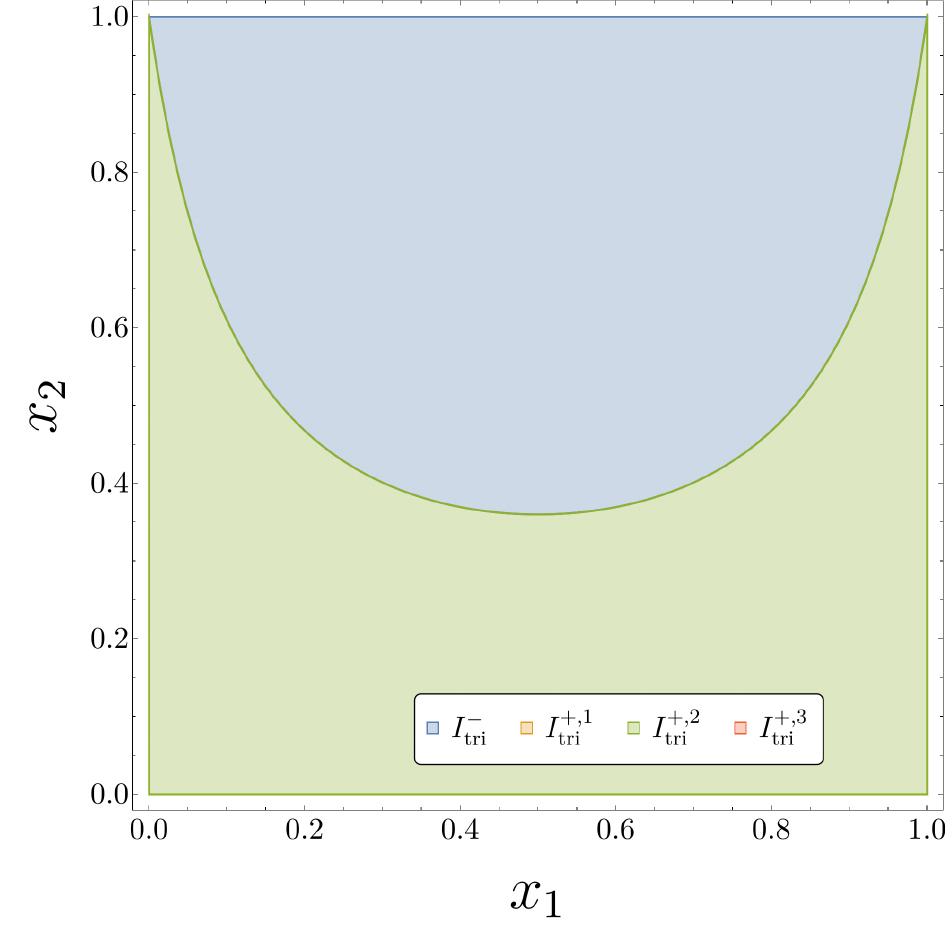}}\hfill $\Rightarrow$\hfill
    \raisebox{-0.5\height}{\includegraphics[width=0.3\textwidth]{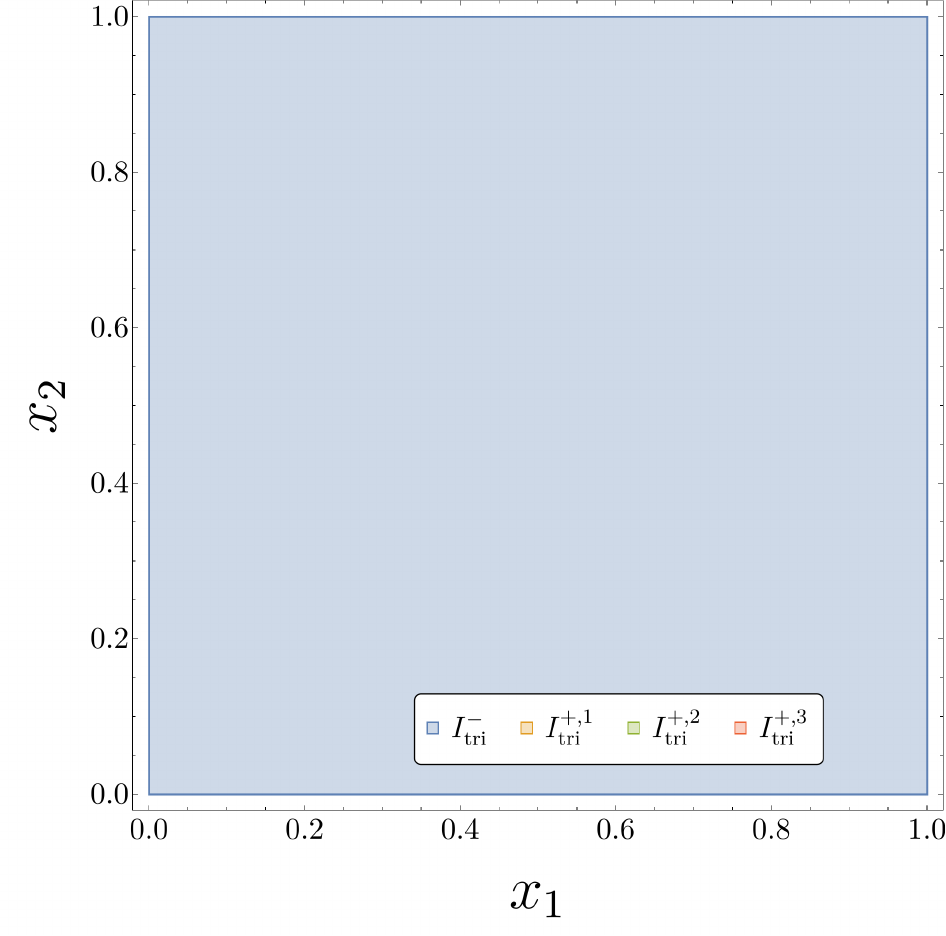}} \\
    \caption{The chain of transformations which maps the negative region of the massive triangle (in blue) to the positive unit square.}
    \label{fig:triangletrans}
\end{figure}
and we show the effect of this transformation in the transition from the first to the second panel of Fig.~\ref{fig:triangletrans} with $\mathcal{F}\rightarrow1-\beta^{2}-(1-(1-2x_1)^{2}\beta^{2})\ x_2$. The geometric visualisation helps us to deduce the next logical step which is to map the variety of this transformed $\mathcal{F}$ to the boundary at $x_2=0$ while keeping the boundary at $x_2=1$ fixed (so as not to map in any points from outside the integration domain). In order to do this, we must first solve $\mathcal{F}=0$ as $x_2=f\left(x_1\right)$. Trivially, we find:
\begin{equation}
    \mathcal{F}=0\quad\Rightarrow\quad x_2=\frac{1-\beta^2}{1-\left(1-2x_1\right)^{2}\beta^{2}}=f\left(x_1\right).
\end{equation}
In this form, we can directly construct the transformation which satisfies the demands above:
\begin{equation}
    x_1'\overset{!}{=}x_1,\quad x_2'\overset{!}{=}\frac{x_2-f\left(x_1\right)}{1-f\left(x_1\right)}\quad\Rightarrow\quad x_1\rightarrow x_1,\quad x_2\rightarrow x_2+\left(1-x_2\right)f\left(x_1\right).
\end{equation}
This transformation generates the transition between the second and third panels in Fig~\ref{fig:triangletrans} and completes the mapping of the negative region to the positive unit square as required. The final result for the negative contribution after taking into account the Jacobian determinants of the transformations is
\begin{equation}
    \tilde{I}_{\mathrm{tri}}^{-}=2^{-1-2\epsilon}\beta^{1-2\epsilon}\int_{0}^{1}\!\!\mathrm{d}x_1\mathrm{d}x_2\left(1-x_1\right)^{-\epsilon}x_1^{-\epsilon}x_2^{-1-\epsilon}\left(1-\left(1-2x_1\right)\beta\right)^{-1}
\end{equation}
where we have already factored out $\left(-1-i\delta\right)^{-1-\epsilon}$. 
We note that this integral is $\mathcal{O}\!\left(\frac{1}{\epsilon}\right)$ whereas the full integral $I_{\mathrm{tri}}$ is finite. This pole cancels exactly with a corresponding pole in $I_{\mathrm{tri}}^{+,2}$ and, since $I_{\mathrm{tri}}^{+,1}$ and $I_{\mathrm{tri}}^{+,3}$ are free of poles in $\epsilon$, the construction consistently reproduces the full result. 

It is illuminating to consider the general structure of this cancellation; if the full result is finite and also has an imaginary part at the leading order (which is true of $I_{\mathrm{tri}}$), the negative contribution must necessarily have a pole in $\epsilon$ to generate the imaginary part from multiplying the $\epsilon^{1}$ term in the expansion of $\left(-1-i\delta\right)^{a+b\epsilon}$. Furthermore, the total positive contribution must have an equal pole in $\epsilon$ to cancel the pole of the negative contribution leaving a finite result. This type of analysis can be fruitful in predicting a priori the pole structure of the constituent integrals in the decomposition.

For completeness, we state the result for the positive contributions:
\begin{align}
    \tilde{I}_{\mathrm{tri}}^{+,1}&=\frac{1}{4}\left(1-\beta\right)^{-\epsilon}\int_{0}^{1}\!\!\mathrm{d}x_1\mathrm{d}x_2\left(2-\left(1-\beta\right)x_1\right)\left(1+\beta-\left(2-\left(1-\beta\right)x_1\right)x_1x_2\right)^{-1-\epsilon}\\
    \tilde{I}_{\mathrm{tri}}^{+,2}&=\frac{\beta}{2}\left(1-\beta^2\right)^{-\epsilon}\int_{0}^{1}\!\!\mathrm{d}x_1\mathrm{d}x_2\left(1-x_2\right)^{-1-\epsilon}\left(1-\left(1-2x_1\right)\beta\right)^{-1}\\
    \tilde{I}_{\mathrm{tri}}^{+,3}&=\frac{1}{4}\left(1-\beta\right)^{1-\epsilon}\int_{0}^{1}\!\!\mathrm{d}x_1\mathrm{d}x_2\left(1-x_1\right)\left(1+\beta-x_2\left(1-x_1\right)\left(1+\beta+\left(1-\beta\right)x_1\right)\right)^{-1-\epsilon}.
\end{align}
This gives the total result
\begin{equation}
    I_{\mathrm{tri}}=\sum_{n_+=\ \!\!1}^{3}I_{\mathrm{tri}}^{+,n_+}+\left(-1-i\delta\right)^{-1-\epsilon}I_{\mathrm{tri}}^{-},
    \label{trisum}
\end{equation}
where we stress again that each integrand in the constituent integrals of \eqref{trisum} is manifestly non-negative in the Minkowskian kinematic region defined by $\beta\in\left(0,1\right)$ throughout the entire integration domain. Additionally, all singularities have been mapped to the endpoints where they can be dealt with using standard techniques such as sector decomposition instead of applying a contour deformation prescription as would usually be required in a numerical calculation.

\subsubsection{2-Loop Elliptic Sunrise}
\label{sec:sunrise}
In this section, we analyse the equal mass sunrise integral, the simplest integral which involves a function class beyond polylogarithms. The sunrise has been studied extensively, see e.g.~\cite{Broadhurst:1993mw,Bloch:2013tra,Adams:2015gva,Remiddi:2017har}, and is known to evaluate to elliptic integrals. 
Using the sunrise integral as an example, we demonstrate that there is no fundamental obstruction to avoiding contour deformation for massive integrals beyond polylogarithms.
Contrary to previous examples, however, we find that the sunrise leads us naturally to algebraic transformations of the Feynman parameters instead of purely rational function transformations. 
To the best of our current knowledge, it seems not to be possible to find a resolution which avoids square roots in the transformations; however, this does not present a fundamental obstacle for numerical evaluation.
Indeed, the resolved integral can be numerically evaluated, and for various kinematic configurations we either obtain results with a smaller relative error or significantly faster than when using contour deformation, as demonstrated in Section~\ref{sec:performance}.

\begin{figure}[t]
\centering
\includegraphics[width=0.5\textwidth]{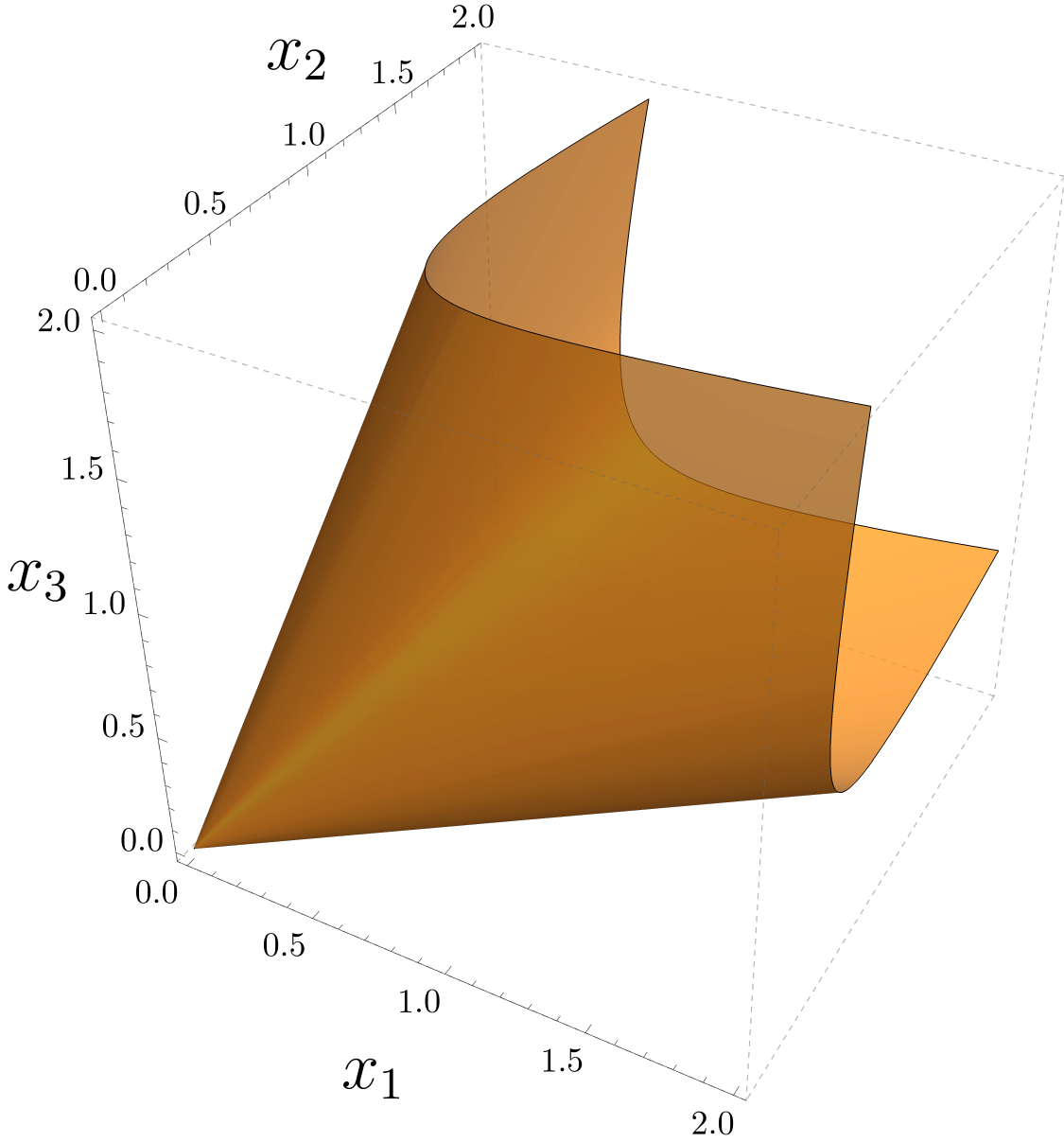}%
    \caption{The $\mathcal{F}=0$ surface of the equal mass sunrise in $\mathbb{R}^3_{>0}$ with the caveat that this should properly be understood projectively. The kinematic regime is given by $\beta^2\in(0,1)$ (with $m^2>0$).}
    \label{elliptic 3d}
    \label{fig:sundiagandvariety}
\end{figure}

The analysis of the 2-loop massive sunrise follows similarly to the 1-loop massive triangle detailed in Section~\ref{sec:triangle} (they are both cases of integrals with three equal-mass propagators), as a result of this, we focus mainly on the differences to that case in this exposition. The integral we wish to resolve without contour deformation is
\begin{align}
J_{\mathrm{sun}}&=\lim_{\delta\to0^+}-\Gamma\left(-1+2\epsilon\right)I_{\mathrm{sun}}\\I_{\mathrm{sun}}&=\int_{\mathbb{R}^3_{\geq0}}\!\!\mathrm{d}x_1\mathrm{d}x_2\mathrm{d}x_3\frac{\left(x_1x_2+x_2x_3+x_1x_3\right)^{-3+3\epsilon}\delta\left(1-\alpha(\mathbf{x})\right)}{\left(-sx_1x_2x_3+\left(x_1x_2+x_2x_3+x_1x_3\right)m^2\left(x_1+x_2+x_3\right)-i\delta\right)^{-1+2\epsilon}}
\end{align}
and is depicted in Fig.~\ref{eqmasssun}. A plot of the sunrise's $\mathcal{F}=0$ surface in $\mathbb{R}^3_{>0}$ above the threshold $s > 9m^2$ (which should be rigorously understood in the context of the projective integral, parameterised with the $\delta$-function, for example) is given in Fig.~\ref{elliptic 3d}. It is the shape of this surface that will end up determining the resolution of the integral and we see that its distorted conic-esque geometry divides the integration into an inside and outside; this appears to be intrinsically linked to our inability to use purely rational functions for the resolution transformations. As we will see after a choice of hyperplane in the $\delta$-function, this will force the use of square roots and guide us towards algebraic transformations.



\begin{figure}[t]
    \centering
    \raisebox{-0.5\height}{\includegraphics[width=0.3\textwidth]{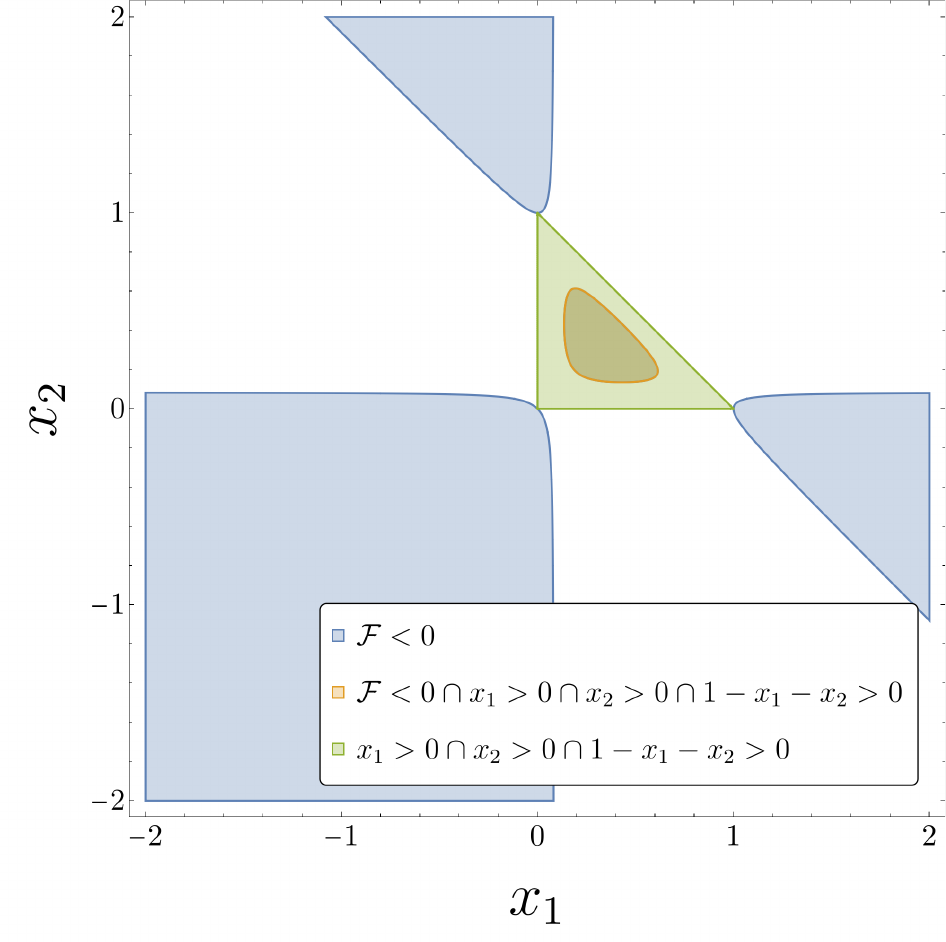}}
    \hfill $\Rightarrow$ \hfill
    \raisebox{-0.5\height}{\includegraphics[width=0.3\textwidth]{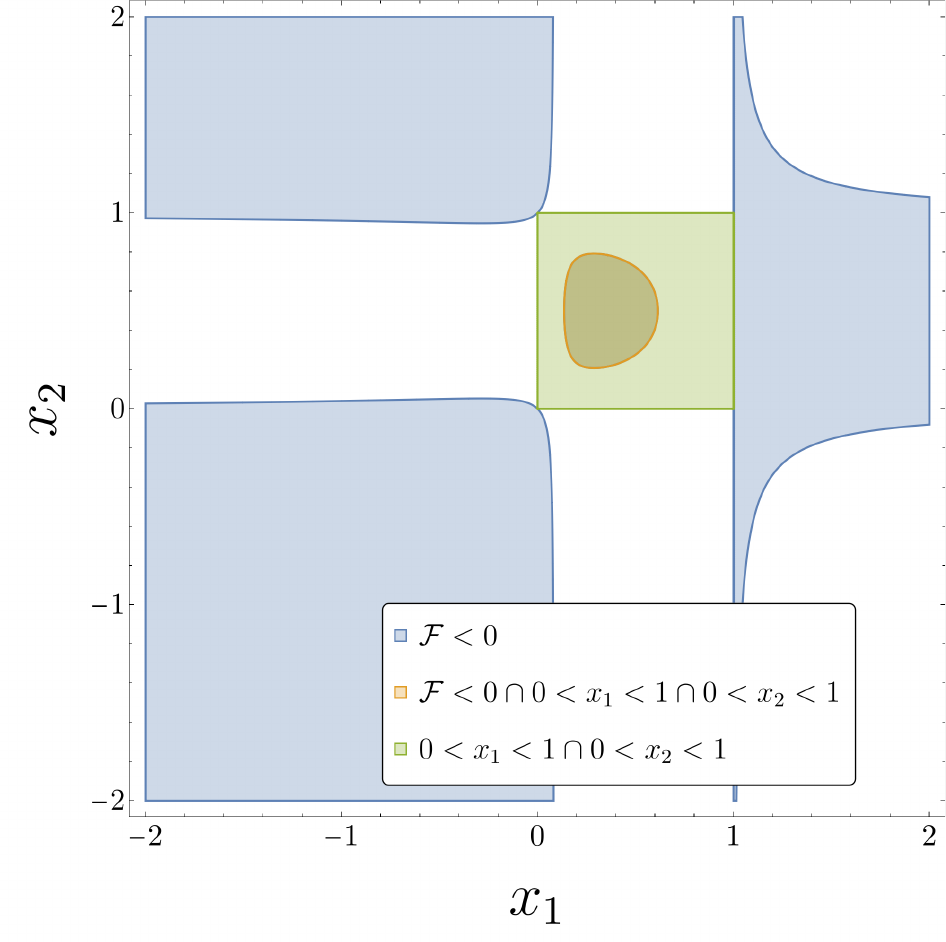}}\hfill $\Rightarrow$\hfill
    \raisebox{-0.5\height}{\includegraphics[width=0.3\textwidth]{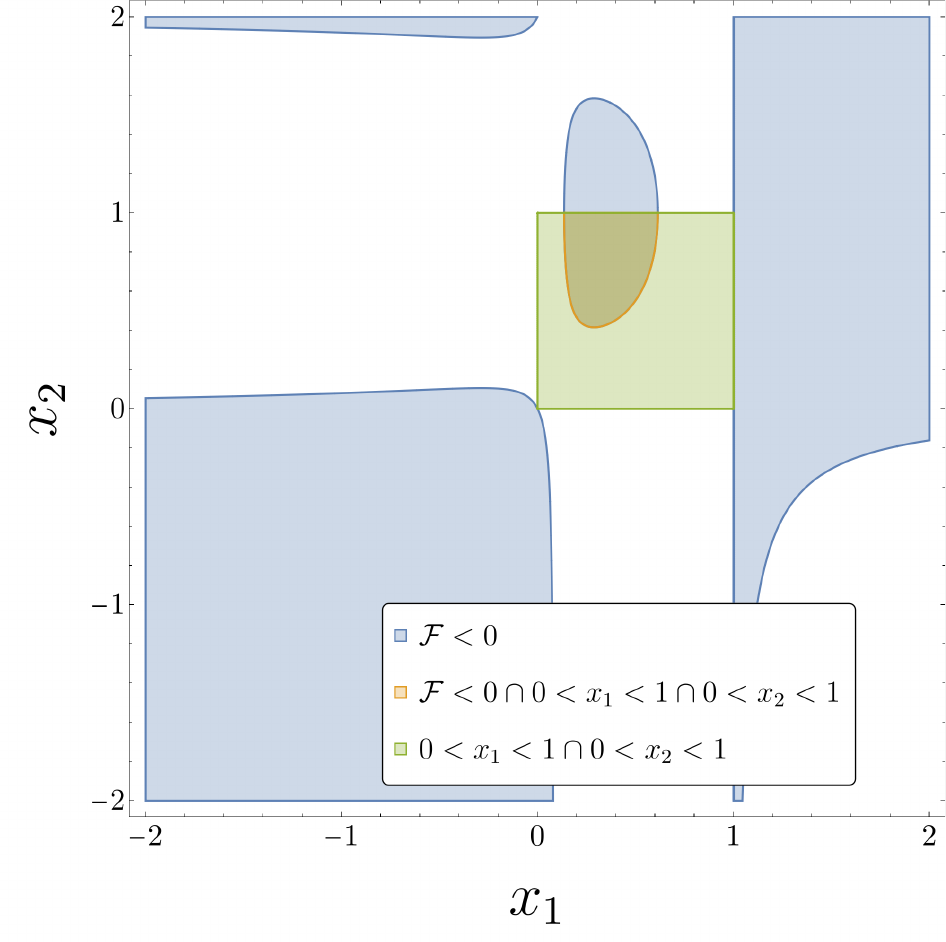}} \\
    \caption{Remapping the simplex integration region of the elliptic sunrise (in green) to the positive unit square in $\mathbb{R}^2_{\geq0}$ then exploiting the symmetry about $x_2=\frac{1}{2}$. Here, $\mathcal{F}$ is to be understood as $\mathcal{F}$ after the $\delta$-function has been integrated out and in the second and third panels, after their respective remapping transformations as well.}
    \label{fig:ellipticdomains}
\end{figure}

Motivated once again by the symmetry of the problem, we choose $\delta\left(1-x_1-x_2-x_3\right)$ and integrate out $x_3$ after having parameterised the Minkowskian kinematic regime we are interested in with $\beta^2=\frac{s-9m^2}{s}\in\left(0,1\right)$. We remap the resulting simplex integration domain to the positive unit square (as shown in the transition between the first and second panels of Fig.~\ref{fig:ellipticdomains}) in the same way as for the massive triangle in Section~\ref{sec:triangle}. These manipulations give us,
\begin{equation}
I_{\mathrm{sun}}=\left(\frac{\tilde{\beta}}{m^2}\right)^{-1+2\epsilon}\!\!\int_{0}^{1}\!\!\mathrm{d}x_1\mathrm{d}x_2\frac{\left(1-x_1\right)^{-1+\epsilon}\left(x_1+\left(1-x_1\right)\left(1-x_2\right)x_2\right)^{-3+3\epsilon}}{\left(\left(\tilde{\beta}-9x_1\right)\left(1-x_1\right)\left(1-x_2\right)x_2+\tilde{\beta}x_1-i\delta\right)^{-1+2\epsilon}},
\label{ellipticfirstremapping}
\end{equation}
with $\tilde{\beta}=\!1-\beta^2$.
By direct inspection of \eqref{ellipticfirstremapping}, the integrand can be seen to enjoy a symmetry under ${x_2\rightarrow 1-x_2}$ (analogously to the equal-mass bubble in Section~\ref{sec:bubble-eq}). This allows us to integrate $x_2$ from $0$ to $\frac{1}{2}$ instead and then double the result. We remap this halved integration domain back to the positive unit square (as shown in the transition between the second and third panels of Fig.~\ref{fig:ellipticdomains}) and benefit from exploiting this symmetry which reduces the number of regions we will need to resolve. Of course, were this symmetry not present, we would still be able to resolve the integral with more regions. Noting that the Jacobian factor of $\frac{1}{2}$ in this second transformation cancels the doubling from the symmetry, we have
\begin{equation}
    I_{\mathrm{sun}}=4^{2-\epsilon}\left(\frac{\tilde{\beta}}{m^2}\right)^{-1+2\epsilon}\!\!\int_{0}^{1}\!\!\mathrm{d}x_1\mathrm{d}x_2\frac{\left(1-x_1\right)^{-1+\epsilon}\left(4x_1+\left(1-x_1\right)\left(2-x_2\right)x_2\right)^{-3+3\epsilon}}{\left(\left(\tilde{\beta}-9x_1\right)\left(1-x_1\right)\left(2-x_2\right)x_2+4\tilde{\beta}x_1-i\delta\right)^{-1+2\epsilon}}.
    \label{ellipticsecondremapping}
\end{equation}
\begin{figure}[t]
    \centering
\includegraphics[width=0.3\textwidth]{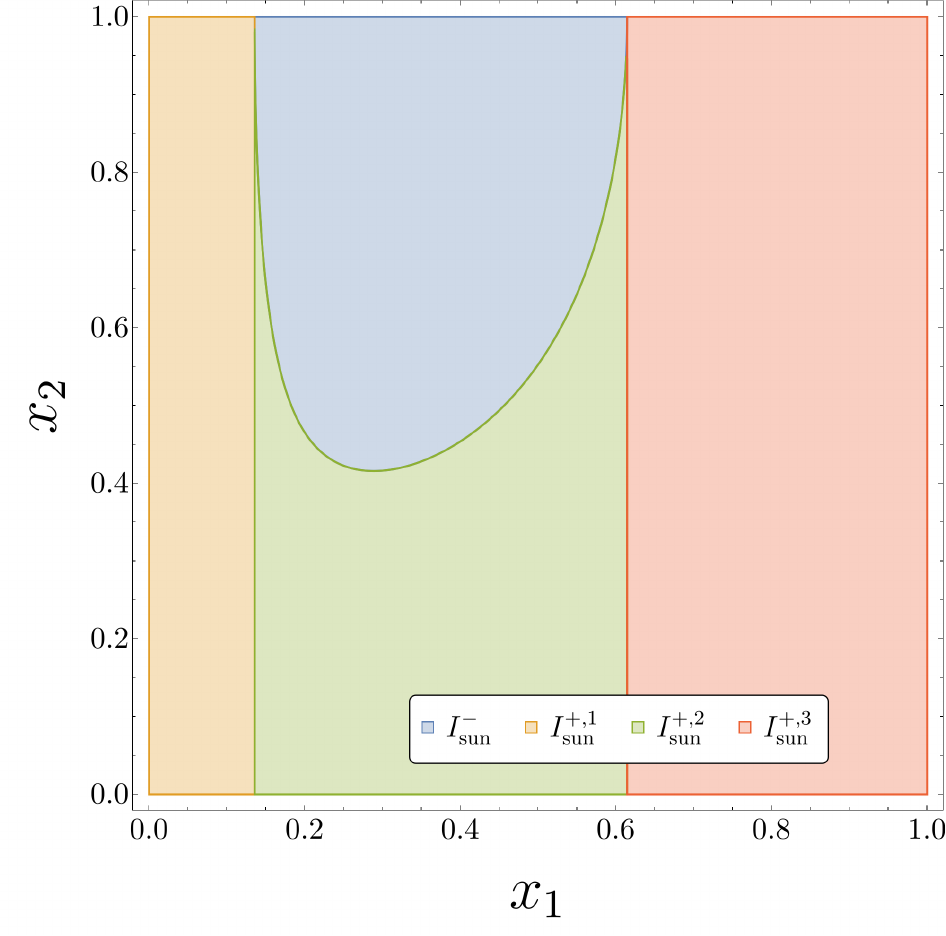}
    \caption{The integration domain of the elliptic sunrise separated into one negative and three positive regions.}
    \label{fig:ellipticregions}
\end{figure}

We can define a set of three positive regions where the denominator of \eqref{ellipticsecondremapping}, which we will loosely refer to as $\mathcal{F}$ (even though we have factored out $1-x_1$), is positive (i.e. $\mathcal{F}>0$) and one where it is negative (i.e. $\mathcal{F}<0$) which we show in Fig.~\ref{fig:ellipticregions}. We will resolve the negative region in detail once more as it plays the special role in solely generating the imaginary part of the full integral, the positive regions can be resolved similarly and will be included in our numerical studies in Section~\ref{sec:performance}.

\begin{figure}[t]
    \centering
    \raisebox{-0.5\height}{\includegraphics[width=0.3\textwidth]{figures/ellipticregions.pdf}}
    \hfill $\Rightarrow$ \hfill
    \raisebox{-0.5\height}{\includegraphics[width=0.3\textwidth]{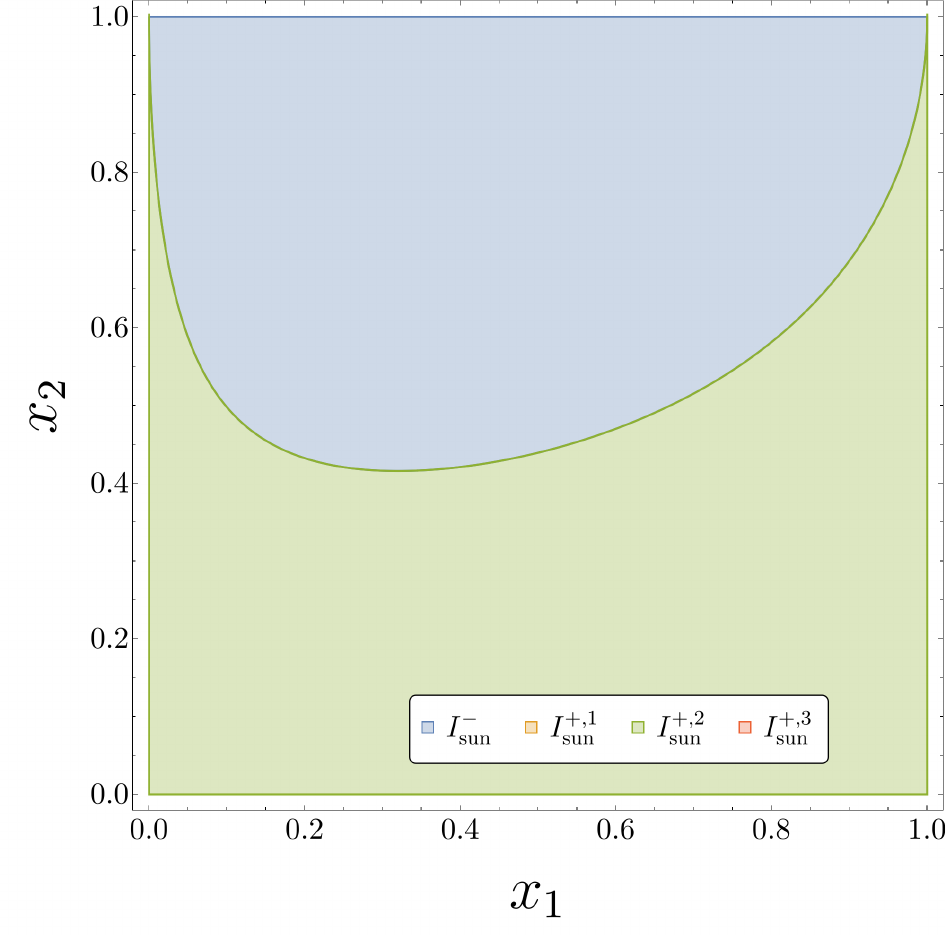}}\hfill $\Rightarrow$\hfill
    \raisebox{-0.5\height}{\includegraphics[width=0.3\textwidth]{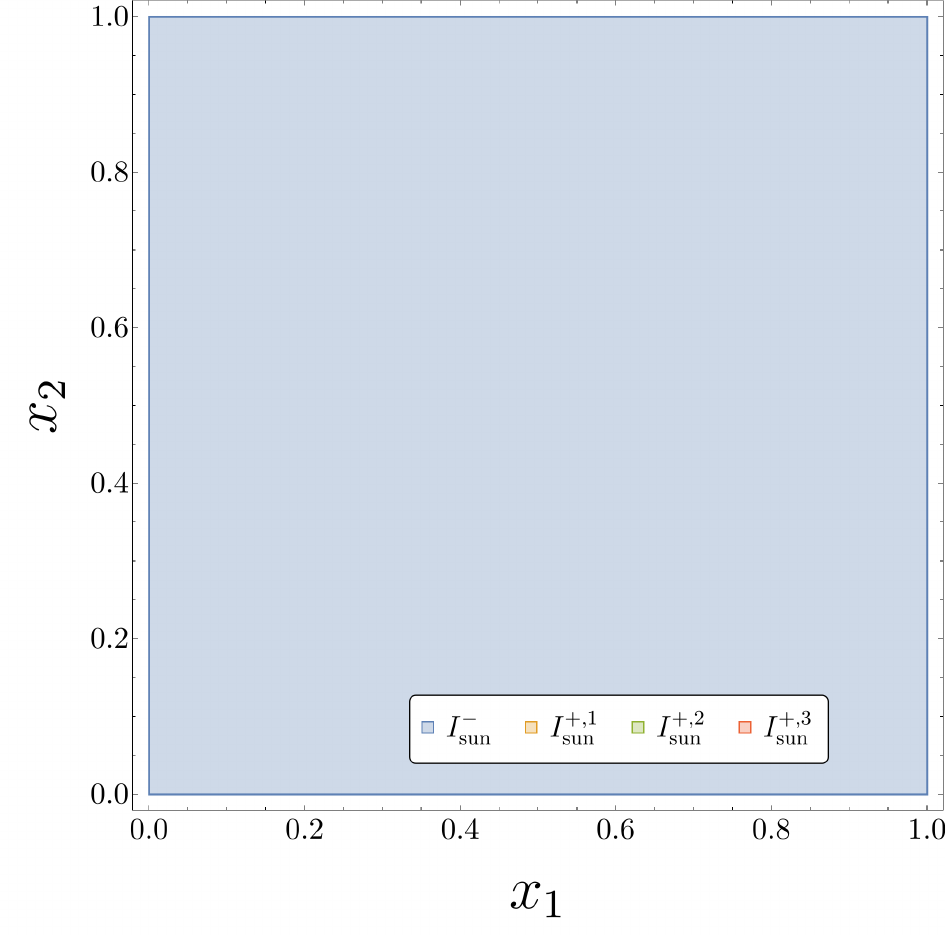}} \\
    \caption{The chain of transformations which maps the negative region of the elliptic sunrise (in blue) to the positive unit square.}
    \label{fig:elliptictrans}
\end{figure}

To resolve the negative region, we first must map the sides of the variety $\mathcal{F}=0$ to the boundary. Using the intersection of the variety with the boundary at $x_2=1$ to find the associated $x_1$-values as in the case of the massive triangle in Section~\ref{sec:triangle}, we find the relevant transformation to be
\begin{gather}
x_1'\overset{!}{=}\frac{x_1-\frac{1}{6}\left(2+\beta^2-\beta\bar{\beta}\right)}{\frac{1}{6}\left(2+\beta^2+\beta\bar{\beta}\right)-\frac{1}{6}\left(2+\beta^2-\beta\bar{\beta}\right)}\nonumber\\\Downarrow\nonumber\\
        x_1\rightarrow\frac{1}{6}\left(2+\beta^2-\left(1-2x_1\right)\beta\bar{\beta}\right),
\end{gather}
with $\bar{\beta}=\!\sqrt{8+\beta^2}$.
The effect of this transformation is shown in the transition from the first to the second panel in Fig.~\ref{fig:elliptictrans}. It is clear from this geometric picture what the next (and final) step of the resolution should be: mapping the variety to $x_2=0$ keeping the boundary at $x_2=1$ fixed. The main distinction between this case and the massive triangle in Section~\ref{sec:triangle} enters here; we want to solve the variety as $x_2=f\left(x_1\right)$ but this gives us a function involving a square root containing $x_1$ instead of a purely rational function. We find that, picking the relevant square root solution for the transformed $\mathcal{F}$,
\begin{gather}
    \mathcal{F}=0\nonumber\\\Downarrow\nonumber\\
    x_2=1-\frac{3\left(1-x_1\right)x_1\beta \bar{\beta}^2}{\sqrt{3\!\left(1\!-\!x_1\right)\!x_1 \bar{\beta}^2 \left(4\!-\!\beta^2\!\left(2\left(1\!+\!\beta^2\right)\!-\!3\left(1\!-\!x_1\right)\!x_1 \bar{\beta}^2 \right)\!-\!2\!\left(1\!-\!2x_1\right)\!\beta \tilde{\beta} \bar{\beta}\right)}}=f\left(x_1\right).
\end{gather}
Applying the same set of transformations to the $\mathcal{U}$ polynomial and accounting for the corresponding Jacobian determinants, we obtain for our final result for the negative contribution,
\begin{align}
    I_{\mathrm{sun}}^-= & \ 2^{7-6 \epsilon } 3^{\frac{1}{2}-\epsilon } \left(\beta ^2\right)^{2-2 \epsilon } \left(\bar{\beta}^2\right)^{2-2 \epsilon } \left(\frac{\tilde{\beta}}{m^2}\right)^{-1+2\epsilon} \nonumber \\
    &\!\!\int_{0}^{1}\!\!\mathrm{d}x_1\mathrm{d}x_2\left(1-x_1\right){}^{\frac{3}{2}-2 \epsilon } x_1^{\frac{3}{2}-2\epsilon } x_2^{1-2 \epsilon }  R_{\mathrm{sun}}^-\left(x_1, x_2; \beta\right),
\label{isunminus}
\end{align}
 where the finite remainder function is given by,
\begin{align}
&R_{\mathrm{sun}}^-\left(x_1, x_2; \beta\right) =
R_1(x_2;\beta)
R_2(x_1;\beta) 
R_3(x_1;\beta) 
R_4(x_1,\beta) 
R_5(x_1,x_2;\beta), \\
&R_1(x_2;\beta) = \bar{x}_2^{1-2 \epsilon }, \\
&R_2(x_1;\beta) = \left[-\beta ^2+\beta  \bar{\beta}  \tilde{x}_1+4\right]{}^{3 \epsilon -2}, \\
&R_3(x_1;\beta) =\left[4-\beta  \left(2 \beta  \left(\beta ^2+1\right)-3 \beta  \bar{\beta} ^2 x_1 \bar{x}_1+2 \tilde{\beta}  \bar{\beta}  \tilde{x}_1\right)\right]{}^{\frac{3}{2}-\epsilon }, \\
&R_4(x_1;\beta) = \left[\beta ^2 \bar{\beta} ^2 x_1 \bar{x}_1 \left(-11 \beta ^2+3 \beta  \bar{\beta}  \tilde{x}_1+20\right)+4 \tilde{\beta} ^2 \left(\beta ^2-\beta  \bar{\beta}  \tilde{x}_1+4\right)\right]{}^{1-2 \epsilon }, \\
&R_5(x_1,x_2;\beta) =\Bigl[\beta ^2 \bar{\beta} ^2 x_1 \bar{x}_1 \left(x_2 \bar{x}_2 \left(-\beta ^2+\beta  \bar{\beta}  \tilde{x}_1+4\right)+4 \beta  \left(3 \beta -\bar{\beta}  \tilde{x}_1\right)\right)\nonumber\\&+4 \tilde{\beta}  \left(\beta ^4+7 \beta ^2-\left(\beta ^2+3\right) \beta  \bar{\beta}  \tilde{x}_1+4\right)\Bigr]{}^{3 \epsilon -3},
\end{align} 
with $\bar{x}_1=1-x_1,\ \tilde{x}_1=1-2x_1,\  \bar{x}_2=2-x_2$ and $\bar{\beta}=\!\sqrt{8+\beta^2},\  \tilde{\beta}=\!1-\beta^2$.
Each of the factors in the integrand of \eqref{isunminus} can be shown to be positive for ${0<x_1<1}$, ${0<x_2<1}$ and ${0<\beta<1}$, thereby removing the need for a contour deformation.

We can perform similar resolutions for the positive regions to obtain the overall construction:
\begin{equation}
    I_{\mathrm{sun}}=\sum_{n_+=\ \!\!1}^{3}I_{\mathrm{sun}}^{+,n_+}+\left(-1-i\delta\right)^{1-2\epsilon}I_{\mathrm{sun}}^{-}.
    \label{sunsum}
\end{equation}
The increase in efficiency of instead integrating the non-negative integrands of the constituent integrals in \eqref{sunsum} with \pysecdec (as opposed to the standard contour-deformed setup) can be seen in Section~\ref{sec:performance}.

\subsubsection{3-Loop Hyperelliptic Banana}
\label{sec:banana}
In this section, we investigate the 3-loop equal mass banana integral. One may associate a geometry to the $L$-loop massive banana defined by the variety of the $\mathcal{F}$ polynomial in the complex projective space, $\mathbb{CP}^L$; for $L\geq2$, the corresponding geometry is that of a Calabi-Yau $(L-1)$-fold (see, for example, \cite{P_gel_2023}). 
In the previous section, we demonstrated that it is possible to resolve the sunrise integral (in this language, a 2-loop massive banana) which is associated to an elliptic curve (i.e. a Calabi-Yau 1-fold) at the expense of seemingly inescapably introducing square roots involving the Feynman parameters. 
In this section, we show that an integral associated to the more complex geometric structure of a K3 surface (i.e. a Calabi-Yau 2-fold) provides no obstruction to our resolution procedure. 
We make no further comment on the underlying geometry associated to these integrals except to say that this motivates studying these particular examples by proving that these geometric properties are not inherently prohibitive to this procedure. 
We also acknowledge that it is known that additional non-trivial geometric structure appears for this family of integrals with $L\geq4$  but resolving fully-massive integrals with more than four propagators is beyond the scope of this current work. \\

\begin{figure}[t]
    \centering
    \raisebox{-0.5\height}{\includegraphics[width=0.3\textwidth]{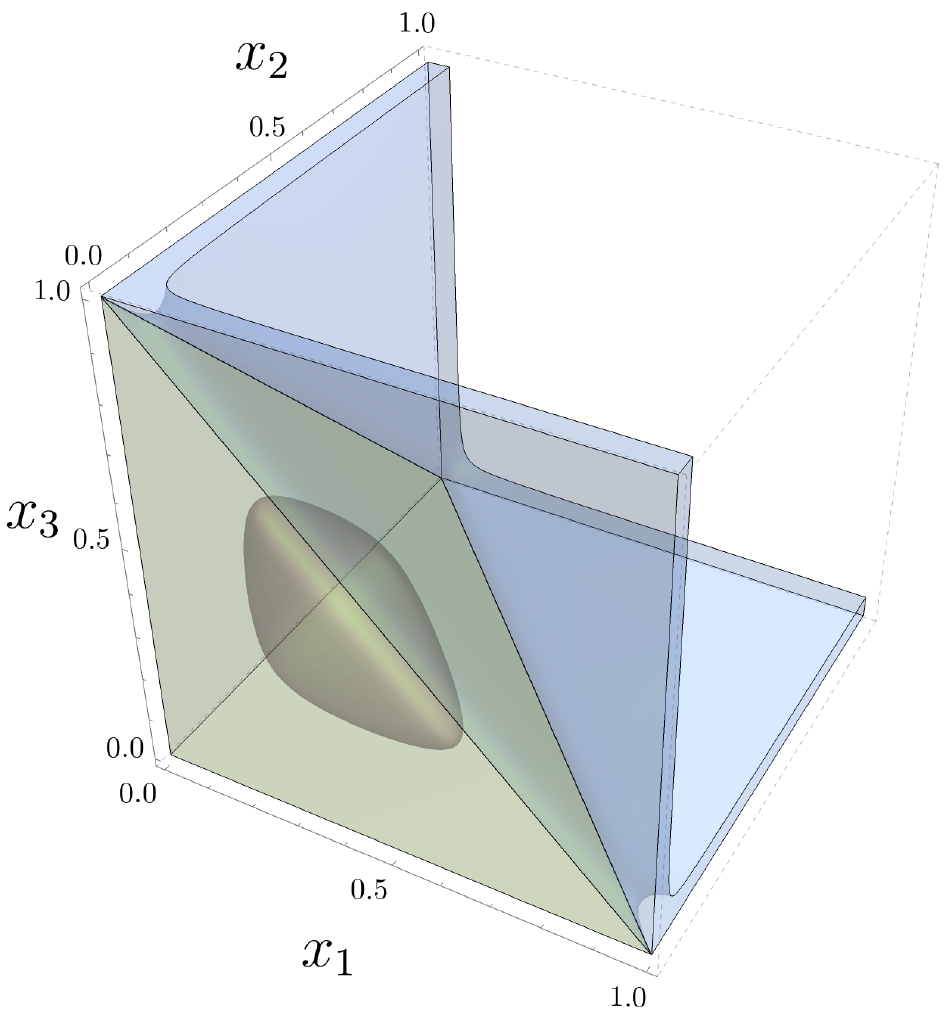}}
    \hfill $\Rightarrow$ \hfill
    \raisebox{-0.5\height}{\includegraphics[width=0.3\textwidth]{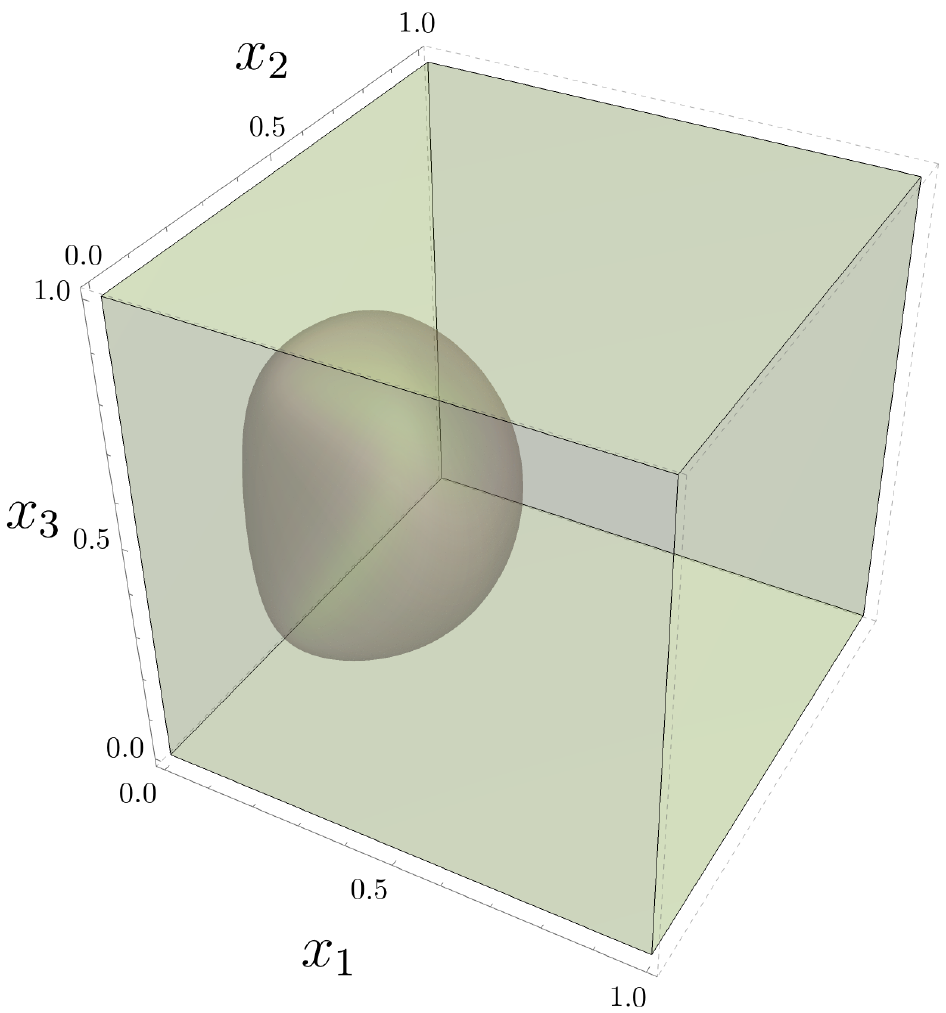}}\hfill $\Rightarrow$\hfill
    \raisebox{-0.5\height}{\includegraphics[width=0.3\textwidth]{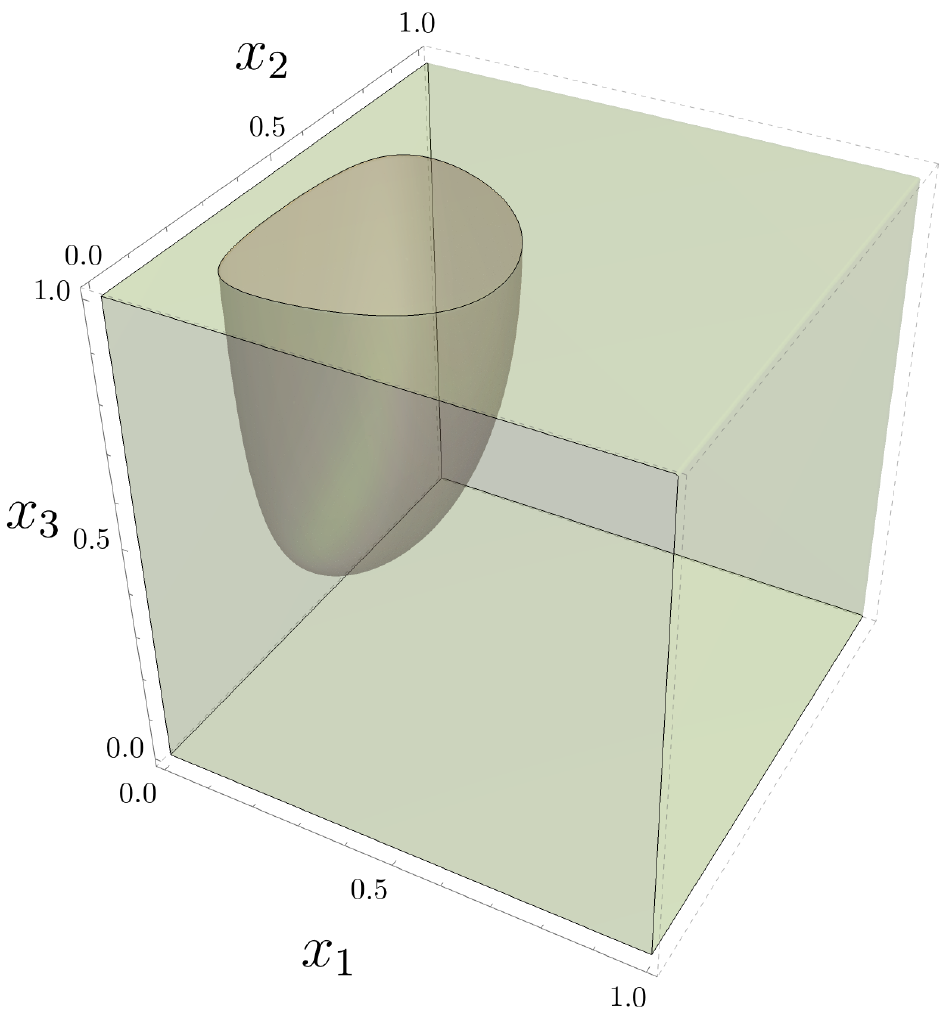}} \\
    \caption{Remapping the simplex integration region of the banana (in green) to the positive unit cube in $\mathbb{R}^3_{\geq0}$ then exploiting the symmetry about $x_3=\frac{1}{2}$. We omit the legend for clarity but this figure should be understood analagously to Fig.~\ref{fig:ellipticdomains} with the $\mathcal{F}<0$ region given in blue outside the domain of integration and in orange within. Here, $\mathcal{F}$ is to be understood as $\mathcal{F}$ after the $\delta$-function has been integrated out and in the second and third panels, after their respective remapping transformations as well. The kinematic regime is given by $\beta^2\in(0,1)$ (with $m^2>0$).} 
    \label{fig:bananadomains}
\end{figure}

The 3-loop equal mass banana integral (shown in Fig.~\ref{banana-diag}) is given by
\begin{align}
J_{\mathrm{ban}}&=\lim_{\delta\to0^+}\Gamma\left(-2+3\epsilon\right)I_{\mathrm{ban}}\\I_{\mathrm{ban}}&=\int_{\mathbb{R}_{\geq0}^{4}}\prod\limits_{i=1}^{4}\mathrm{d}x_{i}\frac{\left(x_1 x_2 x_3+x_1 x_3 x_4+x_2 x_3 x_4+x_1 x_2 x_4\right)^{-4+4\epsilon}\delta\left(1-\alpha(\mathbf{x})\right)}{ \left(\mathcal{F}\left(\mathbf{x},\mathbf{s}\right)-i\delta\right)^{2-3\epsilon}}
\end{align}
where the $\mathcal{F}$ polynomial is
\begin{equation}
    \mathcal{F}\left(\mathbf{x},\mathbf{s}\right)=-s x_1 x_2 x_3 x_4+\left(x_1 x_2 x_3+x_1 x_3 x_4+x_2 x_3 x_4+x_1 x_2 x_4\right)m^2 \left(x_1+x_2+x_3+x_4\right).
\end{equation}
Having presented the geometric resolution procedure a number of times already, we detail this example more schematically. We begin by integrating out $x_4$ using the $\delta$-function with the symmetric choice of hyperplane (i.e. ${x_1+x_2+x_3+x_4}$) and we parameterise the Minkowskian kinematic regime with ${\beta^2=\frac{s-16m^2}{s}\in\left(0,1\right)}$. Analagously to the elliptic sunrise example, we remap the resulting simplex integration domain to the positive unit hypercube (in this case, the positive unit cube in $\mathbb{R}^3$). The symmetry of the resulting integrand under $x_3\rightarrow 1-x_3$ allows us once again to reduce the number of integrals into which we will eventually decompose $I_{\mathrm{ban}}$ by integrating instead from $x_3=0$ up to $x_3=\frac{1}{2}$, remapping this back to the unit positive cube and doubling the result (however, we remark once more that this is not strictly necessary and one could perform the entire resolution without exploiting this). This sequence of domain-remapping transformations is depicted in Fig.~\ref{fig:bananadomains} (where this time we focus solely on the positive unit cube) and should be compared and contrasted with the corresponding set of transformations for the elliptic sunrise in Fig.~\ref{fig:ellipticdomains}.

\begin{figure}
    \centering
    \begin{subfigure}{0.16\textwidth}
        \centering
        \includegraphics[width=\linewidth]{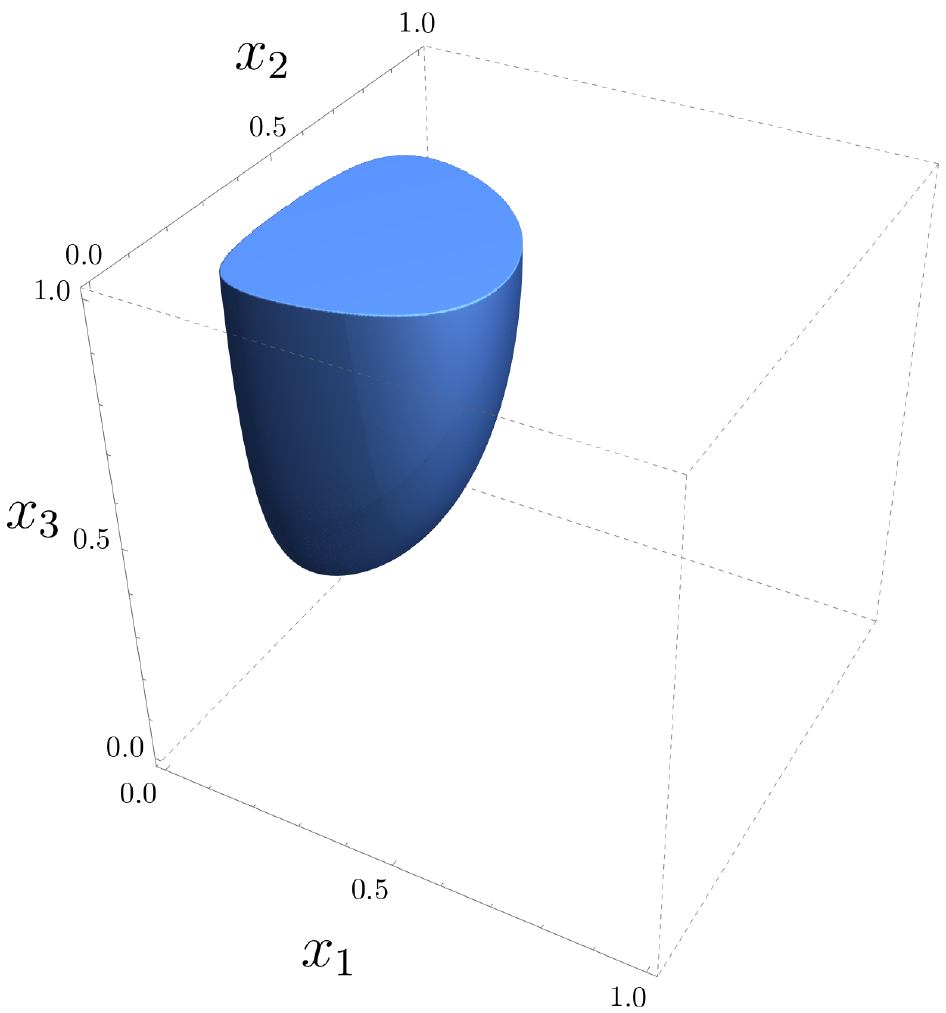}
        \caption{$I_\mathrm{ban}^-$}\label{banneg}
        
    \end{subfigure}
    \hfill
    \begin{subfigure}{0.16\textwidth}
        \centering
        \includegraphics[width=\linewidth]{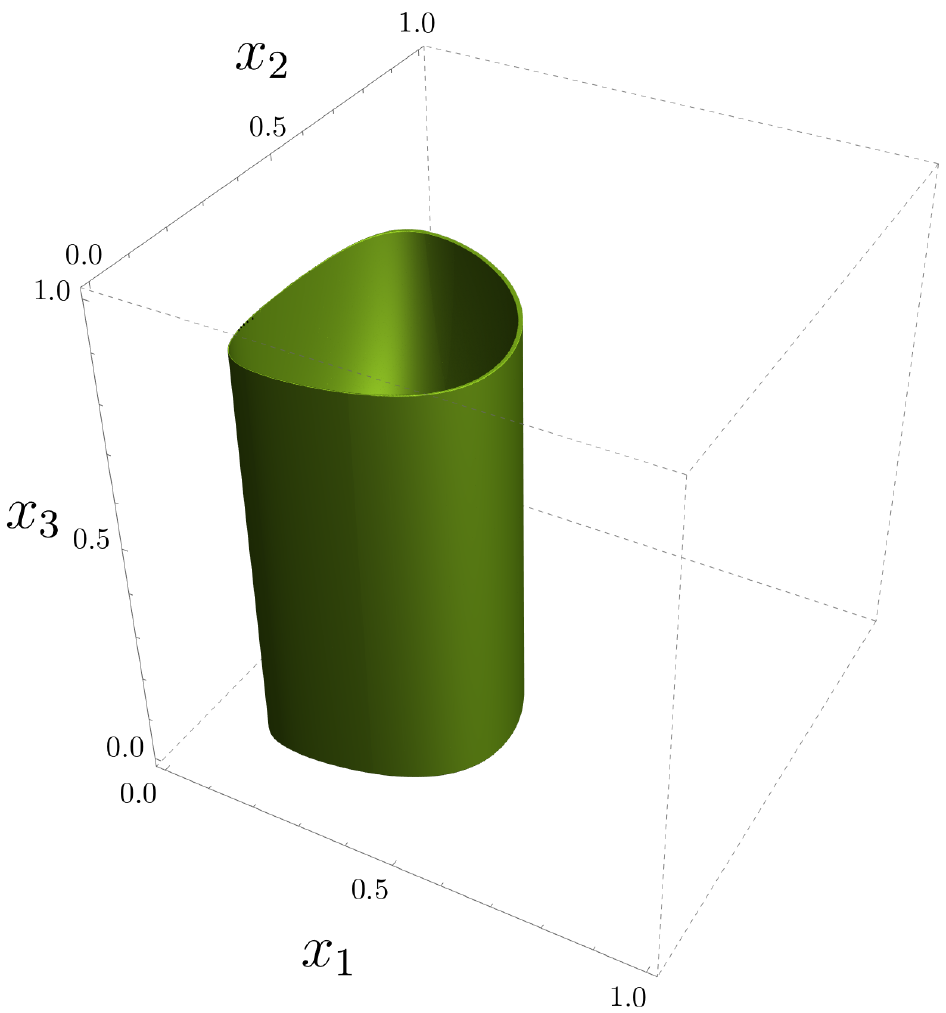}
        \caption{$I_\mathrm{ban}^{+,1}$}\label{banpos1}
    \end{subfigure}
    \hfill
    \begin{subfigure}{0.16\textwidth}
        \centering
        \includegraphics[width=\linewidth]{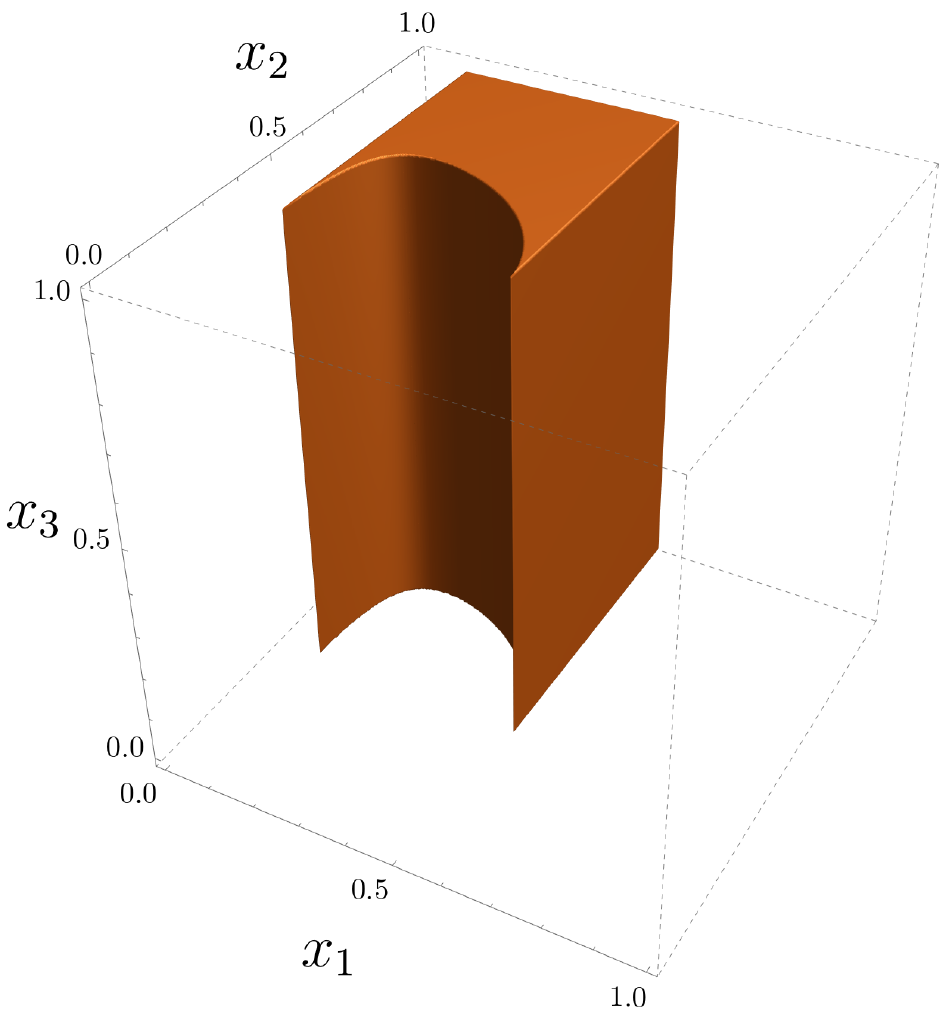}
        \caption{$I_\mathrm{ban}^{+,2}$}\label{banpos2}
    \end{subfigure}
    \hfill
    \begin{subfigure}{0.16\textwidth}
        \centering
        \includegraphics[width=\linewidth]{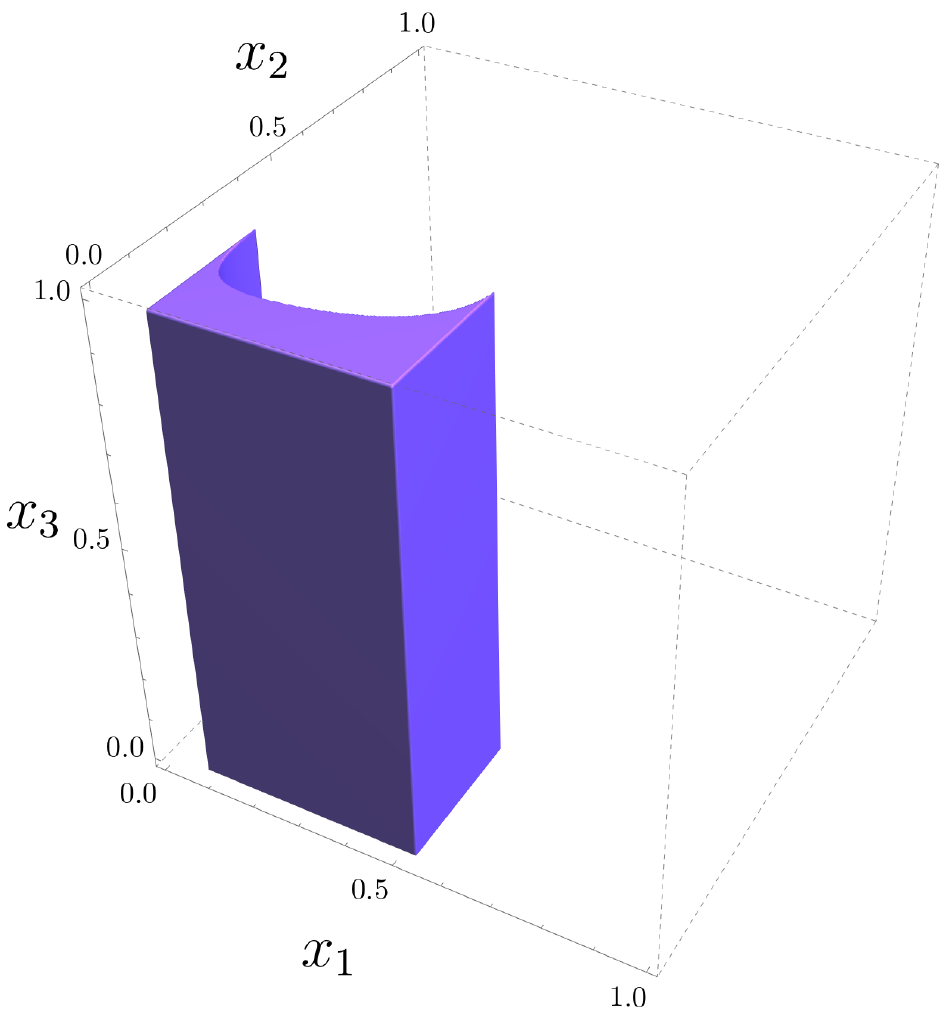}
        \caption{$I_\mathrm{ban}^{+,3}$}\label{banpos3}
    \end{subfigure}
    \hfill
    \begin{subfigure}{0.16\textwidth}
        \centering
        \includegraphics[width=\linewidth]{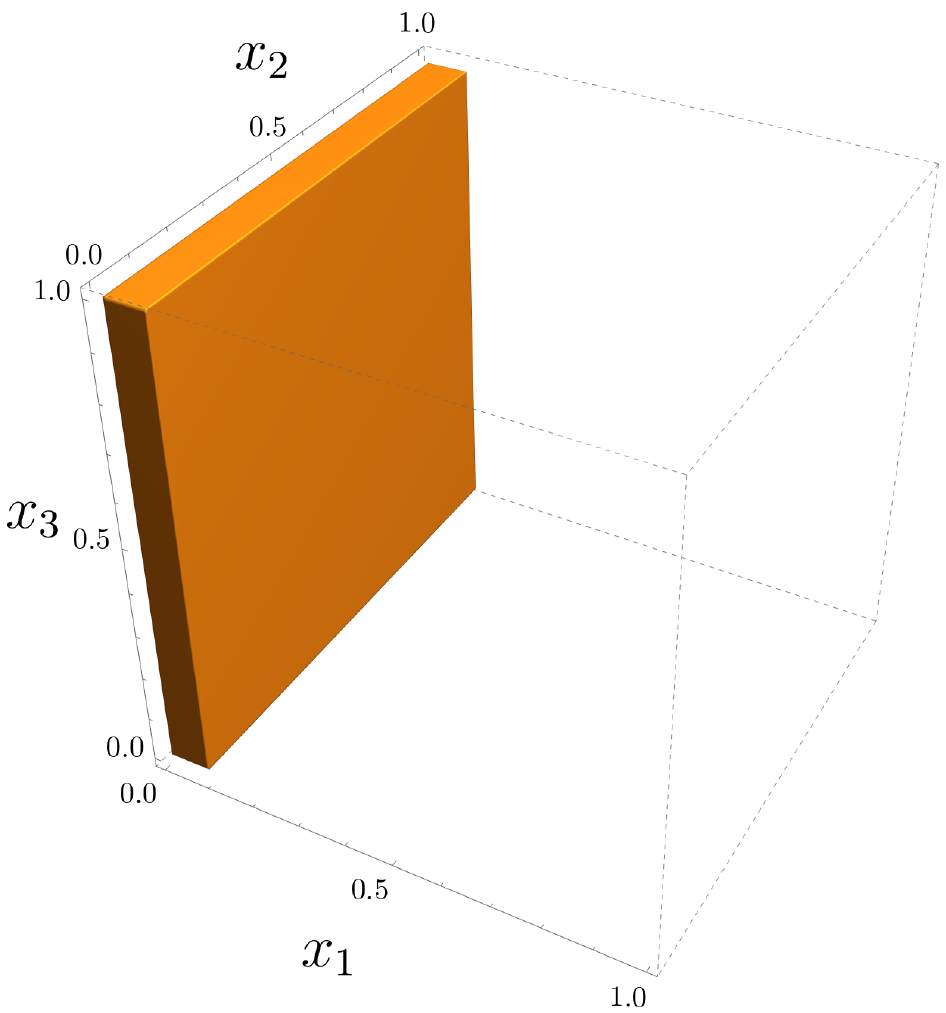}
        \caption{$I_\mathrm{ban}^{+,4}$}\label{banpos4}
    \end{subfigure}
    \hfill
    \begin{subfigure}{0.16\textwidth}
        \centering
        \includegraphics[width=\linewidth]{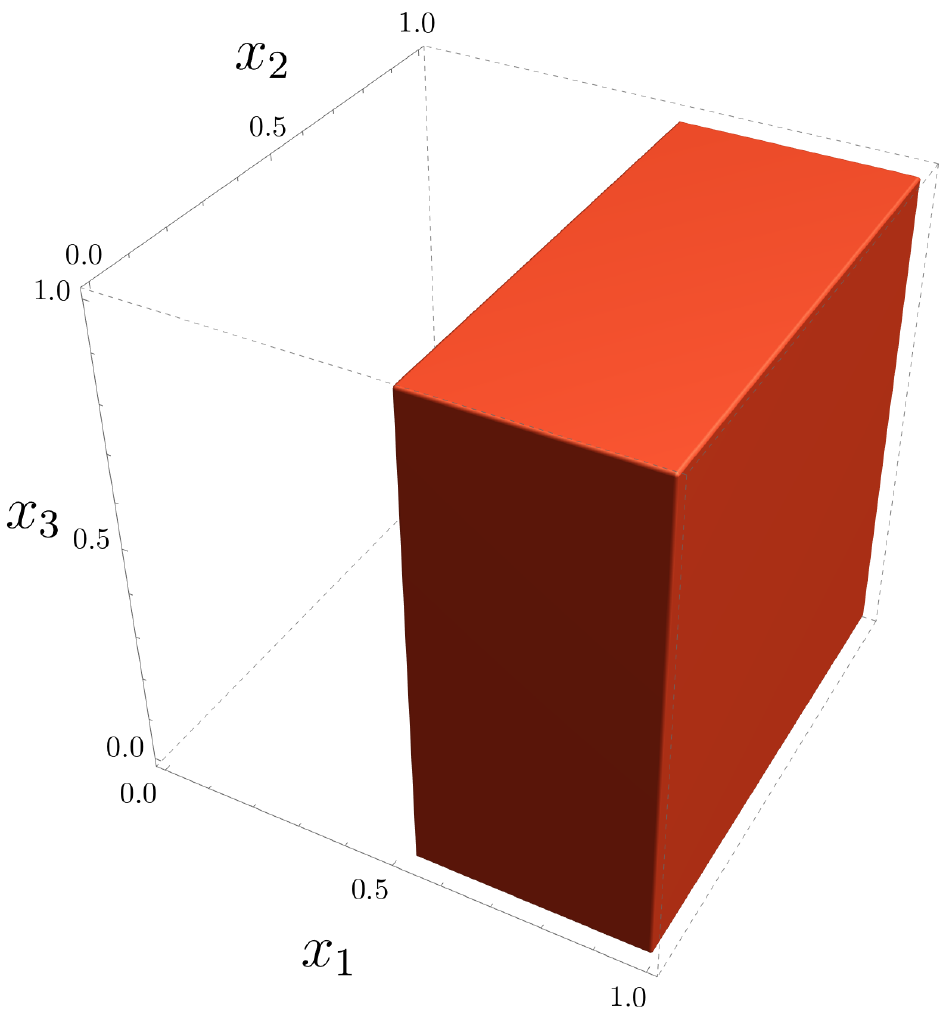}
        \caption{$I_\mathrm{ban}^{+,5}$}\label{banpos5}
    \end{subfigure}
    \caption{The six regions (one negative, \ref{banneg}, and five positive, \ref{banpos1} -- \ref{banpos5}) into which the integration domain of the banana is partitioned in this resolution.}
    \label{fig:bananaregions}
\end{figure}

Once the domain has been successfully remapped to the unit positive cube, it may be (somewhat arbitrarily) dissected into regions where the transformed $\mathcal{F}$ polynomial is positive and negative. The art of partitioning the integration domain in general, as we currently understand it, revolves around the balance between minimising the number of integrals in the decomposition (minimally one negative region and one positive region) and producing regions which can all be successfully mapped back to the positive unit hypercube. We show, in Fig.~\ref{fig:bananaregions}, an example partitioning of the positive unit cube in $\mathbb{R}^3$ for the resolution of the banana which generates five positive contributions and one negative contribution. 
\begin{figure}[t]
    \centering
    \raisebox{-0.5\height}{\includegraphics[width=0.2\textwidth]{figures/bananaregion2.pdf}}
    \hfill $\Rightarrow$ \hfill
    \raisebox{-0.5\height}{\includegraphics[width=0.2\textwidth]{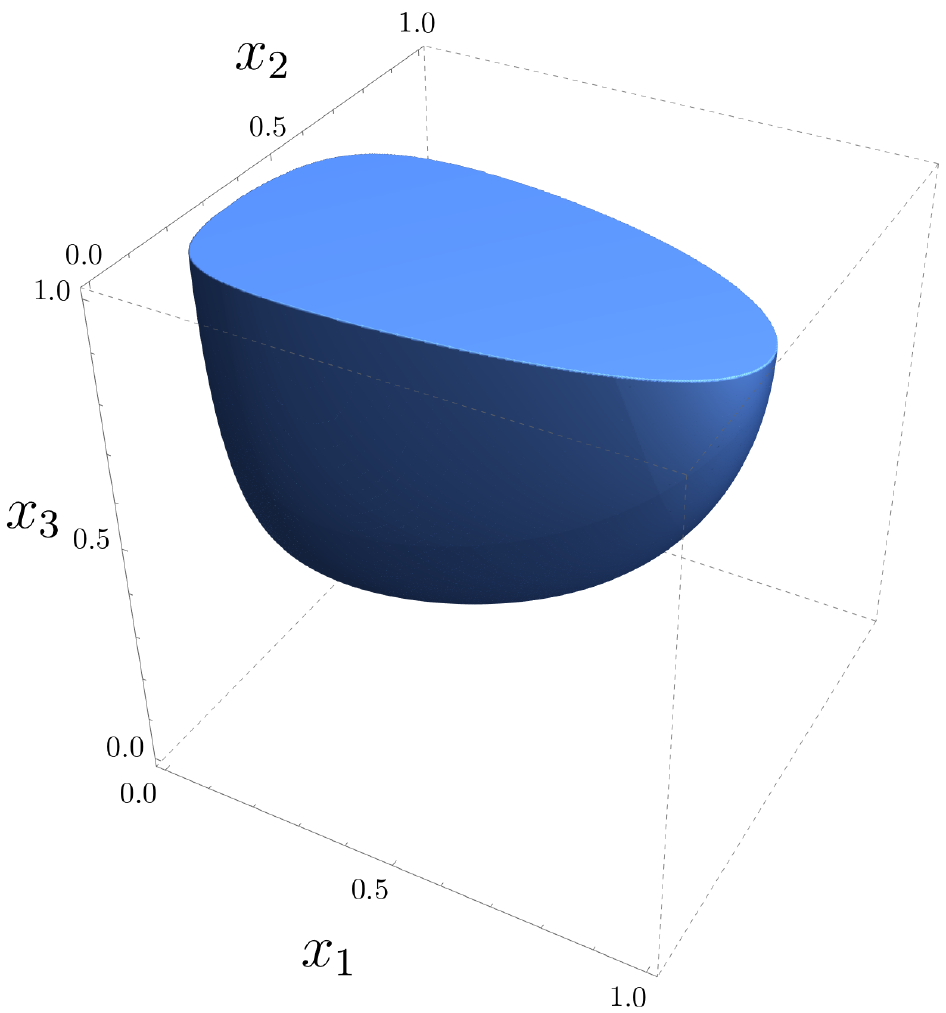}}\hfill $\Rightarrow$\hfill
    \raisebox{-0.5\height}{\includegraphics[width=0.2\textwidth]{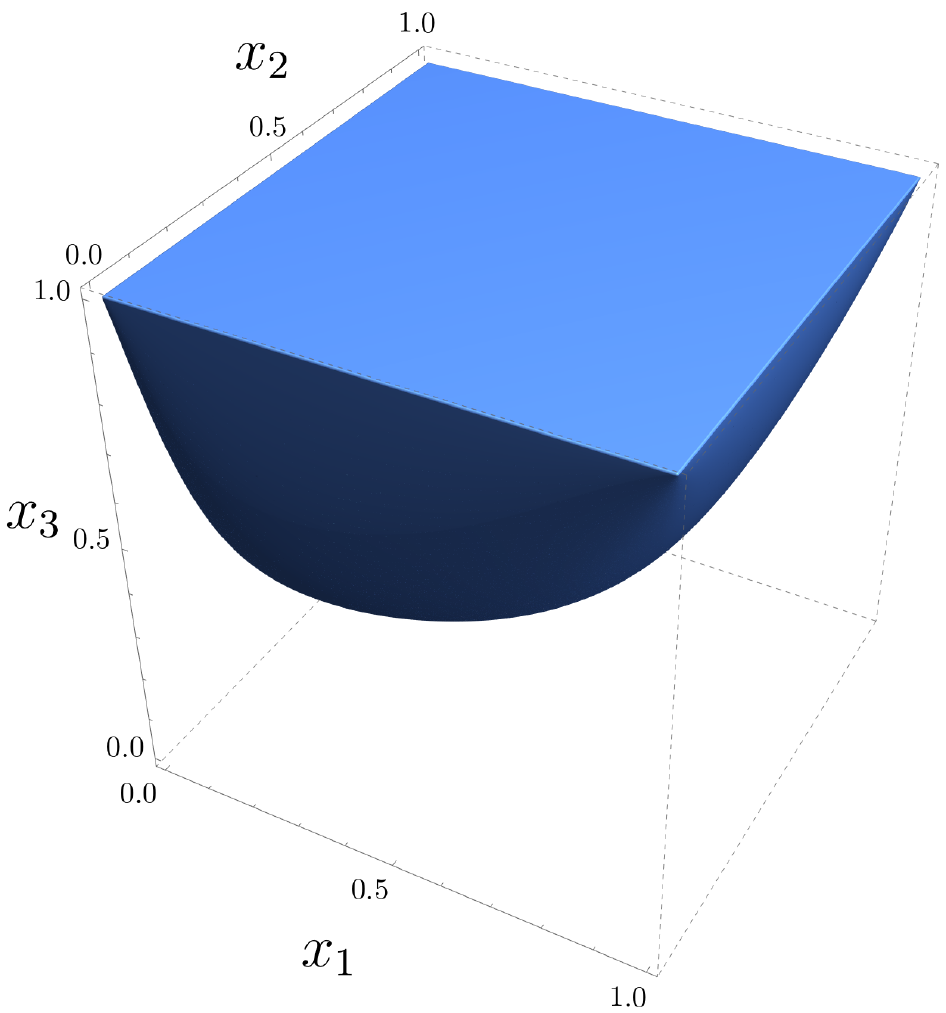}}\hfill $\Rightarrow$\hfill
    \raisebox{-0.5\height}{\includegraphics[width=0.2\textwidth]{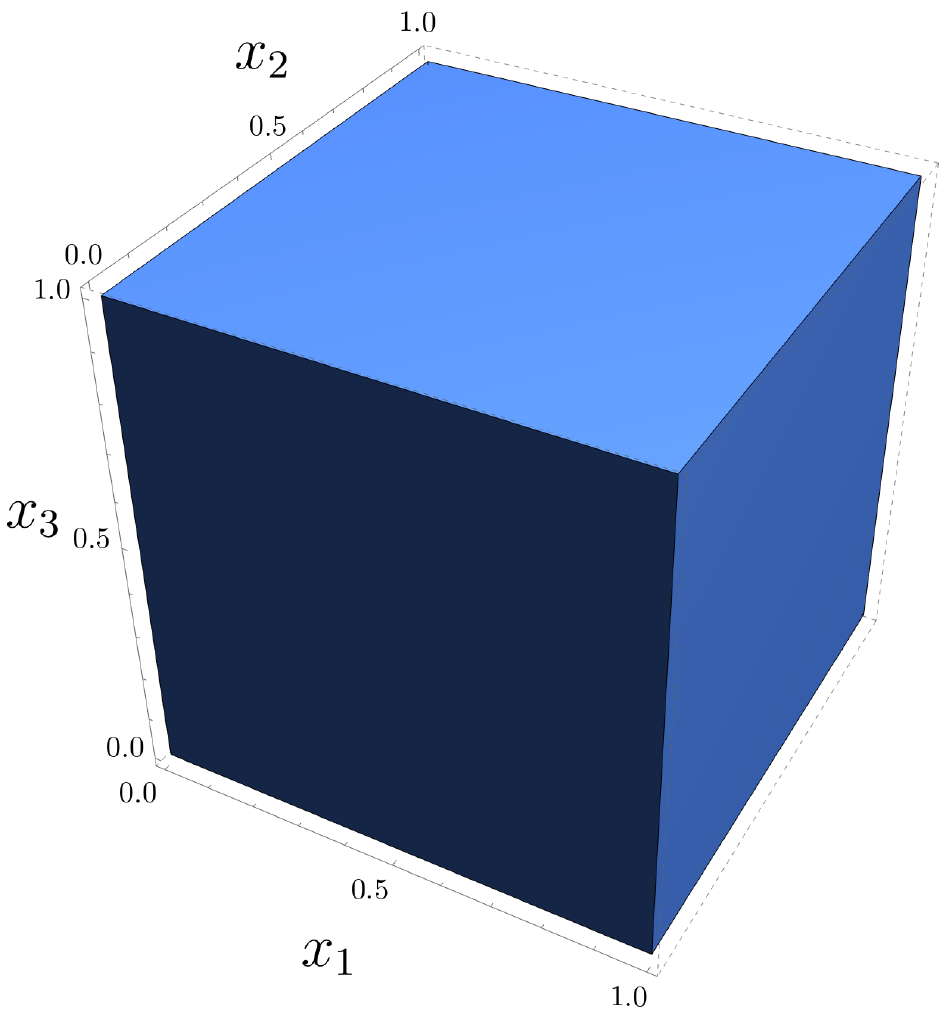}} \\
    \caption{The chain of transformations which maps the negative region of the banana (in blue) to the positive unit cube.}
    \label{fig:bananatrans}
\end{figure}

We will focus on the set of transformations which maps the negative region to the original integration domain as shown in Fig~\ref{fig:bananatrans}. First, we look at the intersection of the variety of the transformed $\mathcal{F}$ polynomial with the plane $x_3=1$ and analyse the resulting closed curve. 
It is clear from focusing on the plane $x_3=1$ in the first panel of Fig.~\ref{fig:bananatrans} that this curve has four turning points (by which we mean the four points on the curve of maximal and minimal $x_1$ and $x_2$ respectively). In the resolution of the negative region shown in Fig.~\ref{fig:bananatrans}, we arbitrarily choose to first identify the points of maximal and minimal $x_1$ on the curve using standard turning point analysis and then map the planes $x_1=x_1^\mathrm{min}$ and $x_1=x_1^\mathrm{max}$ to $x_1=0$ and $x_1=1$ respectively. This transformation is shown in the transition between the first and second panels of Fig.~\ref{fig:bananatrans}. On the curve, both $x_1^\mathrm{min}$ and $x_1^\mathrm{max}$ have a corresponding $x_2$-value of $x_2=\frac{1}{3}$ (independent of $\beta$) and taking a slice through the integration domain at this value of $x_2$ (combining all regions in Fig.~\ref{fig:bananaregions}) generates a plot which is, superficially, very similar to Fig.~\ref{fig:ellipticregions}. The next step is to solve the transformed curve at $x_3=1$ as $x_2=f\left(x_1\right)$. Clearly, from the second panel of Fig.~\ref{fig:bananatrans}, this will have two solutions (the curve is quadratic in $x_2$) and a square root involving the Feynman parameters -- specifically, $x_1$ -- enters at this point. The map which takes $x_2=f_1\left(x_1\right)$ to the plane $x_2=1$ and $x_2=f_2\left(x_1\right)$ to the plane $x_2=0$ is trivial to construct once $f_1$ and $f_2$ are known and we demonstrate the effect of this transformation in the transition between the second and third panels of Fig.~\ref{fig:bananatrans}. Finally, we have to solve the variety of the transformed $\mathcal{F}$ as $x_3=g\left(x_1,x_2\right)$ (which also has two solutions due to the quadratic appearance of $x_3$ but only one solution which is relevant within the integration domain) and map this to the plane $x_3=0$ while keeping $x_3=1$ fixed. This transformation is shown in the transition between the third and fourth panels of Fig.~\ref{fig:bananatrans} and concludes the mapping of the original negative region to the positive unit cube.

A similar analysis is carried out for the remaining five positive regions in Fig.~\ref{fig:bananaregions}, resulting in\footnote{For our numerical evaluation in \pysecdec, we perform a manual sector decomposition of one of the positive contributions into two constituent integrals but this is implementation-dependent and orthogonal to the resolution procedure for avoiding contour deformation.}
\begin{equation}
    I_{\mathrm{ban}}=\sum_{n_+=\ \!\!1}^{5}I_{\mathrm{ban}}^{+,n_+}+\left(-1-i\delta\right)^{1-2\epsilon}I_{\mathrm{ban}}^{-}
    \label{bansum}
\end{equation}
where all of the integrands appearing in \eqref{bansum} are manifestly positive within the domain of integration (and away from the boundary).

\section{Numerical Benchmarks}
\label{sec:performance}
The procedure described above yields manifestly positive integrands from
parameter integrals that lack a definite sign.
This greatly reduces the complexity of numerically evaluating the remaining integrals and is likely to be beneficial for a variety of parameter space-based approaches to evaluating Feynman integrals as discussed further in Section~\ref{sec:eval}. For finite integrals, we can numerically integrate the resulting integrands either directly or using FeynTrop~\cite{Borinsky_2023}, for example.
For divergent integrals, we still need to perform a subtraction to obtain finite integrals, for which we can employ sector decomposition as implemented in tools such as FIESTA~\cite{Smirnov:2008py,Smirnov:2009pb,Pak:2010pt,Smirnov:2013eza,Smirnov:2015mct,Smirnov:2021rhf} or \pysecdec~\cite{Borowka:2017idc,Borowka:2018goh,Heinrich:2021dbf,Heinrich:2023til}.

In this work, we use the public \pysecdec v1.6.4 program to benchmark the integration time of the resolved integrals and compare to the timings when using contour deformation. 
We emphasise that \pysecdec is not optimised for this new approach and its efficiency could be greatly improved for resolved integrals.
For example, parsing prefactor expressions can dominate the evaluation time, primarily due
to the inefficient \texttt{SymPy} routines used. In contrast, the same expressions load
almost instantly in systems like \texttt{Mathematica}.
These expressions appear as a result of the resolution procedure, particularly in the massive examples, and have therefore not been a bottleneck in \pysecdec before. 
We subtract the loading time of all prefactors from the results, and perform our comparisons strictly on the actual \textit{integration time}. 
This issue is particularly relevant in the case of the 1-loop massive triangle.
Additional inefficiencies in the current implementation of \pysecdec include the re-evaluation of common sub-expressions appearing across multiple $\epsilon$ orders and the unnecessary expansion of finite expressions (especially those appearing at higher orders in $\epsilon$ in finite integrals).
Parts of the current implementation also unnecessarily utilise complex numbers, even for the real integrands we obtain after resolution.
For this reason, the speed-up factors presented in this section should be interpreted as a pessimistic estimate of the impact of the resolution procedure.

Unless otherwise specified, the integration was performed using the \texttt{Disteval} integrator, which was run on an NVIDIA A100 80G GPU. The \pysecdec integration libraries were compiled\footnote{Using the following make command: \texttt{make disteval SECDEC\_WITH\_CUDA\_FLAGS=-arch=sm\_80 CXXFLAGS="-O3 -mfma -mavx2" CXX=g++-12.}} using \cuda 12.4.131. 
The exceptions are the 2L sunrise and 3L banana integrals, where an older integrator was used in order to access the feature of user-defined \texttt{C++} functions. 
The resolved versions of these examples contain large positive remainder functions raised to high integer powers, that by default get expanded into very large expressions by \texttt{FORM} routines within \pysecdec. 
Future versions of \pysecdec should therefore offer the option to prevent certain functions from being expanded. 
For now, this problem can be circumvented by manually defining the remainder expressions as symbolic functions, and providing them directly as \texttt{C++} functions. 
The integration libraries for these two examples were compiled\footnote{Using the simple make command: \texttt{make pylink}.} with \texttt{gcc}~7.5.0 and run on one core of an AMD EPYC 7352 CPU. 

We now proceed by presenting benchmarks for each resolved integral, comparing integration times with and without contour deformation. Both massless and massive integrals are evaluated for increasing centre-of-mass energies with other kinematics fixed. Additionally, the massive integrals are benchmarked in the small-mass regime, where the centre-of-mass energy is fixed and the internal mass $m$ is pushed to very small values. We emphasise that for the massive examples shown here, results are not obtained using Algorithm~\ref{alg:univariate} and instead various possible generalisations of this algorithm were explored. 
In particular, results are obtained after integrating out the Dirac-delta functional with a particular choice of hyperplane.
We do not expect our resolution to be optimal; instead, our goal is to demonstrate that contour deformation can, in principle, be avoided for massive integrals, and that correct numerical results can still be obtained in a reasonable amount of time.
In future work, we will explore avenues for improving the resolution procedures, especially for massive integrals. 

\subsection{1-Loop Box with One Off-Shell Leg}\label{subsec:1-loopboxtimings}
\begin{figure}[h!]
\centering
    \includegraphics[width=0.9\textwidth]{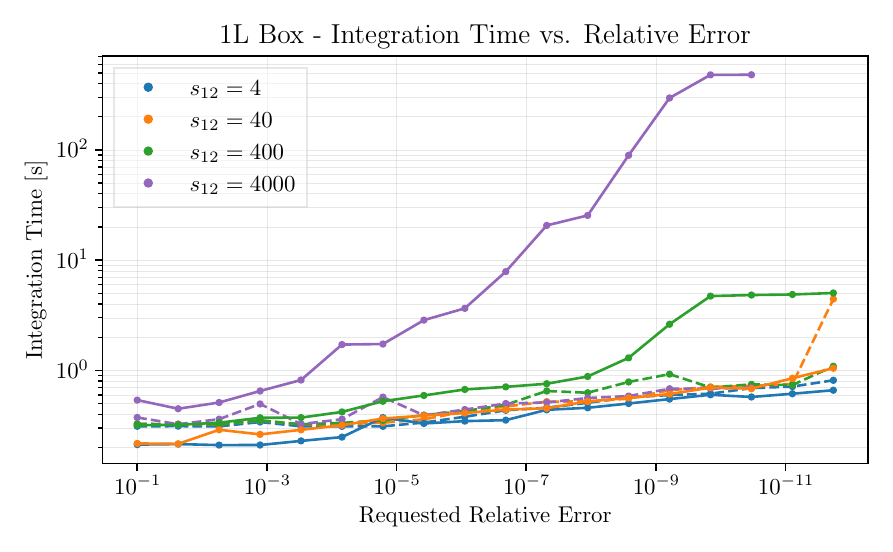}
    \caption{Timings with (solid lines) and without (dashed lines) contour deformation for the massless 1-loop box with off-shell legs, expanded up to the finite order. Evaluated for different values of $s_{12}$ with $s_{13}=-1$ and $p_1^2=2$ fixed.}
    \label{fig:box1L_timings}
\end{figure}

Figure~\ref{fig:box1L_timings} shows the integration times for the 1-loop box with one off-shell leg, with and without contour deformation, for increasing values of the centre-of-mass energy $s_{12}$, keeping $s_{13}=-1$ and $p_1^2=2$ fixed. For non-extreme values of $s_{12}$, we do not observe significant differences between the two versions, which is expected since the 1-loop box is a very simple integral and the time is dominated by fixed overheads, e.g. generating the Quasi Monte Carlo sampling lattice. However, the timings for large $s_{12}$ reveal a general feature that persists in more complicated integrals: the relative performance of the contour deformation is much worse in extreme kinematic configurations. For example, to obtain 10 digits of precision with $s_{12}=4000$, avoiding contour deformation yields a factor of $\frac{480s}{0.7s} \approx 686 \, \times$ speed-up. 

\subsection{1-Loop On-Shell Pentagon}\label{subsec:1-looppenttimings}
\begin{figure}[h!]
\centering
    \includegraphics[width=0.9\textwidth]{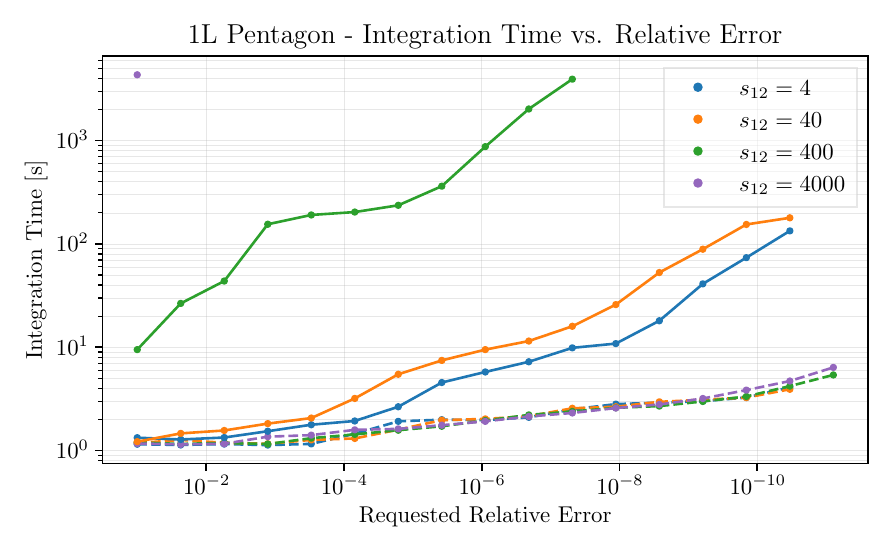}
    \caption{Timings with (solid lines) and without (dashed lines) contour deformation for the massless 1-loop pentagon, expanded up to the finite order. Evaluated for different values of $s_{12}$ while the other kinematics are fixed at $(s_{23},s_{34},s_{45},s_{51}) = (-3,2.5,-3,5)$.}
    \label{fig:pentagon1L_timings}
\end{figure}

Figure~\ref{fig:pentagon1L_timings} shows the integration times for the 1-loop pentagon, with and without contour deformation, for increasing values of the centre-of-mass energy $s_{12}$, keeping the other kinematics fixed as $(s_{23},s_{34},s_{45},s_{51}) = (-3,2.5,-3,5)$. We see similar behaviour to the 1-loop box; the performance differences are more significant for large $s_{12}$. In this case, however, in addition to the aforementioned behaviour being more extreme, the resolved integrands reach high precision more quickly even for non-extreme $s_{12}$ values. For example, to obtain 7 digits of precision with $s_{12}=400$, avoiding contour deformation yields a factor of $\frac{2021}{2.21} \approx 914 \, \times$ speed-up. For the same number of digits with $s_{12} = 4$, we instead get a factor of $\frac{11.5}{2.15} \approx 5.3 \, \times$ speed-up. For $s_{12} \geq 4000$, it is difficult to obtain any digits at all with contour deformation, while the resolved integral is essentially insensitive to this increase. 

\subsection{2-Loop Non-Planar Boxes}
\begin{figure}[h!]
\centering
    \includegraphics[width=0.9\textwidth]{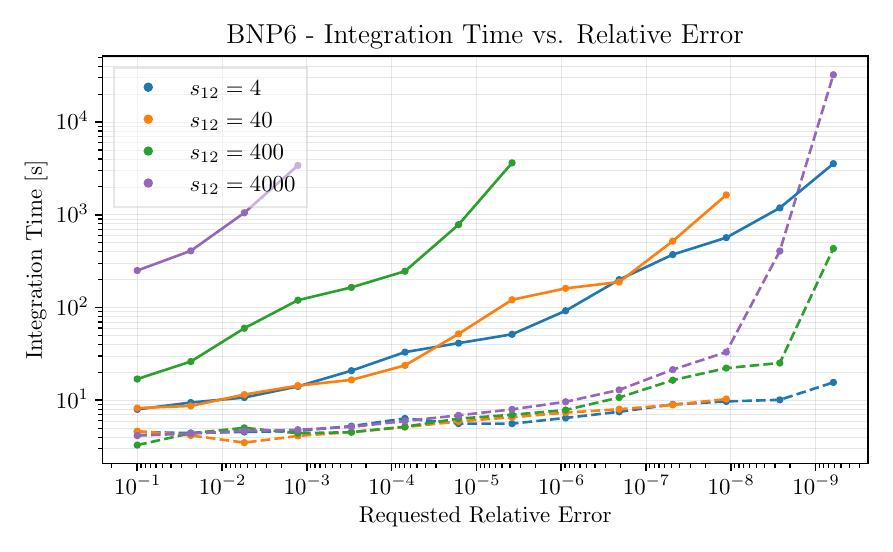}
    \caption{Timings with (solid lines) and without (dashed lines) contour deformation for the 2-loop non-planar box with 6 propagators, expanded up to the finite order. Evaluated for different values of $s_{12}$ with $s_{23}=-1$ fixed.}
    \label{fig:BNP6_timings}
\end{figure}

\begin{figure}[h!]
\centering
    \includegraphics[width=0.9\textwidth]{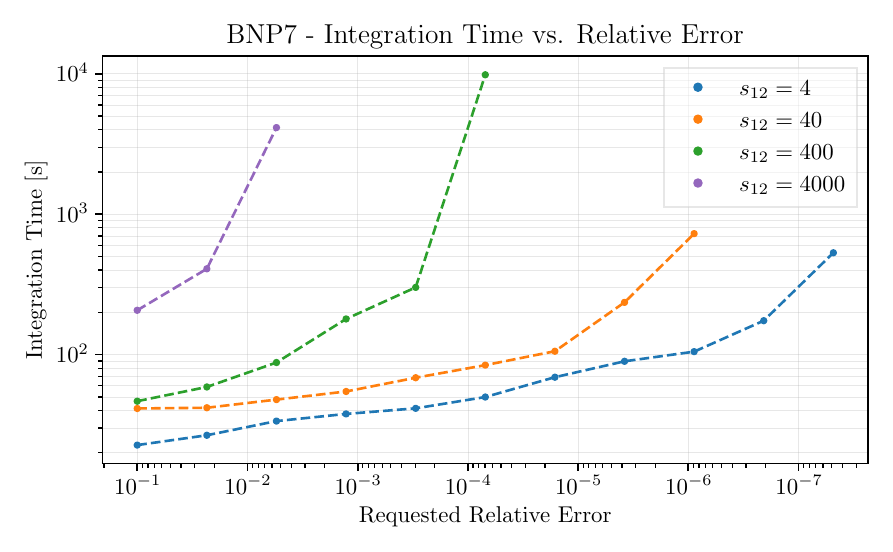}
    \caption{Timings with (solid lines) and without (dashed lines) contour deformation for the 2-loop non-planar box with 7 propagators, expanded up to the finite order. Evaluated for different values of $s_{12}$ with $s_{23}=-1$ fixed. In this example no digits could be obtained with contour deformation within a time limit of 10 hours.}
    \label{fig:BNP7_timings}
\end{figure}

Figures~\ref{fig:BNP6_timings} and \ref{fig:BNP7_timings} show the integration times for the 2-loop non-planar boxes with 6 and 7 propagators respectively (BNP6 and BNP7), with and without contour deformation, for increasing values of the centre-of-mass energy $s_{12}$, keeping $s_{23}=-1$ fixed. These integrals have quite a similar structure, but BNP7 is much more challenging numerically. 
In fact, for BNP7, numerical integration with contour deformation failed to yield any
digits of precision across all tested kinematic configurations, within a
10-hour time limit. 
The resolved BNP7, however, converges with reasonable precision across the entire range in $s_{12}$, though the integration time does deteriorate for large values of $s_{12}$. 
The results for the resolved integral have therefore instead been checked and validated against the analytic expression from Ref.~\cite{Tausk:1999vh}, in which both integrals were computed. 
The integration with contour deformation eventually converged to 1 digit of precision for $s_{12} = 4$ after 192 hours. By avoiding contour deformation, the time to obtain this digit is therefore accelerated by a factor of $\frac{689858}{23}\approx 29994 \times$. The cost to compute additional digits using contour deformation is prohibitive, however, for the resolved integral, we can obtain 6 digits of precision in about 100 seconds.
For BNP6 we see a similar behaviour as for the other massless examples. For example, to obtain about 3 digits of precision with $s_{12}=4000$ we see a factor of $\frac{3404}{4.8} \approx 709 \, \times$ speed-up from avoiding contour deformation. To obtain about 6 digits of precision, but with $s_{12}=4$ instead, avoiding contour deformation yields a factor of $\frac{92}{6.4} \approx 14 \, \times$ speed-up.

\subsection{3-loop Non-Planar Box}

\begin{figure}[h!]
\centering
    \includegraphics[width=0.9\textwidth]{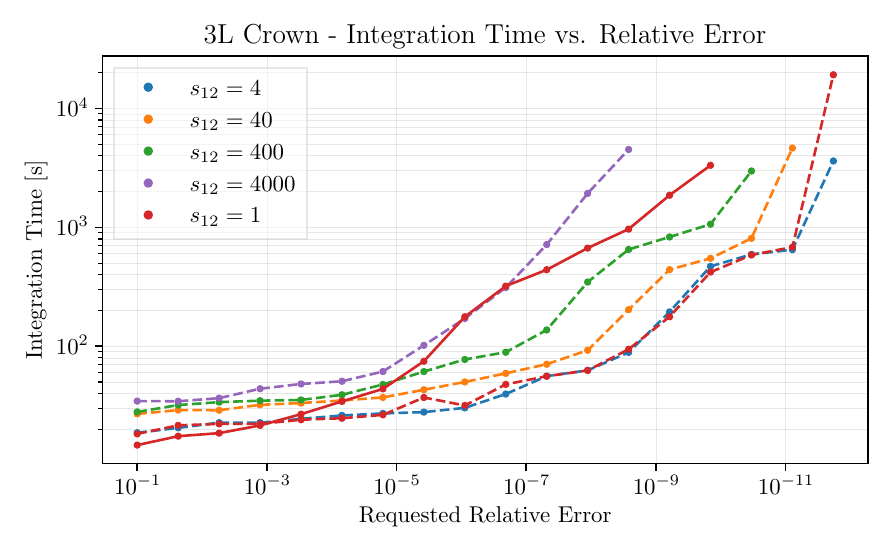}
    \caption{Timings with (solid lines) and without (dashed lines) contour deformation for the 3-loop non-planar box, with only the leading $\epsilon^{-4}$ pole included. Evaluated for different values of $s_{12}$ with $s_{13}=-1$ fixed, except for the point $s_{12}=1$, where $s_{13}=-0.2$. For the benchmarks where $s_{12}\geq4$ no digits could be obtained with contour deformation.}
    \label{fig:crown3L_timings}
\end{figure}

Figure~\ref{fig:crown3L_timings} shows the integration times for the 3-loop non-planar box, only without contour deformation, for increasing values of the centre-of-mass energy $s_{12}$ with $s_{13}=-1$ fixed. Additionally, it includes the point $s_{12}=1, s_{13}=-0.2$, both with and without contour deformation, since this is the only test point we evaluated where a valid contour could be found with \pysecdec. 
For this point, to obtain about 8 digits of precision, we observe a factor of $\frac{670}{62.7} \approx 11 \, \times$ speed-up from avoiding contour deformation. For all other points, integrating with contour deformation fails to provide any result while the resolved integrals achieve high precision for the full range in $s_{12}$. This is different from BNP7, since there valid contours are found but the integration instead converges too slowly to yield any digits.
This means that no amount of computing resources would yield results for the 3-loop box when integrating with contour deformation (as implemented in \pysecdec). 

\subsection{1-Loop Triangle with an Off-Shell Leg}

\begin{figure}[h!]
\centering
    \includegraphics[width=0.95\textwidth]{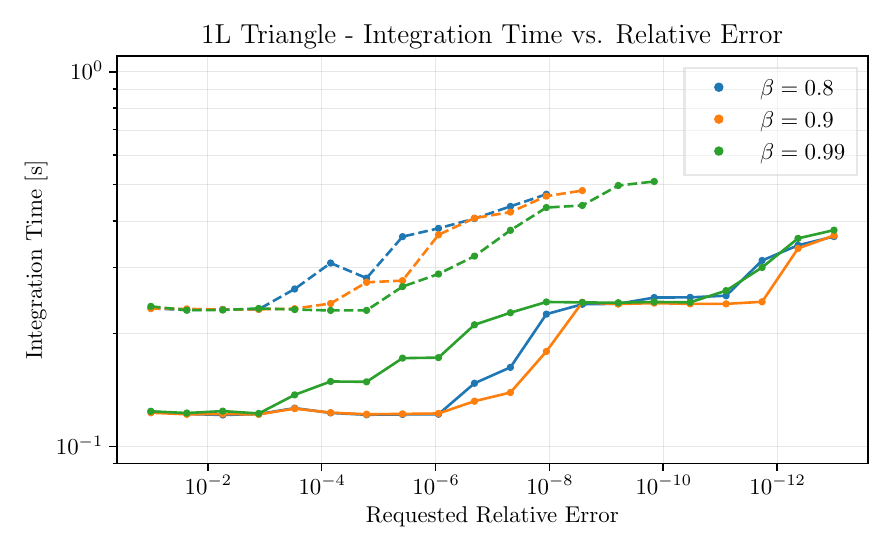}
    \caption{Timings with (solid lines) and without (dashed lines) contour deformation for the all-massive 1-loop triangle, expanded up to order $\epsilon^4$. Evaluated for different values of $\beta$ with $m=1$ fixed.}
    \label{fig:triangle1L3m_timings}
\end{figure}

Figure~\ref{fig:triangle1L3m_timings} shows the integration times for the 1-loop triangle, with and without contour deformation, for increasing values of the centre-of-mass energy (parameterised by $\beta$ in this case), with the internal mass fixed as $m^2=1$. 
For most phase-space points, the triangle can be evaluated to high precision in short time, both with and without contour deformation. The minimal lattice size of the QMC integrator is almost enough to reach about 10 digits of precision, and the example is essentially too simple to see any gain from removing contour deformation. 
In fact, the integrand evaluations are not the bottleneck of this integral, which motivates us to look strictly at the integration time for the benchmarks, and ignore the overhead of loading prefactors. Despite these measures, we still see that the resolved version of the integral evaluates more slowly than when using contour deformation. The reason is that, as a consequence of our chosen resolution procedure,
the resolved integrands have singularities at both the upper and the lower integration boundary. To handle this, \pysecdec splits the integrals and remaps all singularities to the lower boundary, which results in more sectors and thereby increased evaluation time. For the positive contributions, $J^{+,1}_{\text{tri}}, J^{+,2}_{\text{tri}}$ and $J^{+,3}_{\text{tri}}$, the singularities are only present at one of the boundaries, and the splitting could either be turned off, or avoided by applying simple transformations of the type $x_i \to 1-x_i$ before passing them to \pysecdec. This reduced the number of sectors significantly, and brought the integration time of the resolved integral close to the time of integrating with contour deformation. For the negative contribution, $J^{-}_{\text{tri}}$, singularities are present at both boundaries, and splitting the integral is therefore necessary. The small remaining time difference can mostly be attributed to the extra sector this split generates. Usually, dealing with extra sectors due to the resolution procedure is not a major issue, since the resolved integrands typically scale much better. The problem with the triangle, is that it is too easy to integrate in either case, and the scaling with respect to number of lattice points becomes irrelevant. In the other massive examples we also require integral splitting, but then the contour deformed integrals are eventually unable to reach higher precision, and we are able to access more digits with the resolved integrands.

\begin{figure}[h!]
\centering
    \includegraphics[width=0.9\textwidth]{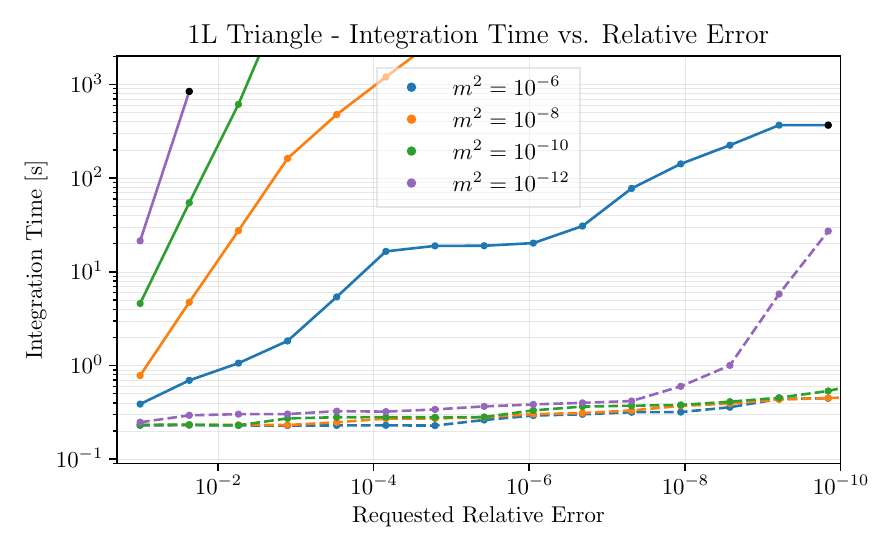}
    \caption{Timings with (solid lines) and without (dashed lines) contour deformation for the all-massive 1-loop triangle, expanded up to order $\epsilon^4$. Evaluated for different values of $m^2$ with $s_{12}=1$ fixed. The black dot on the CD $m^2 = 10^{-12}$ line indicates that no higher accuracy could be reached within 5 hours.}
    \label{fig:triangle1L3m_timings_masses}
\end{figure}

The situation is different when looking at the small-mass regime. 
The three-mass triangle integral is finite, however, in the limit $m^2 \rightarrow 0$ the integral develops poles in $\epsilon$ up to $1/\epsilon^2$. 
Therefore, when the integral is evaluated with small mass values, we numerically approach an end-point singularity which pinches the contour close to the boundaries of integration. And indeed, we observe in Figure~\ref{fig:triangle1L3m_timings_masses} that integrating with contour deformation gets progressively worse for smaller values of the internal mass while the resolved integral is stable across all values of $m^2$. For example, already for $m^2 = 10^{-6}$ we observe a factor of $\frac{368}{0.45} \approx 818 \, \times$ speed-up when avoiding contour deformation. For smaller values of $m^2$ the difference is even larger already for lower precision levels. For $m^2 = 10^{-12}$ we almost get no convergence with contour deformation while the resolved integral remains mostly stable. This demonstrates that the resolution procedure can be useful for simple examples too, albeit more so in extreme kinematic configurations. 

\subsection{2-Loop Elliptic Sunrise}
\begin{figure}[h!]
\centering
    \includegraphics[width=0.9\textwidth]{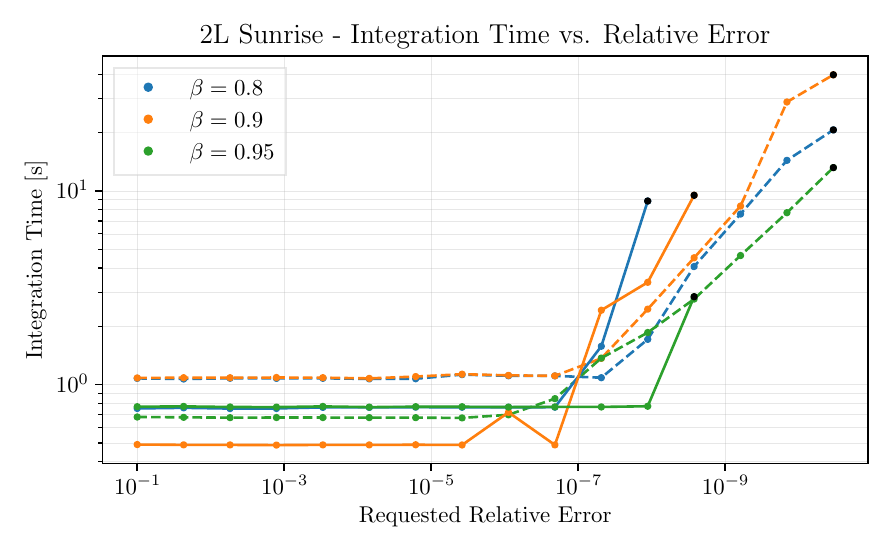}
    \caption{Timings with (solid lines) and without (dashed lines) contour deformation for the all massive 2L elliptic sunrise, expanded up to order $\epsilon^4$. Evaluated for different values of $\beta$ with $m=2$ fixed. The black dots indicate points at which the evaluation time exceeded 10 hours.}
    \label{fig:elliptic-sunrise-timings}
\end{figure}

\begin{figure}[h!]
\centering
    \includegraphics[width=0.9\textwidth]{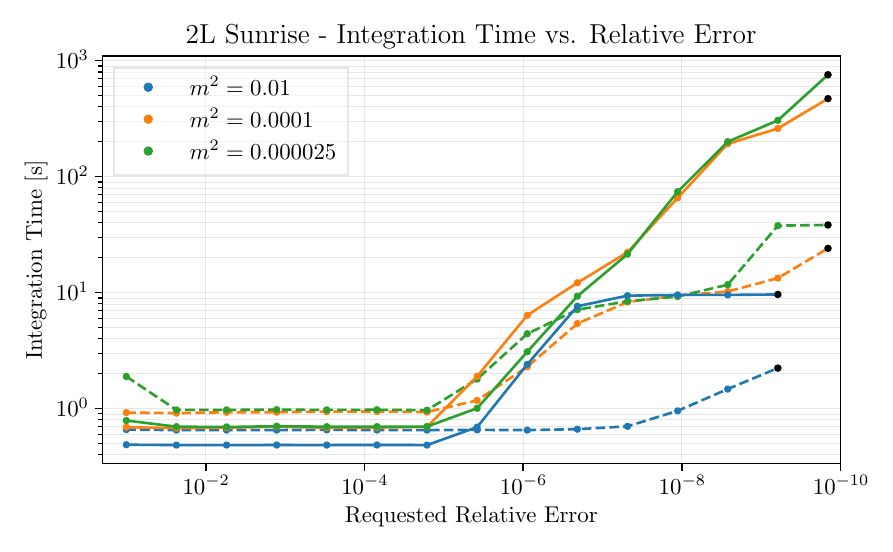}
    \caption{Timings with (solid lines) and without (dashed lines) contour deformation for the all massive 2L elliptic sunrise, expanded up to order $\epsilon^4$. Evaluated for different values of $m^2$ with $s_{12}=1$ fixed. The black dots indicate points after which the evaluation time diverged, and the integration was terminated after $>10$ hours.}
    \label{fig:elliptic-sunrise-timings_masses}
\end{figure}

Figure~\ref{fig:elliptic-sunrise-timings} shows the integration times for the 2-loop massive sunrise, with and without contour deformation, for increasing values of the centre-of-mass energy $\beta$ with $m=2$ fixed. Figure~\ref{fig:elliptic-sunrise-timings_masses} shows the same but for smaller values of the internal mass $m$, keeping $s=1$ fixed instead. This integral is similar to the massive triangle in that the integration time for a small lattice size is not much faster without contour deformation. The difference here is that for a fixed lattice size, using contour deformation does not yield the same precision as the resolved integrals do. This hints at the variance of the resolved integrands being smaller than the variance of the contour deformed integrands. This is more clearly demonstrated in Table~\ref{tab:elliptic} where the actual relative errors at different QMC lattice sizes are displayed. The relative errors are estimated by:
\begin{equation}
    \varepsilon_{\text{rel}} = \sqrt{\frac{(\text{Re}(\varepsilon_{\text{abs}})^2 + \text{Im}(\varepsilon_{\text{abs}})^2)}{(\text{Re}(I)^2 + \text{Im}(I)^2)}},
\end{equation}
where $\varepsilon_{\text{abs}}$ is the absolute error estimated by the QMC integrator from random shifts, and $I$ is the QMC estimate for the integral. The appearance of the timing figures can be understood by looking at the results for the higher $\epsilon$-orders. The results from the contour deformed integrals reach a minimum error that is several digits larger than the results from the resolved integrals, and increasing the lattice size fails to improve the situation. This means that besides improving integration time, the new procedure additionally allows us to obtain more digits than could ever be obtained when integrating with contour deformation. 

\begin{table}
    \centering
    \begin{tabular}{c|c|c c c c}
        \multicolumn{2}{c|}{\textsubscript{$\epsilon$-order}\textbackslash\textsuperscript{points}}
        & $\mathbf{10^3}$ & $\mathbf{10^4}$ & $\mathbf{10^5}$ & $\mathbf{10^6}$ \\
        \toprule
        \multirow{2}{*}{$\mathbf{\epsilon^0}$}
& CD &
\multicolumn{1}{c|}{$3.41\cdot10^{-10}$} &\multicolumn{1}{c|}{$4.84\cdot10^{-14}$} &\multicolumn{1}{c|}{$5.86\cdot10^{-15}$} &\multicolumn{1}{c|}{$8.39\cdot10^{-16}$} \\ 
& No CD &
\multicolumn{1}{c|}{$3.13\cdot10^{-10}$} &\multicolumn{1}{c|}{$2.47\cdot10^{-14}$} &\multicolumn{1}{c|}{$1.72\cdot10^{-16}$} &\multicolumn{1}{c|}{$3.35\cdot10^{-16}$} \\ 
\midrule 
\multirow{2}{*}{$\mathbf{\epsilon^1}$}
& CD &
\multicolumn{1}{c|}{$1.11\cdot10^{-9}$} &\multicolumn{1}{c|}{$2.87\cdot10^{-13}$} &\multicolumn{1}{c|}{$2.10\cdot10^{-13}$} &\multicolumn{1}{c|}{$1.08\cdot10^{-13}$} \\ 
& No CD &
\multicolumn{1}{c|}{$1.13\cdot10^{-9}$} &\multicolumn{1}{c|}{$8.91\cdot10^{-14}$} &\multicolumn{1}{c|}{$6.53\cdot10^{-14}$} &\multicolumn{1}{c|}{$9.88\cdot10^{-14}$} \\ 
\midrule 
\multirow{2}{*}{$\mathbf{\epsilon^2}$}
& CD &
\multicolumn{1}{c|}{$3.49\cdot10^{-9}$} &\multicolumn{1}{c|}{$1.86\cdot10^{-12}$} &\multicolumn{1}{c|}{$7.57\cdot10^{-13}$} &\multicolumn{1}{c|}{$5.60\cdot10^{-12}$} \\ 
& No CD &
\multicolumn{1}{c|}{$3.24\cdot10^{-9}$} &\multicolumn{1}{c|}{$2.87\cdot10^{-13}$} &\multicolumn{1}{c|}{$4.44\cdot10^{-13}$} &\multicolumn{1}{c|}{$7.99\cdot10^{-13}$} \\ 
\midrule 
\multirow{2}{*}{$\mathbf{\epsilon^3}$}
& CD &
\multicolumn{1}{c|}{$9.22\cdot10^{-9}$} &\multicolumn{1}{c|}{$9.02\cdot10^{-11}$} &\multicolumn{1}{c|}{$7.19\cdot10^{-10}$} &\multicolumn{1}{c|}{$2.47\cdot10^{-8}$} \\ 
& No CD &
\multicolumn{1}{c|}{$6.91\cdot10^{-9}$} &\multicolumn{1}{c|}{$6.05\cdot10^{-12}$} &\multicolumn{1}{c|}{$1.88\cdot10^{-12}$} &\multicolumn{1}{c|}{$5.30\cdot10^{-12}$} \\ 
\midrule 
\multirow{2}{*}{$\mathbf{\epsilon^4}$}
& CD &
\multicolumn{1}{c|}{$1.48\cdot10^{-8}$} &\multicolumn{1}{c|}{$8.75\cdot10^{-10}$} &\multicolumn{1}{c|}{$4.44\cdot10^{-7}$} &\multicolumn{1}{c|}{$1.33\cdot10^{-8}$} \\ 
& No CD &
\multicolumn{1}{c|}{$7.90\cdot10^{-9}$} &\multicolumn{1}{c|}{$1.77\cdot10^{-11}$} &\multicolumn{1}{c|}{$5.31\cdot10^{-12}$} &\multicolumn{1}{c|}{$1.47\cdot10^{-11}$} \\
        \bottomrule
    \end{tabular}
    \caption{Relative errors for the massive 2-loop sunrise integral with and without contour deformation for different QMC lattice sizes, at orders $\epsilon^0$, $\epsilon^1$, $\epsilon^2$, $\epsilon^3$ and $\epsilon^4$, with kinematics $\beta=0.8$ and $m=2$.}
    \label{tab:elliptic}
\end{table}

\subsection{3-Loop Hyperelliptic Banana}
\begin{figure}[h!]
\centering
    \includegraphics[width=0.9\textwidth]{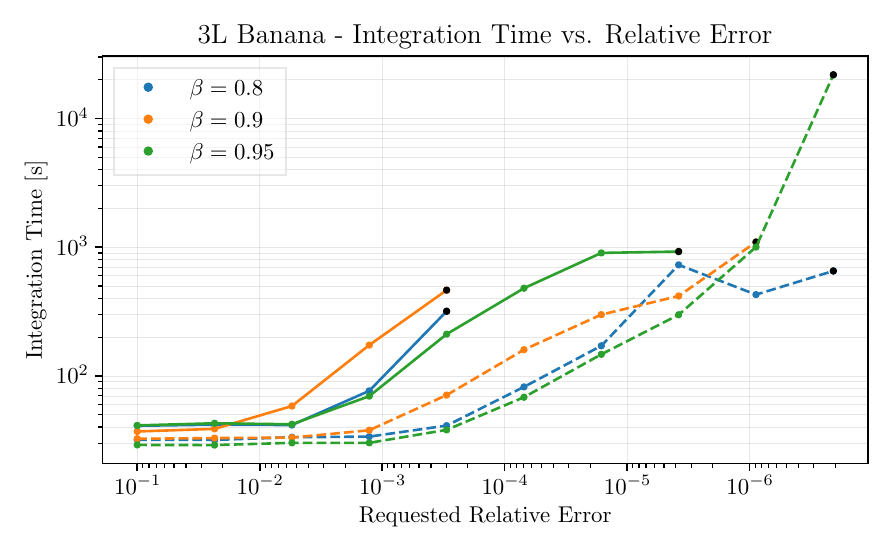}
    \caption{Timings with (solid lines) and without (dashed lines) contour deformation for the all massive 3L banana, expanded up to order $\epsilon^4$. Evaluated for different values of $\beta$ with $m=2$ fixed. The black dots indicate points after which the evaluation time diverged, and the integration was terminated after $>10$ hours.}
    \label{fig:banana-timings}
\end{figure}

\begin{figure}[h!]
\centering
    \includegraphics[width=0.9\textwidth]{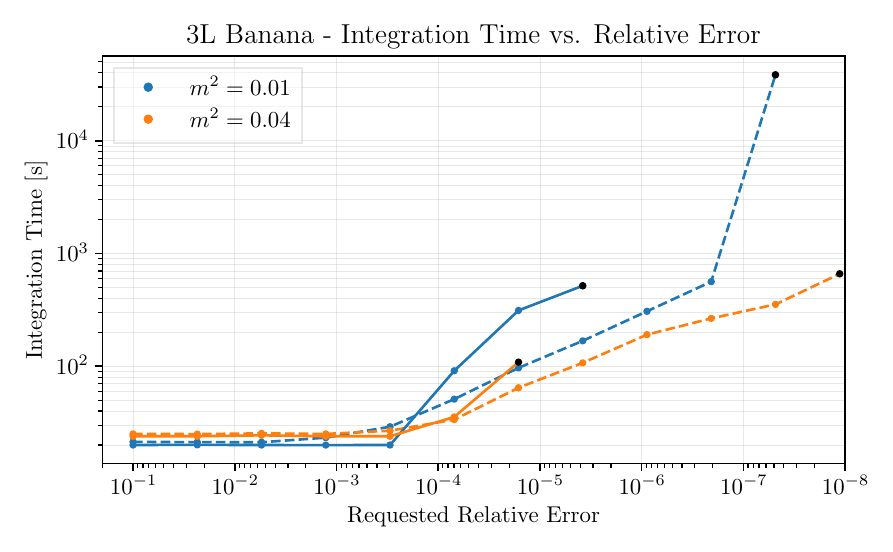}
    \caption{Timings with (solid lines) and without (dashed lines) contour deformation for the all massive 3L banana, expanded up to order $\epsilon^4$. Evaluated for different values of $m^2$ with $s_{12}=1$ fixed. The black dots indicate points after which the evaluation time diverged, and the integration was terminated after $>10$ hours.}
    \label{fig:banana-timings_masses}
\end{figure}

Figure~\ref{fig:banana-timings} shows the integration times for the 3-loop massive banana, with and without contour deformation, for increasing values of the centre-of-mass energy $\beta$ with $m=2$ fixed. Figure~\ref{fig:banana-timings_masses} shows the same but for smaller values of the internal mass $m$, keeping $s=1$ fixed instead. For this integral we essentially observe the same results as for the massive sunrise, with the integration times for initial lattice sizes even being slightly faster for the resolved integrands in this case. Again, we obtain more digits than what is possible with contour deformation, as is demonstrated clearly in Table~\ref{tab:banana}. The situation is the same in the small-mass regime, where we observe that the resolved integrals obtain many more digits than contour deformation manages before hitting the machine precision limit. 

\begin{table}
    \centering
    \begin{tabular}{c|c|c c c c}
        \multicolumn{2}{c|}{\textsubscript{$\epsilon$-order}\textbackslash\textsuperscript{points}}
        & $\mathbf{10^3}$ & $\mathbf{10^4}$ & $\mathbf{10^5}$ & $\mathbf{10^6}$ \\
        \toprule
        \multirow{2}{*}{$\mathbf{\epsilon^0}$}
& CD &
\multicolumn{1}{c|}{$2.37\cdot10^{-4}$} &\multicolumn{1}{c|}{$1.43\cdot10^{-8}$} &\multicolumn{1}{c|}{$7.80\cdot10^{-11}$} &\multicolumn{1}{c|}{$1.96\cdot10^{-11}$} \\ 
& No CD &
\multicolumn{1}{c|}{$3.92\cdot10^{-5}$} &\multicolumn{1}{c|}{$2.88\cdot10^{-8}$} &\multicolumn{1}{c|}{$5.96\cdot10^{-10}$} &\multicolumn{1}{c|}{$7.79\cdot10^{-10}$} \\ 
\midrule 
\multirow{2}{*}{$\mathbf{\epsilon^1}$}
& CD &
\multicolumn{1}{c|}{$6.19\cdot10^{-4}$} &\multicolumn{1}{c|}{$3.81\cdot10^{-8}$} &\multicolumn{1}{c|}{$5.15\cdot10^{-9}$} &\multicolumn{1}{c|}{$1.34\cdot10^{-8}$} \\ 
& No CD &
\multicolumn{1}{c|}{$1.01\cdot10^{-4}$} &\multicolumn{1}{c|}{$8.90\cdot10^{-8}$} &\multicolumn{1}{c|}{$1.44\cdot10^{-9}$} &\multicolumn{1}{c|}{$3.15\cdot10^{-9}$} \\ 
\midrule 
\multirow{2}{*}{$\mathbf{\epsilon^2}$}
& CD &
\multicolumn{1}{c|}{$9.26\cdot10^{-4}$} &\multicolumn{1}{c|}{$1.41\cdot10^{-6}$} &\multicolumn{1}{c|}{$6.06\cdot10^{-6}$} &\multicolumn{1}{c|}{$1.06\cdot10^{-5}$} \\ 
& No CD &
\multicolumn{1}{c|}{$1.86\cdot10^{-4}$} &\multicolumn{1}{c|}{$1.98\cdot10^{-7}$} &\multicolumn{1}{c|}{$1.67\cdot10^{-8}$} &\multicolumn{1}{c|}{$1.18\cdot10^{-8}$} \\ 
\midrule 
\multirow{2}{*}{$\mathbf{\epsilon^3}$}
& CD &
\multicolumn{1}{c|}{$1.51\cdot10^{-3}$} &\multicolumn{1}{c|}{$6.89\cdot10^{-6}$} &\multicolumn{1}{c|}{$4.89\cdot10^{-5}$} &\multicolumn{1}{c|}{$7.21\cdot10^{-5}$} \\ 
& No CD &
\multicolumn{1}{c|}{$3.69\cdot10^{-4}$} &\multicolumn{1}{c|}{$4.70\cdot10^{-7}$} &\multicolumn{1}{c|}{$5.64\cdot10^{-8}$} &\multicolumn{1}{c|}{$4.98\cdot10^{-7}$} \\ 
\midrule 
\multirow{2}{*}{$\mathbf{\epsilon^4}$}
& CD &
\multicolumn{1}{c|}{$5.05\cdot10^{-3}$} &\multicolumn{1}{c|}{$1.46\cdot10^{-3}$} &\multicolumn{1}{c|}{$5.21\cdot10^{-4}$} &\multicolumn{1}{c|}{$2.36\cdot10^{-2}$} \\ 
& No CD &
\multicolumn{1}{c|}{$4.74\cdot10^{-4}$} &\multicolumn{1}{c|}{$1.17\cdot10^{-6}$} &\multicolumn{1}{c|}{$6.16\cdot10^{-7}$} &\multicolumn{1}{c|}{$6.41\cdot10^{-7}$} \\ 
        \bottomrule
    \end{tabular}
    \caption{Relative errors for the hyperelliptic banana integral with and without contour deformation for different QMC lattice sizes, at orders $\epsilon^0$, $\epsilon^1$, $\epsilon^2$, $\epsilon^3$ and $\epsilon^4$, with kinematics $\beta=0.8$ and $m=2$.}
    \label{tab:banana}
\end{table}

\subsection{Cancellations}

Since the resolved integrals are split into a sum of several contributions, there could, in principle, be a drop in precision on the full integrals due to cancellations. 
The benchmarks in the previous section already take this into account by considering the relative error on the full integral\footnote{This is automated within the \texttt{sum\_package} module of \pysecdec.}. 
We therefore already implicitly know that if there are any cancellations, they are not severe for the kinematic configurations considered so far.
In this section, we demonstrate that severe cancellations do not occur for non-zero terms in the $\epsilon$ expansion of the integrals as we scan over the phase space.
However, our resolution procedure can (in fact, must in some cases) generate spurious poles in $\epsilon$, which then cancel upon summation. 
These spurious singularities nevertheless do not degrade the performance of our resolved integrals as significantly as using contour deformation.

\begin{figure}[h!]
\centering
    \includegraphics{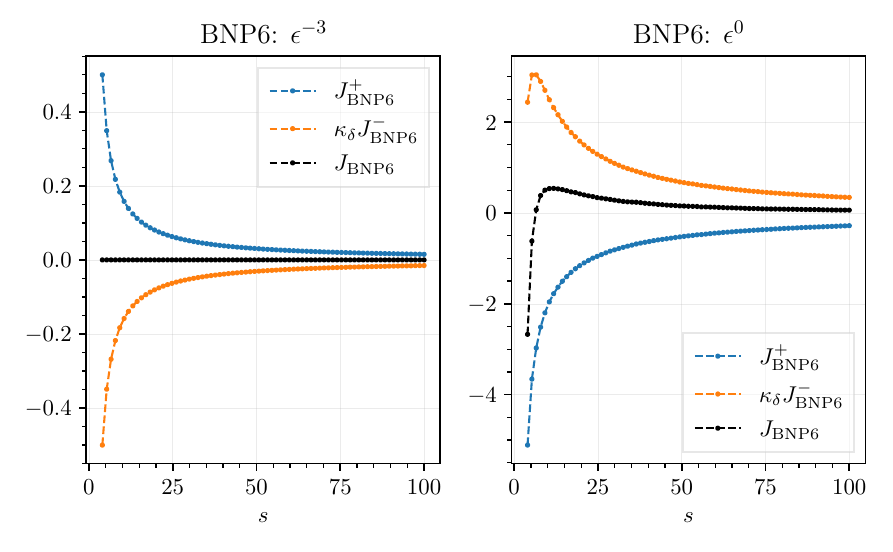}
    \caption{Magnitude of the real part of the positive and negative contributions compared to the total integral for BNP6 at orders $\epsilon^{-3}$ and $\epsilon^0$. $\kappa_{\delta} = \lim_{\delta \to 0^+}(-1-i\delta)^{-2-2\epsilon}$. The $\epsilon^{-3}$ pole is spurious and is a consequence of the cancellation between $J_{\mathrm{BNP6}}^+$ and $\kappa_{\delta} J_{\mathrm{BNP6}}^-$. $s \in (4, 100)$ and $s_{23}=-1$ fixed.}
    \label{fig:BNP6-cancellations}
\end{figure}

\begin{figure}[h!]
\centering
    \includegraphics{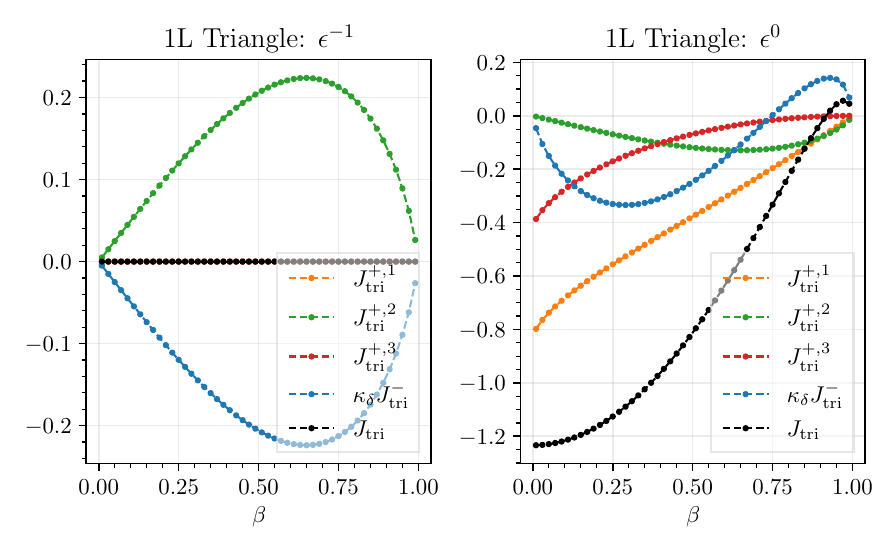}
    \includegraphics{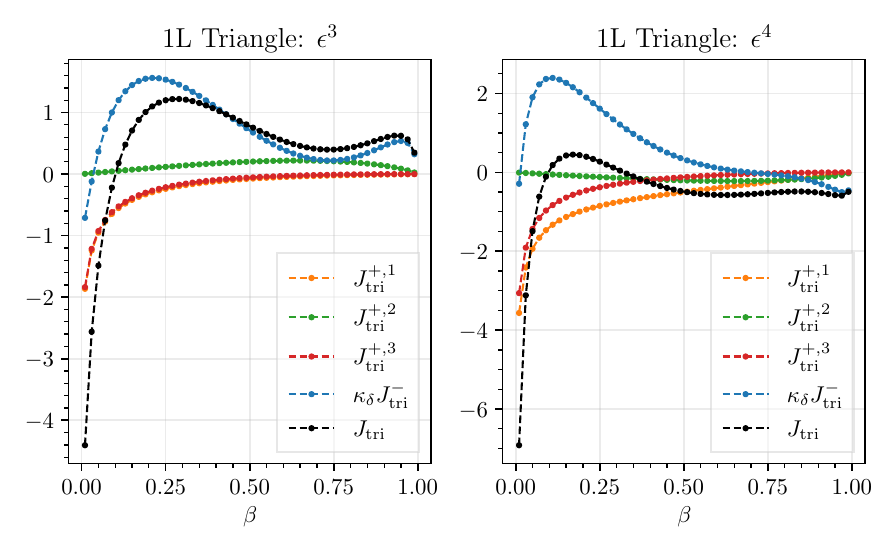}
    \caption{Magnitude of the real part of the positive and negative contributions compared to the total integral for the all massive 1-loop triangle at orders $\epsilon^{-1}$, $\epsilon^0$, $\epsilon^3$ and $\epsilon^4$. $\kappa_{\delta}~=~\lim_{\delta \to 0^+}(-1-i\delta)^{-1-\epsilon}$. The $1,2,3$ indices corresponds to the different positive regions shown in Figure~\ref{fig:triangleregions}. $\beta \in (0.01, 0.99)$ and $m=1$ fixed.}
    \label{fig:triangle1L3m-cancellations}
\end{figure}

In Figure~\ref{fig:BNP6-cancellations}, we show the positive and negative contributions to the real parts of the $\epsilon^{-3}$ and $\epsilon^0$ orders of BNP6 (massless 2-loop non-planar box), for values of $s_{12} \in (4, 100)$ with $s_{23} = -1$ fixed. 
For the $\epsilon^0$ coefficient, the cancellation between the positive and negative pieces is relatively small, remaining $\mathcal{O}(1)$ for most points.
In the small $s$ limit, where we approach the limit of the validity of our resolution, the integral is dominated by $J^+_\mathrm{BNP6}$, with no significant cancellation occurring between the different terms.
Instead, in the high-energy limit for $s \sim 100$, we do observe cancellations of 1 digit of precision.
The final result for the BNP6 integral has poles up to and including $\epsilon^{-2}$. 
However, after resolution, the integral develops a spurious $\epsilon^{-3}$ pole which vanishes due to cancellation between the positive and negative contributions\footnote{In fact, for this integral, additional spurious $\epsilon^{-3}$ poles appear also during the sector decomposition of the original integral when using contour deformation.}.
The same type of cancellation happens for the finite massive triangle, where a spurious $\epsilon^{-1}$ pole appears and is cancelled due to $J_{\mathrm{tri}}^{+,2}$ and $J_{\mathrm{tri}}^{-}$, which can be seen in Figure~\ref{fig:triangle1L3m-cancellations}. 
One can understand why this pole must appear by looking at the expansion of the prefactor to the negative contribution, which for the 1-loop triangle is
\begin{equation}
    \lim_{\delta \to 0^+}(-1-i\delta)^{-1-\epsilon} = -1 - i \,\pi \, \epsilon + \mathcal{O}(\epsilon^2). 
\end{equation}
We know that this integral is finite and complex at $\epsilon^0$. This means that $J_{\mathrm{tri}}^{-}$ must necessarily have a $\epsilon^{-1}$ contribution such that a complex part can be generated at leading order (the positive contributions are purely real). 
It follows that some of the positive contributions must also have a $\epsilon^{-1}$ contribution to cancel the pole and leave the result finite. 
Coefficients of spurious poles can not be numerically integrated to high relative precision (though good absolute precision can be obtained). For the benchmarks in this paper, we skip the integration of the coefficients of poles that we know are spurious. In practice, the order of the leading pole might not be known, and in such cases, the integration must be terminated based on an absolute error instead.
In future work, it would be interesting to explore to what extent the cancellation of spurious singularities between resolved regions can be arranged locally in parameter space.
This research direction is simplified in our approach by the fact that we know these cancellations occur precisely on an integration boundary, and we retain knowledge of which boundary in each resolved integral contributes to the spurious singularity.
Achieving a local cancellation would eliminate the appearance of spurious poles, allowing the finiteness of integrals to be manifestly maintained and removing any associated performance impact. 
We remark that this situation is potentially related to the more general problem of the appearance of spurious poles in the context of sector decomposition.

\begin{figure}[h!]
\centering
    \includegraphics{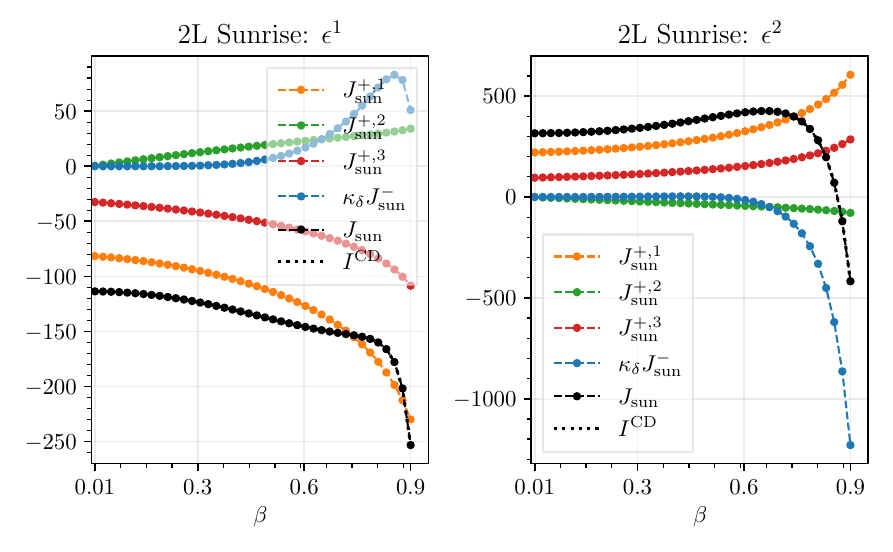}
    \includegraphics{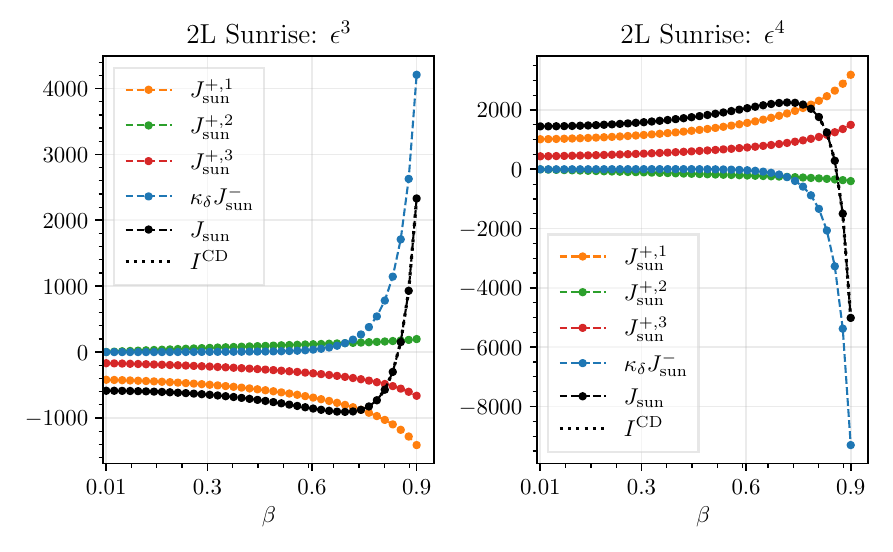}
    \caption{Magnitude of the real part of the positive and negative contributions compared to the total integral for the elliptic sunrise at orders $1,2,3,4$ in the $\epsilon$ expansion. $\kappa_{\delta} = \lim_{\delta \to 0^+}(-1-i\delta)^{-1-2\epsilon}$. The $1,2,3$ indices corresponds to the different positive regions shown in \ref{fig:ellipticregions}. $\beta \in (0.01, 0.9)$ and $m=2$.} 
    \label{fig:sunrise-cancellations}
\end{figure}

Figure~\ref{fig:triangle1L3m-cancellations} also shows the interplay between the contributions at orders $\epsilon^0$, $\epsilon^3$ and $\epsilon^4$ for the 1-loop triangle integral. 
We see explicitly that there are no severe cancellations in this case either. 
Finally, Figure~\ref{fig:sunrise-cancellations} shows the magnitude of the one negative and three positive contributions to the 2-loop massive sunrise. In this case we observe that at each order the value of the full integral diverges as $\beta \to 1$. However, there are still no significant cancellations between the resolved pieces. Additionally, we include the result from using contour deformation to demonstrate that it agrees at each point with the sum of the resolved contributions, also in the high-energy regime where contributions become very large. 

\section{Implication for Integrals in Parameter Space}
\label{sec:eval}
The work presented here, as demonstrated in Section~\ref{sec:performance}, is immediately relevant for numerical approaches relying on the use of parameter space.
So far, we have focused on the impact in the context of sector decomposition-based approaches; however, this concept is also immediately relevant for avoiding contour deformation in tropical geometry approaches to computing Feynman integrals~\cite{Borinsky:2020rqs,
Borinsky:2023jdv}.
Other numerical approaches, such as those using integrable neural networks to approximate the integrand~\cite{Maitre:2022xle}, which are known to work for Euclidean integrals but are not as straightforward to apply to integrals in the Minkowski regime, are also likely to benefit from this work.

We anticipate that our work will have implications beyond the strict application of accelerating the direct numerical computation of Feynman integrals. 
In this section, we summarise some observations and connections to other approaches in the literature that we believe would be interesting to explore.

Other, quite conceptually different, numerical or semi-numerical methods for computing Feynman integrals are also plagued or completely obstructed by the presence of physical thresholds.
For example, in Ref.~\cite{Zeng:2023jek}, an approach to approximating/computing Feynman integrals using positivity constraints was presented.
The method was demonstrated to perform well for integrals in the Euclidean region, relying on the fact that the integrand is guaranteed to have the same sign in this region.
Our work suggests that any Feynman integral (even those without a Euclidean region) can be cast in this form; therefore, it would be interesting to investigate the performance of this technique on our integrands.
In this context, we emphasise that it would be essential (or at least extremely helpful) to have a robust implementation of integration-by-parts identities directly in parameter space~\cite{Bitoun:2017nre,Chen:2019mqc,Chen:2019fzm,Chen:2020wsh,Artico:2023jrc}.
In Ref.~\cite{Borowka:2018dsa}, a method was introduced to compute Feynman integrals both above and below thresholds based on Taylor expansion and subsequent integration in parameter space.
The method applies to a wide variety of integrals and avoids the need to numerically integrate the parameter space representation; however, as demonstrated by the authors of that work, the performance in some cases is significantly reduced by the presence of thresholds.
We speculate that this large performance impact is caused by the presence of $\mathcal{F}=0$ type singularities, which our procedure would remove.

Finally, we observe that the representation we target, see Eq.~\eqref{eq:decomp}, makes the analytic continuation of any integral manifest; it is controlled entirely by a factor of $(-1-i\delta)$ raised to the power of the second Symanzik polynomial, or, equivalently, an overall phase for each Feynman integral.
If we imagine writing an entire amplitude in terms of real resolved integrals, then the analytic continuation of any scattering process becomes trivial~\cite{Hannesdottir:2022bmo}.
Furthermore, at the leading order in the $\epsilon$ expansion for finite integrals, the integrands appearing in our resolution are strictly non-negative.
It would be interesting to explore if being able to obtain such a representation for a complete amplitude has any connection to concepts such as positive geometries, see e.g. Refs.~\cite{Henn:2024qwe,Herrmann:2022nkh}, and if any of the additional requirements important to this programme, such as complete monotonicity, could be satisfied.
In Ref.~\cite{Britto:2023rig}, it was argued that taking $\mathcal{F}(\mathbf{x};\mathbf{s})=0$ as a boundary of integration leads to maximal (and non-maximal) cuts/discontinuities, in general, it would be interesting to explore the interpretation of the integrals obtained from our approach.
We also remark that, naively, a modification of our resolution procedure could be applicable in the context of the Method of Regions~\cite{Beneke:1997zp,Jantzen:2011nz}.
Specifically, it has been shown that dissecting the domain of integration on the Landau singularities can reveal regions that are not visible using geometric methods based on Newton polytopes, which neglect the possibility of cancellations between monomials~\cite{Jantzen:2012mw,Gardi:2024axt} whereas our method can naturally account
for such cancellations by construction.
Having a systematic method for mapping mixed-sign polynomials to same-sign polynomials for which the geometric approach to the Method of Regions applies provides a potentially useful step in fully systematising the Method of Regions in parameter space.

\section{Conclusions}
\label{sec:conclusions}
In this work, we propose a representation of dimensionally regulated parameter integrals in the Minkowski regime as a sum of non-negative parameter integrals, which can be evaluated directly through numerical integration without deformation into the complex domain.
This provides an avenue for removing the need for contour deformation in the context of Feynman parameter integration, avoiding one of the most significant obstacles to obtaining precise and stable numerical results.
In this representation, the analytic continuation of Feynman integrals becomes trivial; it is entirely encoded in a simple prefactor.

We have provided an algorithm for explicitly constructing our proposed representation for a class of integrals which we refer to as \emph{univariate bisectable} integrals.
We consider example Feynman integrals at 1-loop up to 5-point and 2- and 3-loops up to 4-point, demonstrating both that our positive integrand representation can be obtained and that it can dramatically improve the performance of traditional numerical approaches to computing such integrals in the Minkowski regime (above physical thresholds), achieving an acceleration of many orders of magnitude and allowing previously intractable integrals to be straightforwardly evaluated numerically.
We also consider a set of massive Feynman integrals up to 3-loops which are not \emph{univariate bisectable}, and we demonstrate that positive integrands can be obtained for elliptic and hyperelliptic cases, indicating that these underlying integral geometries do not necessarily obstruct the construction of positive integrands.

The method presented here is immediately applicable in the context of sector decomposition (as implemented in e.g. FIESTA~\cite{Smirnov:2008py,Smirnov:2009pb,Pak:2010pt,Smirnov:2013eza,Smirnov:2015mct,Smirnov:2021rhf} and \pysecdec~\cite{Borowka:2017idc,Borowka:2018goh,Heinrich:2021dbf,Heinrich:2023til}) and tropical geometry~\cite{Borinsky:2020rqs} (as implemented in FeynTrop~\cite{Borinsky:2023jdv}).
With the obstacle of contour deformation removed, several new bottlenecks in the standard numerical computation of Feynman
integrals become apparent, we expect that addressing these bottlenecks without the added complexity previously introduced by contour deformation will allow existing tools to be further improved.
We also anticipate that the techniques presented here may be beneficial for other approaches to computing/approximating Feynman integrals, such as TAYINT~\cite{Borowka:2018dsa} and positivity constraints~\cite{Zeng:2023jek}.

The present work serves as a proof-of-principle.
However, as noted in Section~\ref{ssec:method_beyond_ub}, a general algorithm capable of obtaining the representation we target exists.
In future work, it would be interesting to explore automation focusing on the massless (linear) case and then the massive (quadratic) case, such that the procedure presented here could be fully automated and implemented within public codes for computing Feynman integrals.
In this context, it may prove fruitful to further investigate potential complications in the procedure, for example, by studying integrals with different underlying geometries (as given in e.g. Ref.~\cite{Duhr:2025lbz}).
The connection between the integrals that we obtain and maximal/non-maximal cuts, which can be obtained by taking $\mathcal{F}(\mathbf{x};\mathbf{s})=0$ as a boundary of integration~\cite{Britto:2023rig}, is also an interesting direction to explore.

\acknowledgments

We thank all members of the \pysecdec collaboration, Bakar Chargeishvili, Gudrun Heinrich, Matthias Kerner, Vitaly Magerya, and Johannes Schlenk for their constructive input and support of this work.
It is our pleasure to thank Bakul Agarwal, Einan Gardi, Franz Herzog, Yao Ma, Eric Panzer, Melih Ozcelik, and Bahriye Ilhan Jones for their helpful discussions and comments.
Furthermore, we thank Bakar Chargeishvili, Einan Gardi and Gudrun Heinrich for valuable comments on our initial draft of this work.
This research was supported in part by the Deutsche Forschungsgemeinschaft (DFG, German Research Foundation) under grant 396021762 - TRR 257, and by the UK Science and Technology Facilities Council under contract ST/X000745/1. 
SJ is supported by a Royal Society University Research Fellowship (Grant URF/R1/201268).
SJ and TS are grateful to the Galileo Galilei Institute for hospitality and support during the scientific program on ``Theory Challenges in the Precision Era of the Large Hadron Collider'', where part of this work was conducted.

\clearpage


\bibliographystyle{JHEP}
\bibliography{main}

\end{document}